%% file: main.tex
\documentclass{article}
\usepackage[english]{babel}
\usepackage{amsmath}
\usepackage{amssymb}
\usepackage{graphicx}
\usepackage{listings}
\usepackage{pdfpages}
\usepackage{tikz}
\usetikzlibrary{calc}
\usepackage{xcolor}
\usepackage{ulem}
\usepackage{subcaption}
\usepackage{geometry}
\usepackage{hyperref}
\usetikzlibrary{shapes.geometric,calc,shadows.blur,decorations.pathmorphing,fit,decorations.pathreplacing}
\newcommand{\blue}[1]{\textcolor{blue}{#1}}

\title{Global cell-cell communication enables spatial segregation of cells in organoids of the inner cell mass}
\author{Simon Schardt and Sabine C. Fischer}
\date{\today}

\begin{document}

\maketitle

\begin{abstract}
    During development, cell fates are determined through a combination of intracellular transcriptional regulations and extracellular signaling. As a result, spatial patterns of different cell types arise. We investigate the decision between epiblast and primitive endoderm cells in the inner cell mass of the preimplantation mouse embryo. Our computational model uses global cell signaling for the pattern formation. By varying the signal dispersion, cell type arrangements ranging from a checkerboard to an engulfing pattern can be generated. Pair correlation functions provide a well-suited way of characterizing the model output. With these, we established a quantitative comparison between the simulation results and experimental data of inner cell mass organoids. We obtained an astonishing agreement. Thus, our model proves its capability to replicate the cell differentiation patterns, making global signaling a strong contender to explain pattern formation in the preimplantation embryo.
\end{abstract}

\maketitle

\section{Introduction}

Two successive cell fate decisions shape the preimplantation phase of mammalian embryos. During the first, cells become either part of the trophectoderm (TE) or the inner cell mass (ICM). The second takes place for cells of the ICM. ICM cells differentiate into either epiblast (Epi) or primitive endoderm cells (PrE). In the course of development, Epi cells give rise to the embryo proper, whereas PrE cells contribute to the yolk sac \cite{Lanner2014,Perez2014,Frum2015,Chazaud2016}.

For the mouse embryo as a model organism, the transcription factors NANOG and GATA6 have been identified as the earliest markers for the segregation of the ICM into Epi and PrE cells, respectively \cite{Mitsui2003, Schrode2014}. In early blastocysts, i.e. embryonic day 3 (E3.0) after fertilization, expressions are high for both NANOG and GATA6 in all cells. Subsequently, they are down-regulated until their expressions become mutually exclusive (E3.0-4.5). In addition to the cell fate decision, PrE and Epi precursor cells segregate spatially. In late blastocysts, PrE cells are mainly found adjacent to the blastocoel cavity, separated from the Epi cells. In a previous study, we described this transition from a local to a global pattern statistically \cite{Fischer2020}. This indicates that cell fates are linked not only to the fates of their neighboring cells alone, but also to their spatial position in the blastocyst. So far, the segregation process has not been fully unveiled. While cell sorting due to differential adhesion could provide one possible explanation \cite{Nissen2017}, another study has found no significant evidence to support this hypothesis \cite{Filimonow2019}.

ICM organoids, a model system based on mouse embryonic stem cells, provide another perspective on this issue \cite{Mathew2019}. Cells in ICM organoids are engineered to express high levels of both NANOG and GATA6. Most importantly, they have the ability to differentiate into PrE-like and Epi-like cells \cite{Schroeter2015}. ICM organoids thus provide a scenario similar to the mouse blastocyst. Due to their large cell number and the symmetry provided by their spherical structure, we focus on the data from ICM organoids.

Computational models have already been used to shed light on cell differentiation from different points of view. Chemical models describe the interactions between NANOG and GATA6, as well as other constituents. Their focus lies on the correct cell type proportions and the formation of a checkerboard pattern, i.e. a pattern in which one cell type avoids adjoining cells of the same type \cite{Bessonnard2014, Tosenberger2017}. We recently introduced a model derived from statistical mechanics that satisfies these properties \cite{Schardt2021}. Cell division models have shown, the capability to explain cluster formation \cite{Liebisch2020}. There have also been approaches where combinations of cell division, cell sorting and apoptosis lead to the desired configuration of cells \cite{Krupinski2011, Nissen2017, Saiz2020}. In summary, several areas have already been covered, but none of them address the spatial segregation using a signal-driven organization of cells.

In this study, we use a computational model to explain the pattern formation induced by cell differentiation in ICM organoids. ICM organoids that matured for 48h hours exhibit a radially distributed pattern of Epi- and PrE-like cells. One potential mechanism to explain the pattern formation is global cell-cell communication. This idea has already been explored to some extent \cite{Stanoev2021} and will constitute the central part of our study here. In recent findings on embryonic stem cells, the ability of fibroblast growth factor 4 (FGF4) to migrate over distances beyond the immediate cell neighbors has been demonstrated \cite{Raina2021}. Generalizing this idea, we describe cell-cell communication via chemical signals that disperse between cells. Our results show that the modification of signal dispersion alone suffices to display the checkerboard as well as engulfing patterns without help of cell sorting. To our knowledge, the resulting patterns have not yet been quantified. To this end, we use an individualized pair correlation function (PCF). The simulated patterns are then brought into comparison with experimental data. For the most part, we succeeded in establishing similarities between simulations and experimental data. Interactive visualizations containing all of the 48h (\blue{\url{https://schardts.github.io/Organoids48h}}) matured organoids from \cite{Mathew2019} provide more insight into the individual data.

\section{Methods}

\subsection{Equation system}

The idea in this section is to describe the regulation of NANOG and GATA6 inside a cell in terms of ordinary differential equations (ODEs). In our gene regulatory network (GRN),  NANOG and GATA6 mutually inhibit each other. At the same time, NANOG is activated by an external signal (Fig. \ref{fig: GRN & signal}). We already developed a mathematical model for cell fate specification in \cite{Schardt2021}. In this model, the interactions between NANOG and GATA6 were derived using ideas from statistical mechanics. At the core of the model are the binding probabilities for NANOG and GATA6. The production of NANOG $\boldsymbol{n}$ and GATA6 $\boldsymbol{g}$ depends on the respective condition of the associated binding site. If a transcription factor is bound, it is reproduced with constant reproduction rate $r_n$ or $r_g$, respectively. Exponential decay guarantees a finite lifetime of transcription factors with constant decay rates $\gamma_n$ and $\gamma_g$. Cell-cell communication occurs through the exchange of chemical signals $\boldsymbol{s}$ and increases the likelihood of NANOG binding. Under these conditions, for $M$ cells interacting with each other, the dynamics can be formulated as a coupled ODE system:
\begin{equation}
\label{eq: ODE system}
\begin{aligned}
    \frac{dn_i}{dt} &= r_n \frac{\eta_n n_i(1+\eta_s \eta_{ns} s_i)}{1 + \eta_n n_i(1+\eta_s \eta_{ns} s_i) + \eta_g g_i + \eta_s s_i} - \gamma_n n_i \\
    \frac{dg_i}{dt} &= r_g \frac{\eta_g g_i}{1 + \eta_n n_i(1+\eta_s \eta_{ns} s_i) + \eta_g g_i + \eta_s s_i} - \gamma_g g_i, \qquad i = 1,...,M.
\end{aligned}
\end{equation}
We decide the cell fate based on the steady state of a cell, i.e. when $\frac{dn_i}{dt} = 0 = \frac{dg_i}{dt}$. In \cite{Schardt2021}, we have already performed an analysis which guarantees that we get one of two different states. These are:
\begin{itemize}
    \item N+G--: High NANOG expression, low GATA6 expression (Epi precursor).
    \item N--G+: Low NANOG expression, high GATA6 expression (PrE precursor).
\end{itemize}

\subsection{Cell graph}

\begin{figure}[htbp]
\centering
\begin{subfigure}{.45\textwidth}
  \centering
  \includegraphics[width=\linewidth]{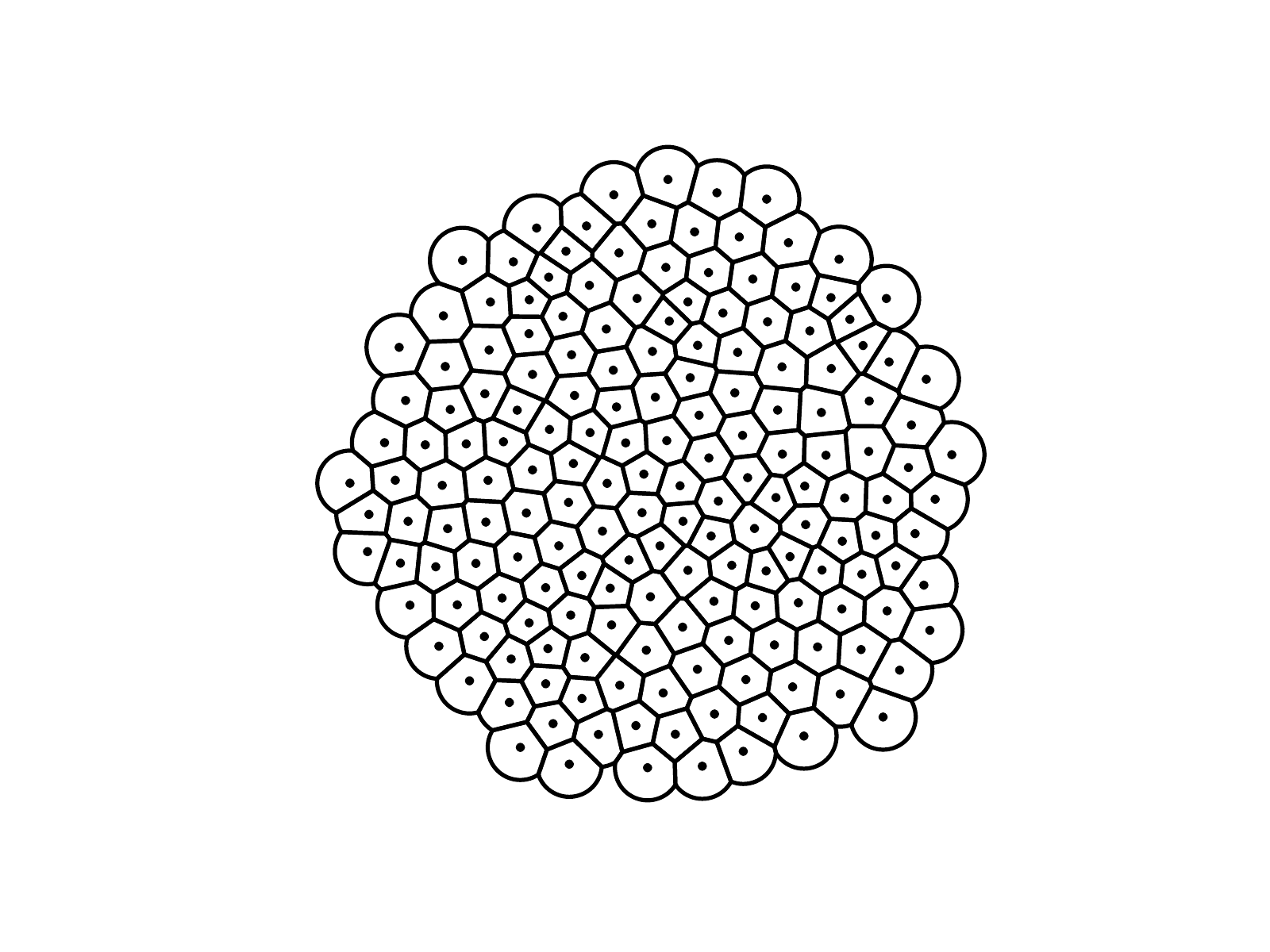}
  \subcaption{Organoid}
\end{subfigure}
\begin{subfigure}{.45\textwidth}
  \centering
  \includegraphics[width=\linewidth]{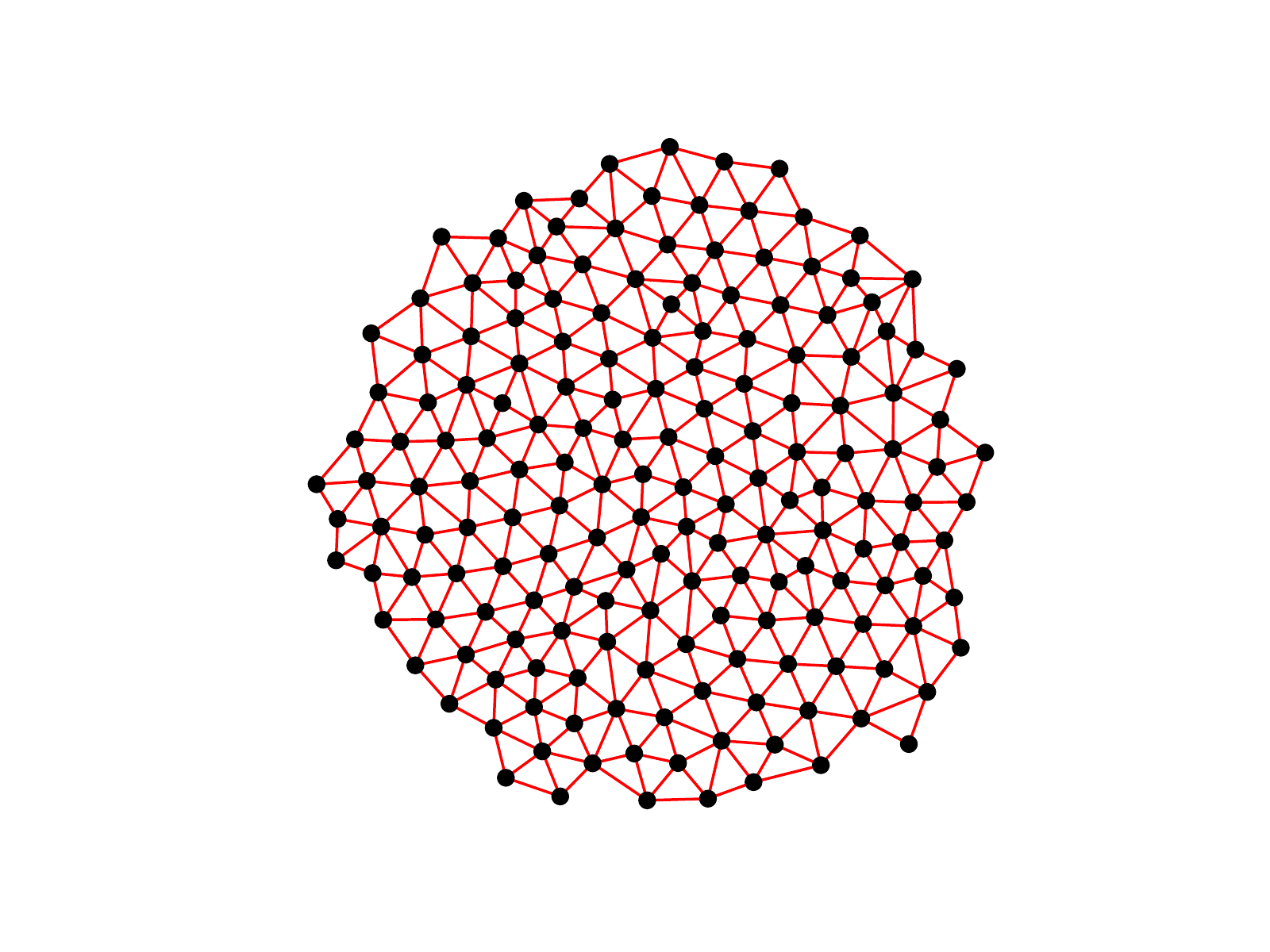}
  \subcaption{Cell graph}
\end{subfigure}
\caption{Visualization of an organoid with $177$ cells (a) and its corresponding cell graph (b). Black lines represent the cell membranes. The cell centroids are shown as black dots in both pictures. Red lines represent the edges, which provide information about which cells are in contact with each other.}
\label{fig: tissue}
\end{figure}

In our context, cells are represented by 2D/3D points in space with a fixed radius which is equal for all cells. The Delaunay cell graph provides a reliable indication of the neighborhood relationships of the cells \cite{Schmitz2017}. Therefore, we initialize our graph $G$ using the Delaunay triangulation. If the Euclidean distance between two cells exceeds the sum of their two radii, then the edge is removed from $G$, i.e. only cells in direct contact with each other are connected via an edge in $G$ (Fig. \ref{fig: tissue}). Edge weights are collectively set to $1$. We then define the cell distance $d_{ij}$ as the length of the shortest path between cells $i$ and $j$.

\subsection{Signal construction}

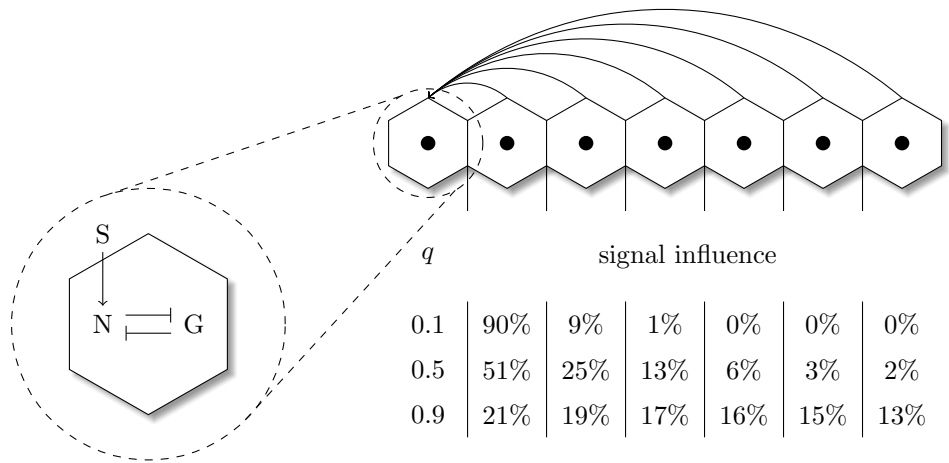
\begin{figure}[htbp]
\centering
\input{GRN.tex}
\caption{Illustration of the GRN represented by our model as well as an exemplary representation of the signaling in a one dimensional cell line. Inside the cell, NANOG and GATA6 mutually inhibit each other. Additionally, NANOG gets activated by an extracellular signal. This signal received by the first cell on the left of the line is the sum of all cell-cell communication between one cell and any other cell in the system. The table highlights how much each cell contributes to the received signal for different dispersions $q \in \{0.1,0.5,0.9\}$. Percentages are rounded to the nearest integer.}
\label{fig: GRN & signal}
\end{figure}

In our setting, internal regulations of cells are influenced by signals emitted by other cells. Depending on how a signal disperses in space, this means that not only directly neighboring cells can have an impact on a cell's fate, but possibly also the collective effect of cells further away. Here, we introduce a purely mathematical construct of the signal $\boldsymbol{s}$ which represents a measure of the influence cells have on each other. Following our previous study \cite{Schardt2021}, we define a signal depending on the GATA6 expression values of a cell. This time however, the effect of any other cell is incorporated. To this end, we define the signal as
\begin{equation}
\label{eq: signal}
    s_i = \left(\sum_{j\neq i}g_j q^{d_{ij}-1}\right)\bigg/\left(\max_k\sum_{j\neq k} q^{d_{kj}-1}\right), \qquad q \in [0,1].
\end{equation}
Here, we use the distances $d_{ij}$ from our cell graph. The weights $q^{d_{ij}-1}$ define the fraction of the signal that gets transported from cell to cell. Let e.g. $q=0.1$, then second nearest neighbors of a cell receive only $10\%$ of the signal of the direct neighbors (Fig. \ref{fig: GRN & signal}). The denominator in \eqref{eq: signal} is used for normalization  (see section \ref{sec: cell type proportions} for further details). The dispersion parameter $q$ enables us to describe the transition from a direct neighbor signal to an equally dispersed signal. For $q=0$, the weights become
\begin{equation}
    q^{d_{ij}-1} = 0^{d_{ij}-1} = \begin{cases}1, \quad \text{for } d_{ij}=1\\0, \quad\text{for } d_{ij}>1
\end{cases},
\end{equation}
meaning that the weights for all cells that are not directly in contact with the respective cell are $0$. Hence, the resulting situation is similar to the local signal in \cite{Schardt2021}. Alternatively, $q=1$ yields
\begin{equation}
    q^{d_{ij}-1} = 1^{d_{ij}-1} = 1.
\end{equation}
This describes the case of every cell having the same impact on other cells independent of the distance between them. In summary, there is a continuous transition from a next neighbor signal at $q=0$, through a distance based global signal for $q \in (0,1)$ to an evenly distributed signal at $q=1$.

\subsection{Pair correlation function}

Cell differentiation patterns in our case are the result of two different cell types arising in an organoid. Patterns with the same premise have already been quantified using pair correlation functions (PCFs) \cite{Binder2013}. We use a similar approach to quantify our patterns with a PCF depending on the cell distances $d_{ij}$. This requires counting different types of cell pairings for certain distances. Therefore, we introduce the sets: 
\begin{align}
S_k &= \left\{(i,j)\in \mathbb{N}^2: d_{ij} = k,\, 1 \leq i,j \leq M\right\} \\
S^n_k &= \left\{(i,j)\in S_k: n_i > g_i,\, n_j > g_j\right\} \\
S^g_k &= \left\{(i,j)\in S_k: g_i \geq n_i,\, g_j \geq n_j\right\} \\
T^n &= \left\{i \in \mathbb{N}: n_i > g_i,\, 1 \leq i \leq M\right\} \\
T^g &= \left\{i \in \mathbb{N}: g_i \geq n_i,\, 1 \leq i \leq M\right\}
\end{align}
The pairings of all N+G-- cells with distance $k$, $S^n_k$, are related to all possible pairings of the same distance $S_k$ by forming their ratios. Analogously, we perform the routine for N--G+ cell pairings $S^g_k$ to get
\begin{equation}
    r_{nn} = \frac{|S^n_k|}{|S_k|}, \qquad r_{gg} = \frac{|S^g_k|}{|S_k|}.
\end{equation}
These ratios alone will not suffice to compare the patterns for varying cell type proportions. Therefore, we normalize these by the probabilities of randomly picking two equal types of cells using the total number of N+G-- cells $T^n$ and N--G+ cells $T^g$
\begin{equation}
\label{eq: PCF normalization}
    p_{nn} = \frac{|T^n|(|T^n|-1)}{M(M-1)}, \qquad p_{gg} = \frac{|T^g|(|T^g|-1)}{M(M-1)}.
\end{equation}
Combined, the PCFs measure the ratios of N+G-- or N--G+ cell pairs within every possible distance normalized by the probability of finding these cell pairs, i.e.
\begin{align}
\label{eq: PCF n}
    \rho_n(k) &= \frac{r_{nn}}{p_{nn}} = \frac{|S^n_k|M(M-1)}{|S_k||T^n|(|T^n|-1)} \\
\label{eq: PCF g}
    \rho_g(k) &= \frac{r_{gg}}{p_{gg}} = \frac{|S^g_k|M(M-1)}{|S_k||T^g|(|T^g|-1)}.
\end{align}
For a uniformly distributed amount of N+G-- or N--G+ cells, the correlation function returns a value close to $1$ for every cell distance $k$. Consequently, deviations from $1$ yield information about how many more or how many fewer equal cell pairs are found in certain ranges.

\subsection{Data pre-processing of experimental data}
\label{sec: data pre-processing}

\begin{figure}[htbp]
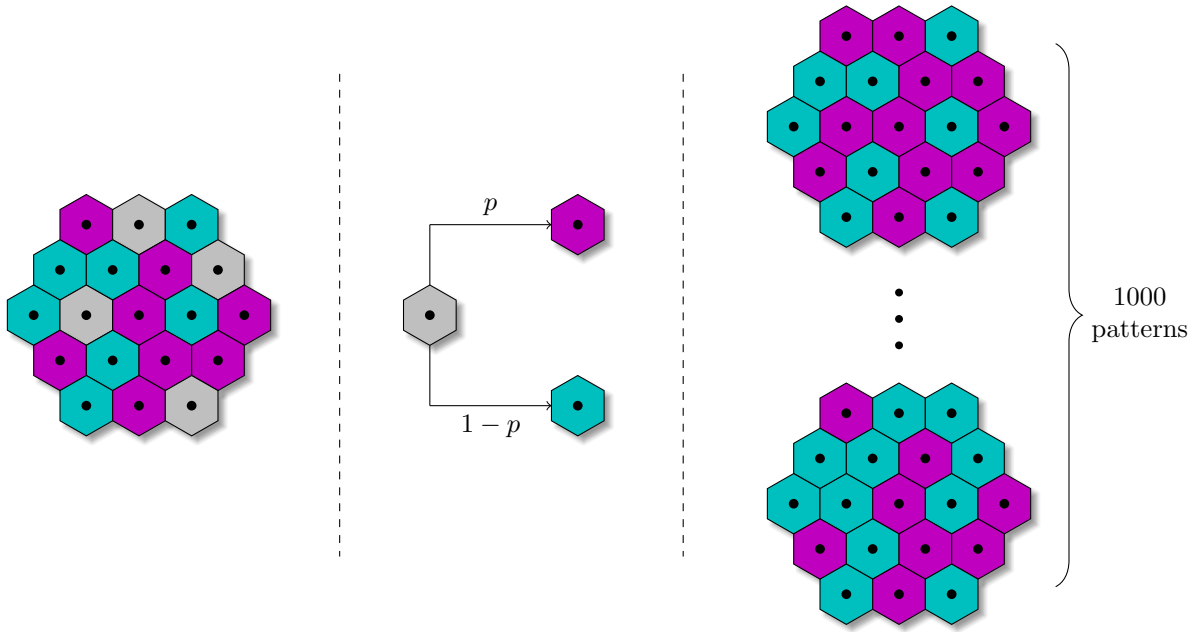

\centering
\include{experimental}
\caption{An illustration of the experimental data (left) shows three different cell types. N+G- cells are shown in magenta, N--G+ cells in cyan. Gray cells represent cells which are either DP or DN and are therefore not further identifiable. In the following (middle), each gray cell is randomly assigned a cell fate. Probability $p$ reflects the proportions of N+G-- cells found in the data. After $1000$ repetitions one obtains a variety of $1000$ different patterns with only two cell types (right).}
\label{fig: experimental data patterns}
\end{figure}

The main objective in this section is to prepare the data from \cite{Mathew2019} so that we can use it to characterize the organoids via PCFs. In this dataset, three-dimensional ICM organoids have been used as an alternative model organism to the mouse embryo. The cells were separated into four different cell types. N+G-- and N--G+ for cells that primarily express one protein. Double positive (DP) and double negative (DN) for cells where expressions of both NANOG and GATA6 are high or low, respectively. DP cells are considered as cells that have not undergone cell fate specification, yet. In contrast to this, DN cells have already moved on in development after cell fate specification \cite{Saiz2016}. Hence, we cannot classify these cells into Epi and PrE cells. For a correct cell neighborhood arrangement, we need to incorporate the DP and DN cells. Therefore, we do not obtain one graph per biological sample, but a range. For a given biological sample, we randomly assign Epi or PrE fate to each DP or DN cell. The probability $p$ to assign a cell with N+G-- fate equals the sum of all N+G-- cells divided by the sum of all N+G-- and N--G+ cells. The probability for assigning cells with N--G+ fate will then be $1-p$ (Fig. \ref{fig: experimental data patterns}). Then, we calculate the corresponding PCFs. We repeat this procedure $1000$ times. Picking the minimum and maximum values of those combinations at each distance provides an envelope of a pair correlation region for each biological sample.
Necessary for this is the distance matrix $D = (d_{ij})_{ij}$ given by the distances in the cell graph. For the cell graph, we again use the Delaunay triangulation. This time, we do not have the information about the cell radii. Therefore, we define a cutoff as the mean distance to neighboring cells of a single organoid plus two times its standard deviation. Any edge longer than this cutoff will be removed. In a normal distribution, this would approximately amount to the $2.28\%$ longest edges. Our distributions are slightly skewed to the right, which means that the percentage of removed edges will be a bit higher. The distances $d_{ij}$ are then computed via Dijkstra's algorithm for any pair of cells $i$ and $j$.

\subsection{Simulations on experimental data}
\label{sec: simulations on data}

The PCFs allow us to relate experimental data and simulation. To this end, we establish a direct comparison of the two scenarios. The data includes the $x$-, $y$- and $z$-coordinates of the centroids taken from each cell nucleus. We use this positional information as input to run simulations on the same cell positions. Our model is in no way affected by the transition from 2D to 3D data. Different values $q \in \{0.2, 0.4, 0.6, 0.8\}$ provide us with an overview of possible cell differentiation patterns. A bisection on the parameter space was used to adjust $-\Delta\varepsilon_g$ such that the cell type ratio (N+G--:N--G+) reflects the same ratio as the experimental data up to a tolerance of $10\%$.\\

\section{Results}

\subsection{Cell type proportions}
\label{sec: cell type proportions}
\begin{figure}[htbp]
\centering
\begin{subfigure}{.32\textwidth}
  \centering
  \includegraphics[width=\linewidth]{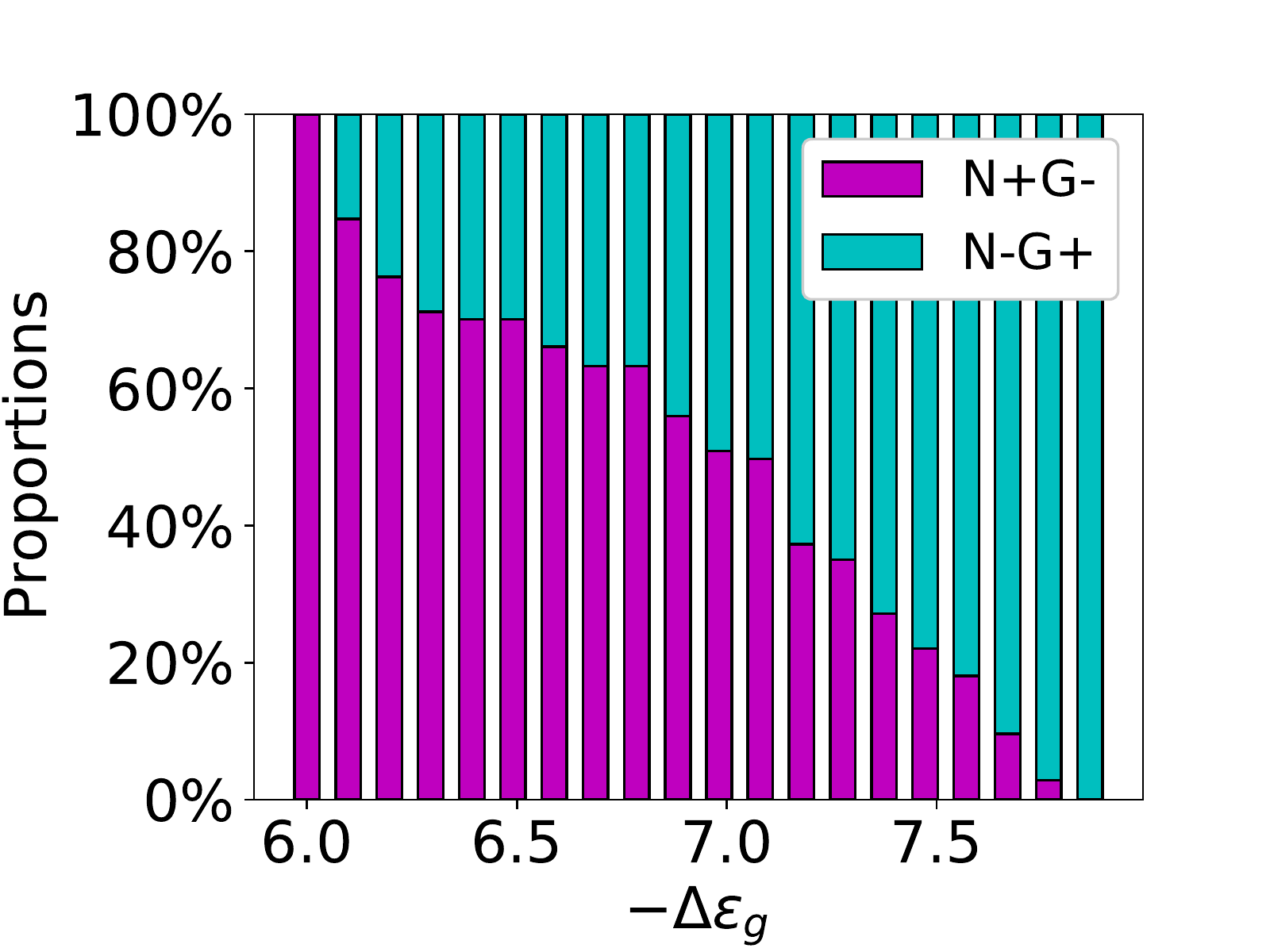}
  \subcaption{$q = 0.1$}
\end{subfigure}
\begin{subfigure}{.32\textwidth}
  \centering
  \includegraphics[width=\linewidth]{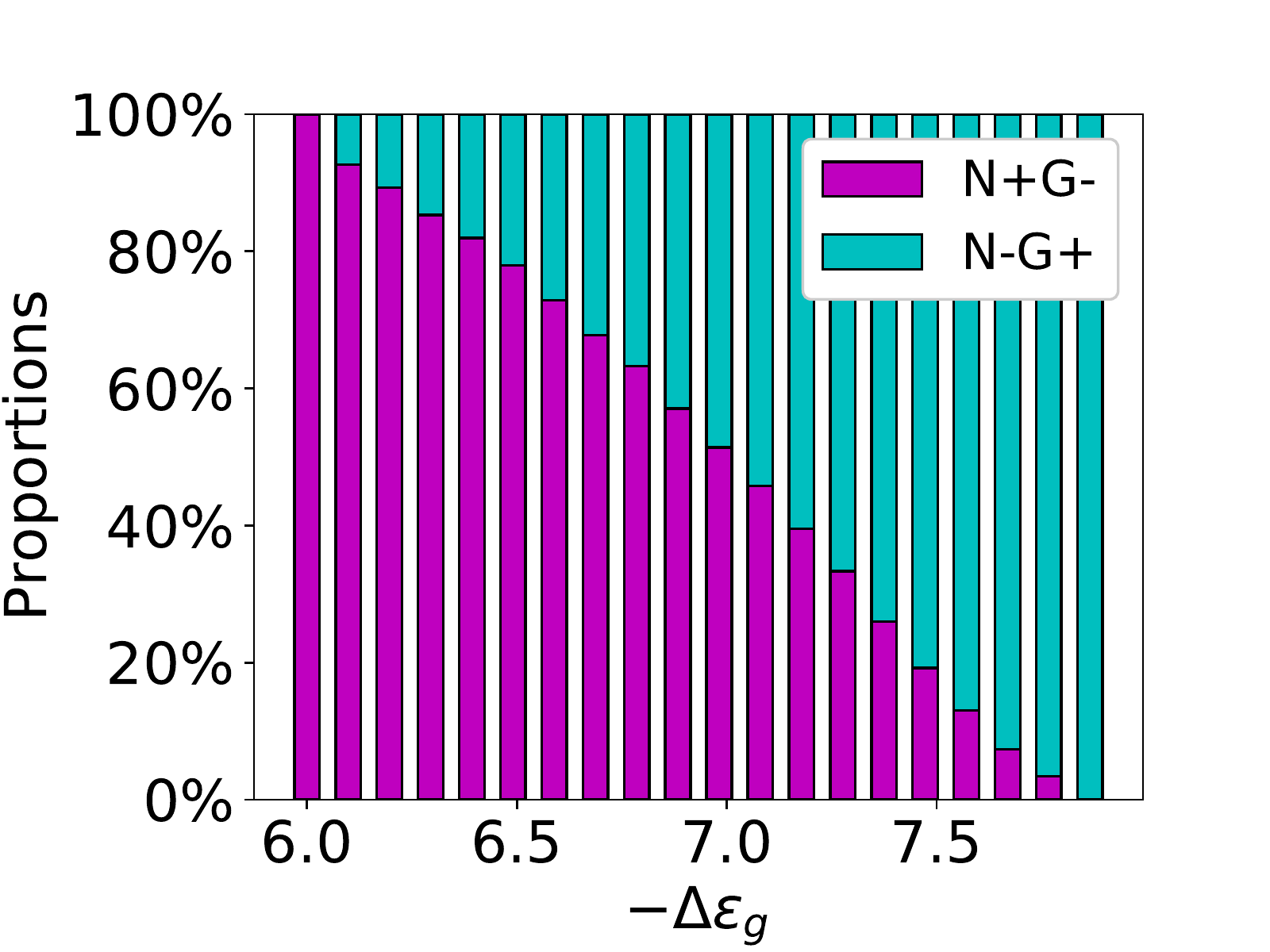}
  \subcaption{$q = 0.5$}
\end{subfigure}
\begin{subfigure}{.32\textwidth}
  \centering
  \includegraphics[width=\linewidth]{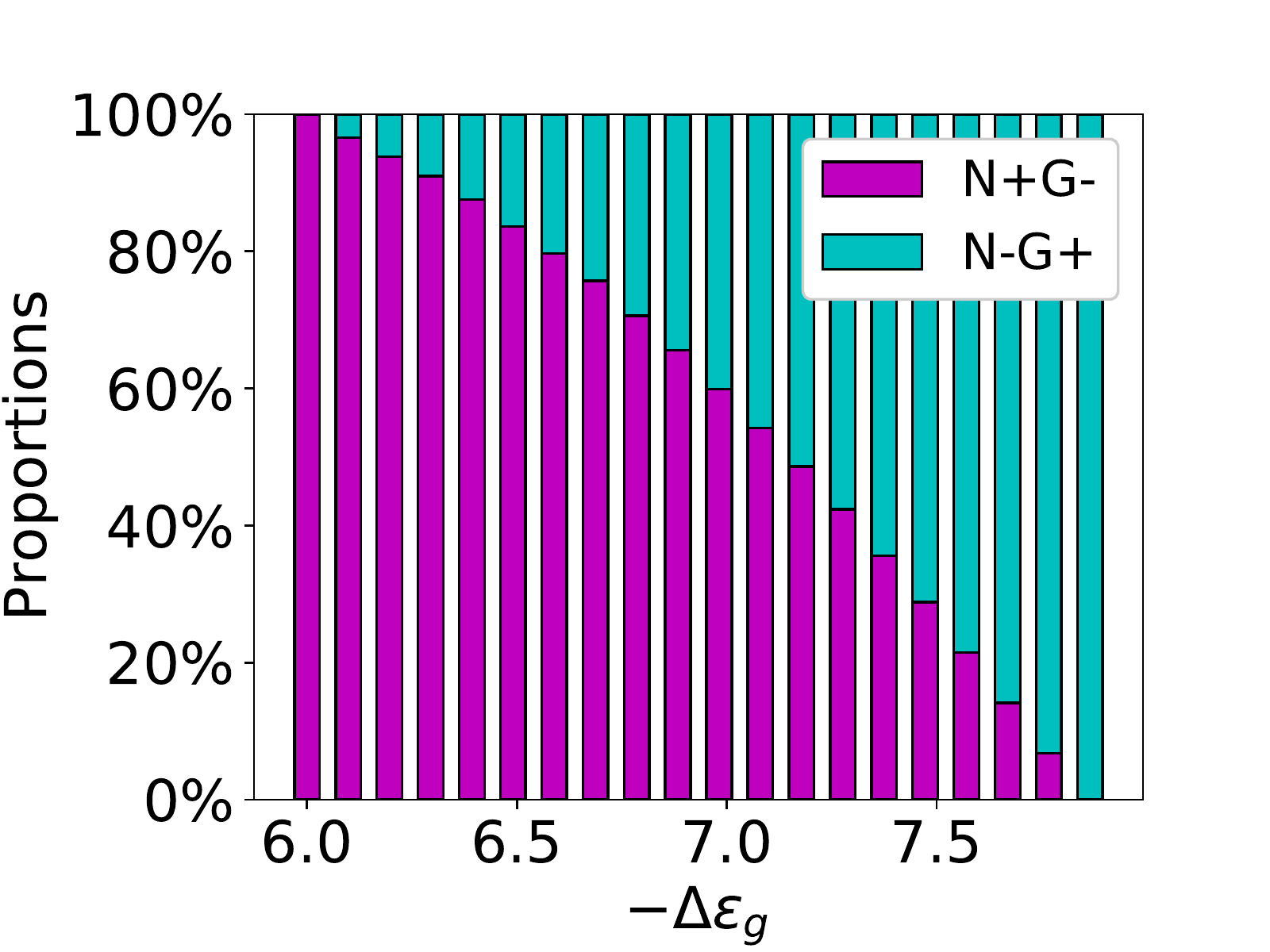}
  \subcaption{$q = 0.9$}
\end{subfigure}
\caption{Cell proportions with respect to $-\Delta\varepsilon_g$ calculated on the model geometry shown in Fig. \ref{fig: tissue}. Simulations were performed by dividing the stability interval for $-\Delta\varepsilon_g$ into $20$ equidistant values. Dispersion parameter $q$ varies increasingly from (a) to (c) resulting in different scenarios.}
\label{fig: proportions non-local}
\end{figure}

Hitting the correct proportions of N+G-- and N--G+ cells using a computational model is an important property, the model must be able to fulfill. This stems from the fact that cell proportions in the embryo are very precise and most likely an essential part of embryonic development \cite{Saiz2016, Saiz2020}. In our previous study \cite{Schardt2021}, we have already shown the model's capability to render different cell type proportions. In a stability analysis, we have found parameter restrictions that would always lead to steady states where either $n_i$ can be high and $g_i$ low (N+G--) or vice-versa (N--G+). This parameter restriction is given by an interval for energy differences $-\Delta \varepsilon_\alpha$, which relate to the coefficients in \eqref{eq: ODE system} as follows
\begin{equation}
    \eta_\alpha = e^{-\Delta\varepsilon_\alpha}, \qquad \alpha \in \{n, g, s, ns\}.
\end{equation}
The stability interval for $-\Delta\varepsilon_g$ then becomes
\begin{equation}
    \label{eq: stability interval}
    \Delta\varepsilon_{min} < -\Delta\varepsilon_g < \Delta\varepsilon_{max}
\end{equation}
with
\begin{align}
    \Delta\varepsilon_{min} &:= -\Delta\varepsilon_n + \ln\left(1+ e^{-\Delta\varepsilon_s-\Delta\varepsilon_{ns}} \min_i s_i\right) + \ln \left(\frac{r_n \gamma_g}{r_g \gamma_n}\right) \\
    \Delta\varepsilon_{max} &:= -\Delta\varepsilon_n + \ln\left(1+ e^{-\Delta\varepsilon_s-\Delta\varepsilon_{ns}} \max_i s_i\right) + \ln \left(\frac{r_n \gamma_g}{r_g \gamma_n}\right).
\end{align}
The steady states of \eqref{eq: ODE system} have been shown to be bound by $0$ and the ratio of reproduction and decay, i.e.
\begin{equation}
    0 \leq n_i \leq \frac{r_n}{\gamma_n}, \qquad 0 \leq g_i \leq \frac{r_g}{\gamma_g}.
\end{equation}
Thus, the normalization of the signal \eqref{eq: signal} allows us to formulate appropriate bounds:
\begin{equation}
     0 \leq \min_i s_i, \qquad \max_i s_i \leq \frac{r_g}{\gamma_g}
\end{equation}
Furthermore, by choosing the parameters exactly like in our previous study \cite{Schardt2021}, $-\Delta\varepsilon_n = 6$, $-\Delta\varepsilon_s = -\Delta\varepsilon_{ns} = 2$, $r_n = r_g = 1$ and $\gamma_n = \gamma_g = 10$, the minimum and maximum signal values are well approximated by their respective bounds. The resulting stability interval is
\begin{equation}
    \label{eq: stability interval values}
    -\Delta\varepsilon_g \in (6, 7.87).
\end{equation}
For different dispersion parameters $q$, the proportions of N+G-- show a monotonous decrease with increasing energy difference $-\Delta\varepsilon_g$ (Fig. \ref{fig: proportions non-local}). For low values of $q$, the proportions show some similarity to the local model in \cite{Schardt2021} due to individual larger jumps (Fig. \ref{fig: proportions non-local} (a)). These jumps become less pronounced for medium (Fig. \ref{fig: proportions non-local} (b)) and high dispersion (Fig. \ref{fig: proportions non-local} (c)). Moreover, the proportions for $q=0.9$ approach the theoretical proportions of an ideal geometry for the local model in \cite{Schardt2021}. Altogether, we have established full control over the cell type proportions.

\subsection{Pattern formation}
\begin{figure}[htbp]
\centering
\input{patterns}
\caption{Different patterns generated by the model on a tissue geometry with $177$ cells. Colors depict the values of $n_i$ in steady state. High values of $n_i$ correspond to low values in $g_i$ and vice-versa, i.e. magenta and cyan represent N+G-- and N--G+ cells, respectively. From left to right, $-\Delta\varepsilon_g$ increases. From bottom to top, the dispersion $q$ increases.}
\label{fig: patterns non-local}
\end{figure}
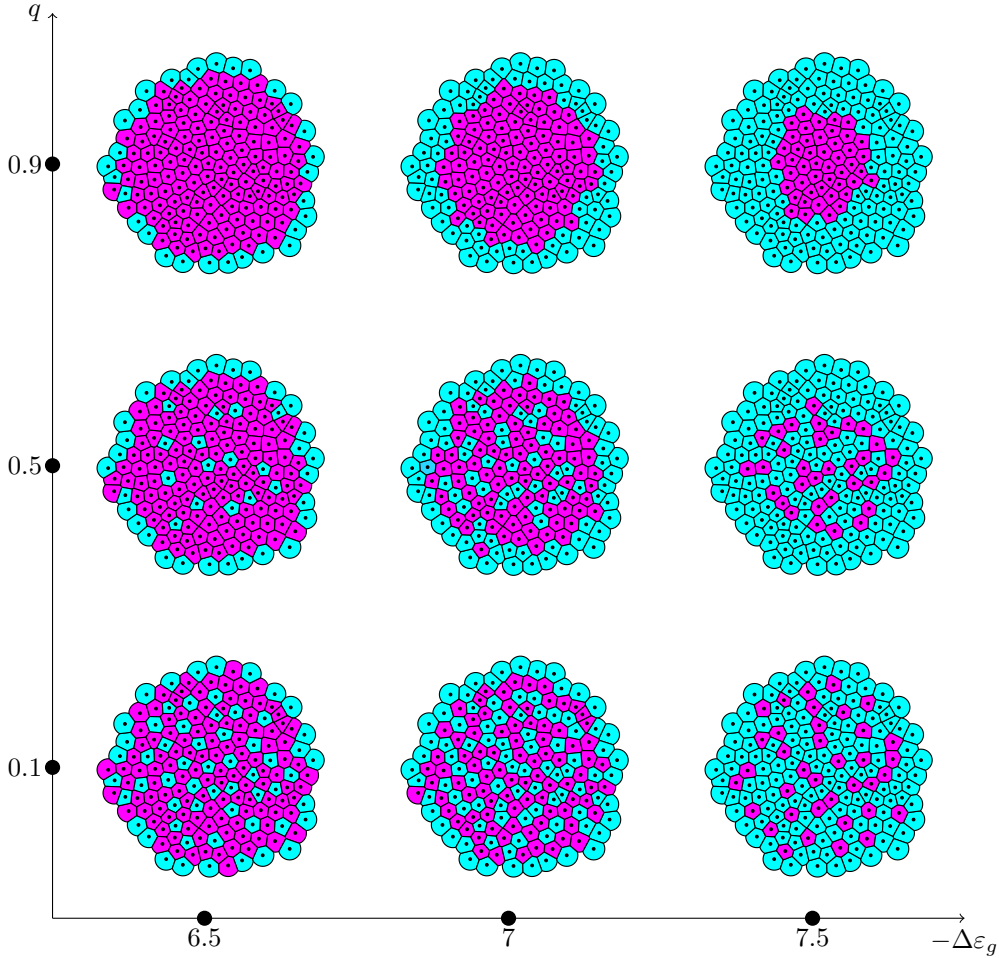

\begin{figure}[htbp]
\centering
\begin{subfigure}{.49\textwidth}
  \centering
  \includegraphics[width=\linewidth]{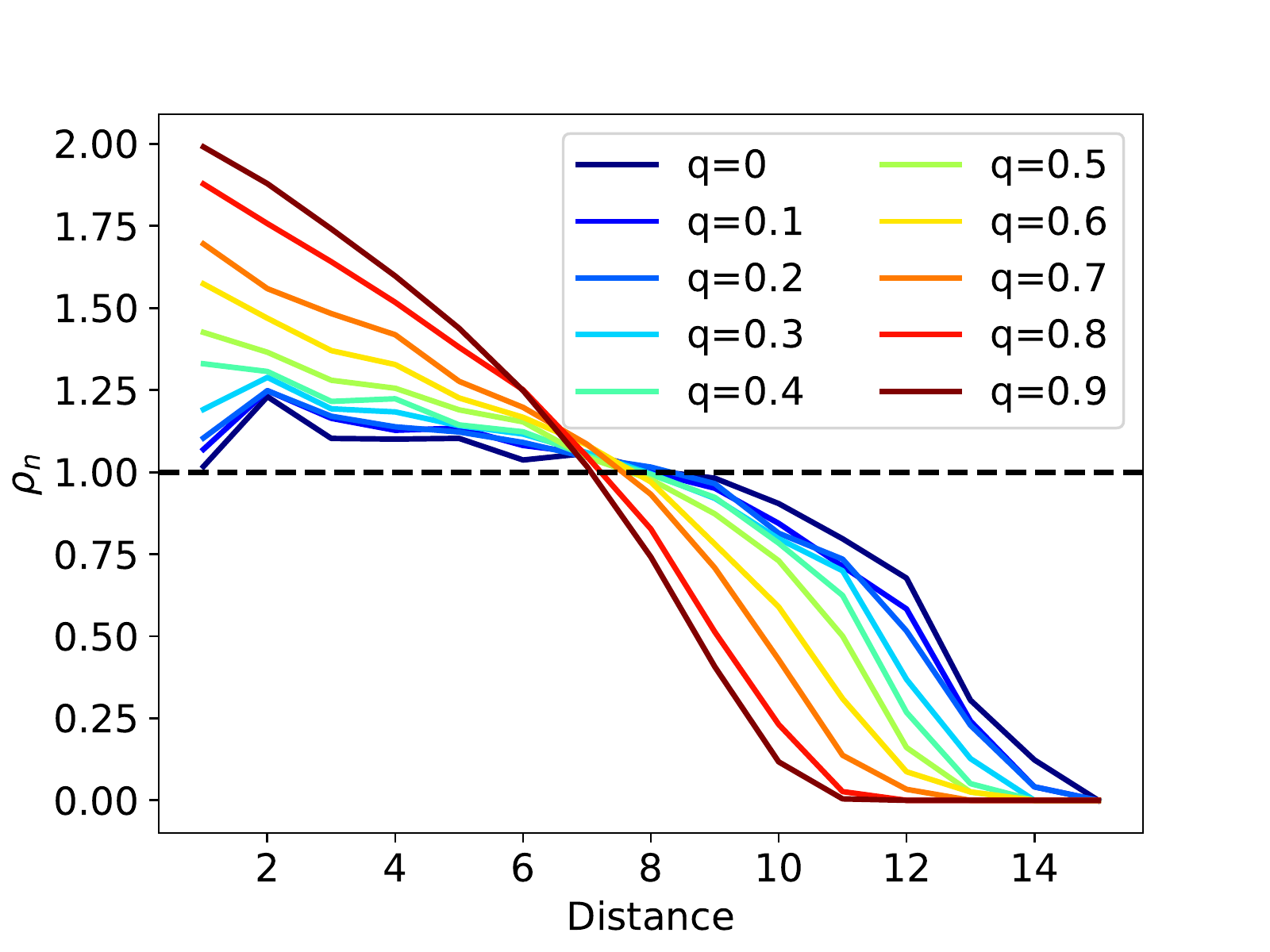}
\end{subfigure}
\begin{subfigure}{.49\textwidth}
  \centering
  \includegraphics[width=\linewidth]{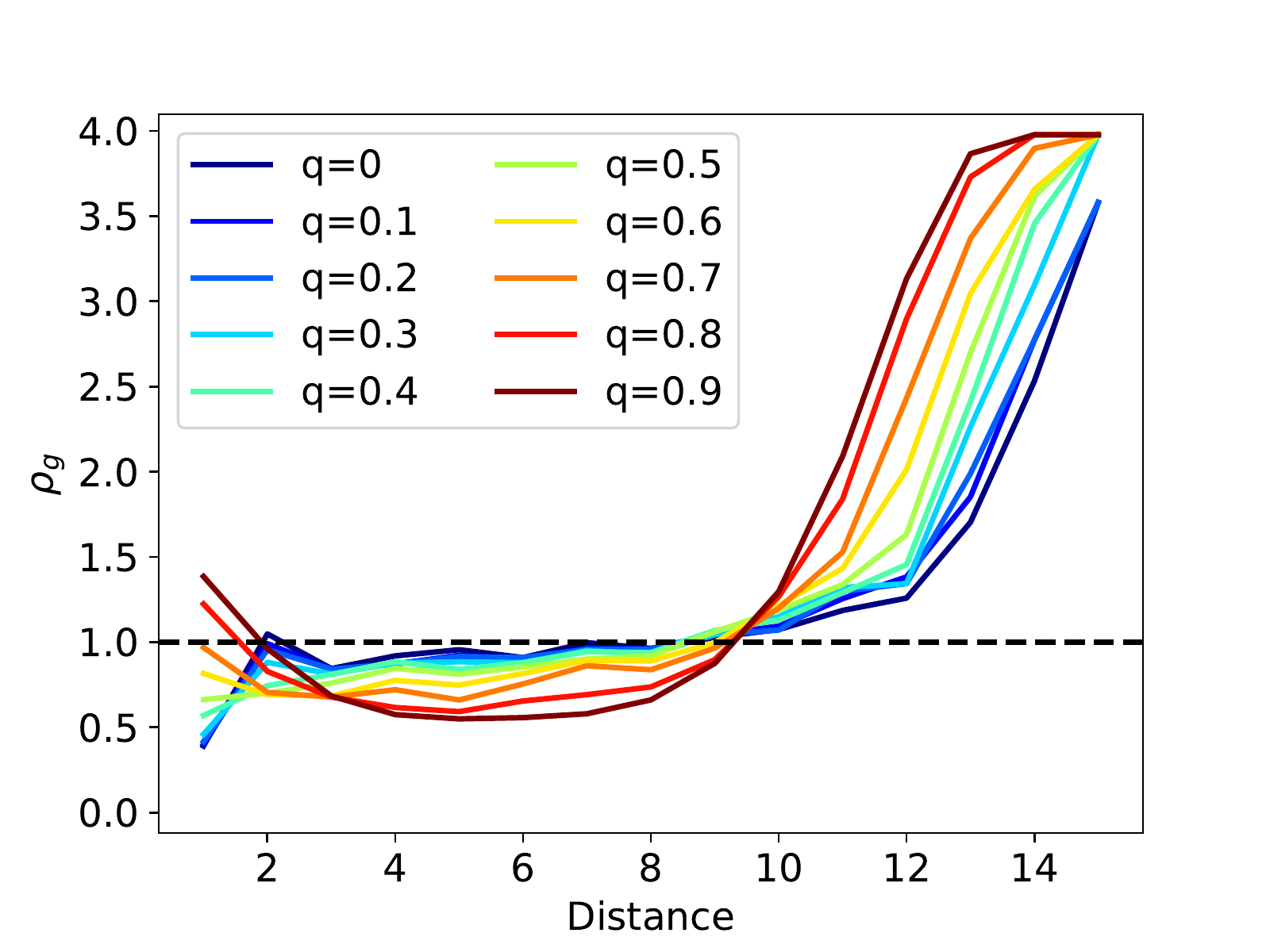}
\end{subfigure}
\caption{PCFs for N+G-- cells (left) and N--G+ cells (right) for different dispersion parameters $q$. Any PCF represents an organoid with a ratio of N+G--:N--G+ $= 88:89$. The dashed black line at $1$ resembles the PCF values of an ideal uniform distribution of two different cell types. If values lie above $1$ this means there are more pairs found at that distance. Consequently, values below $1$ mean fewer pairs.}
\label{fig: pair correlations non-local}
\end{figure}

Models that use an averaged local signal to describe the interaction between cells have already been employed in biological systems to create checkerboard patterns, i.e. two distinct types of cells appearing in alternating fashion \cite{Collier1996}. This idea has been used to generate checkerboard patterns for the mouse embryo \cite{Bessonnard2014, Tosenberger2017} or ICM organoids \cite{Schardt2021}. However, transitions from local to global patterns have been identified in the ICM \cite{Fischer2020}. Global models have only recently been investigated, but already show a promising approach to describing the emergence of these patterns \cite{Stanoev2021}. We build on this and run the simulations on our 2D organoids (Fig. \ref{fig: patterns non-local}). All simulations have been implemented in Python using Euler's method to solve the ODE system numerically until a steady state is established. With the exception of $-\Delta\varepsilon_g$ and $q$, the remaining parameter values remain fixed with $-\Delta\varepsilon_n = 6$, $-\Delta\varepsilon_s = 2$, $-\Delta\varepsilon_{ns} = 2$, $r_n = 0.1 = r_g$ and $\gamma_n = 1 = \gamma_g$. The variation of $-\Delta\varepsilon_g$ now again shows the change in the cell type proportions that we previously described in section \ref{sec: cell type proportions} (Fig. \ref{fig: patterns non-local}). This time, it becomes visible that the nature of the respective patterns is not affected by changing the proportions. The patterns generated for $q=0.1$ can mostly be considered of the checkerboard type. In contrast to \cite{Bessonnard2014, Tosenberger2017, Stanoev2021}, our signal \eqref{eq: signal} is not averaged over the number of neighbors. Cells at the boundary
typically have three to four neighboring cells, whereas cells in the bulk area have a mean of six. Therefore, cells at the boundary will potentially not be able to get the same amount of signal as cells in the bulk area. The received signal however, is the deciding factor with regard to the cell fate decision in our model. The low amounts of signal received at the boundary make them more likely to adopt the N--G+ fate.\\
As $q$ increases, we see a higher accumulation of N--G+ cells near the boundary with a slight clustering behavior in the bulk. For $q=0.9$ the signal disperses strong enough to generate an engulfing pattern, where N+G-- cells are completely surrounded by N--G+ cells. The pattern formation with respect to $q$ can be quantified using the PCFs for both N+G-- and N--G+ cells (Fig. \ref{fig: pair correlations non-local}). For comparison, the ratio of the two cell types are fixed to $89:88$. We discover that an increase in $q$ leads to a decrease of $\rho_n$ for large distances, i.e. less and less pairs of N+G-- cells pairing in the boundary regions. Simultaneously, it increases for small distances due to the cells accumulating in the center. For $\rho_g$, we see a slight increase for small distances and a tremendous one for large distances. The slight increase at small distances comes from the fact that the cells arranging at the boundary usually leave them with more neighbors of the same type just enough to be greater than the mean density of N--G+ cells. The values in the bulk region slightly decrease as the corresponding regions become more and more devoid of N--G+ pairs. In conclusion, the signal \eqref{eq: signal} generates patterns ranging from checkerboard to engulfing by increasing the dispersion parameter $q$. Additionally, the PCFs capture the characteristics of these patterns making it a powerful tool for pattern identification and comparison.

\subsection{Independence of cell number}
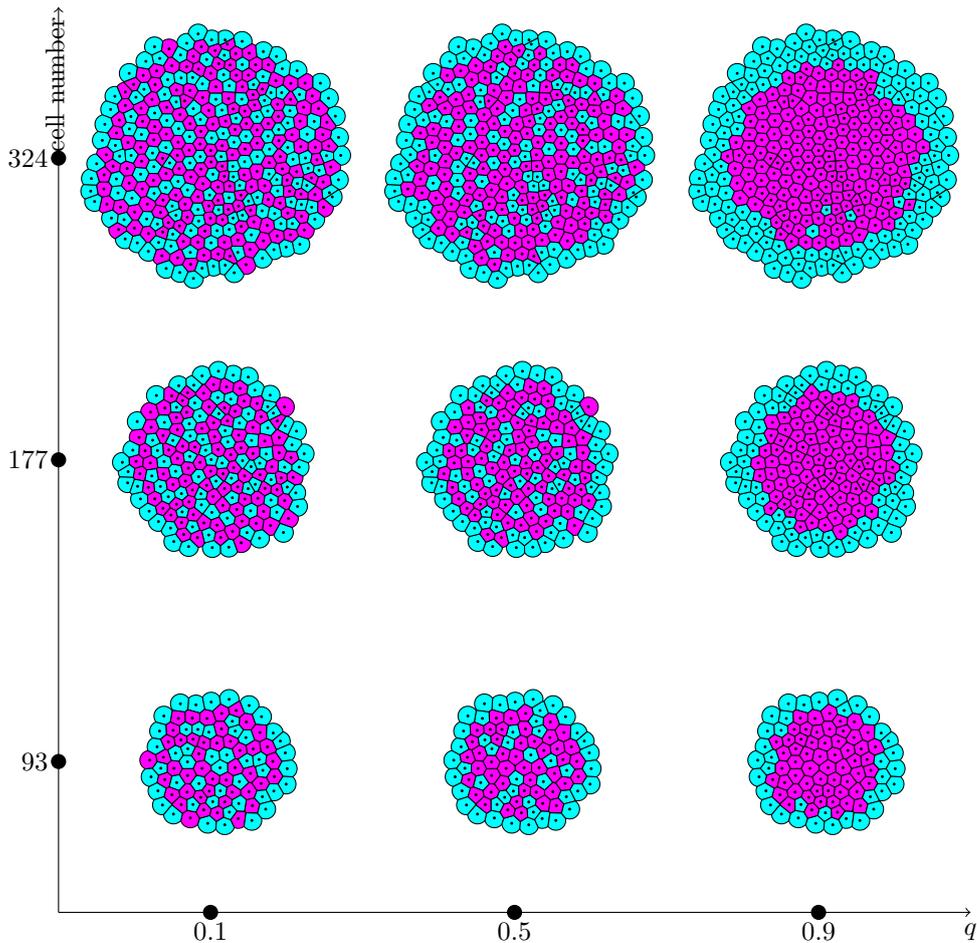
\begin{figure}[htbp]
\centering
\input{patterns_cell_number}
\caption{Patterning of two different cell types for different cell numbers $M$ and dispersion parameters $q$. From left to right, $q$ increases. Cell number $M$ increases from bottom to top. Each organoid resembles the result of an individual simulation up to steady state. N+G-- cells are colored in magenta, N--G+ cells in cyan.}
\label{fig: non local size comparison}
\end{figure}

\begin{figure}[htbp]
\centering
\begin{subfigure}{.32\textwidth}
  \centering
  \includegraphics[width=\linewidth]{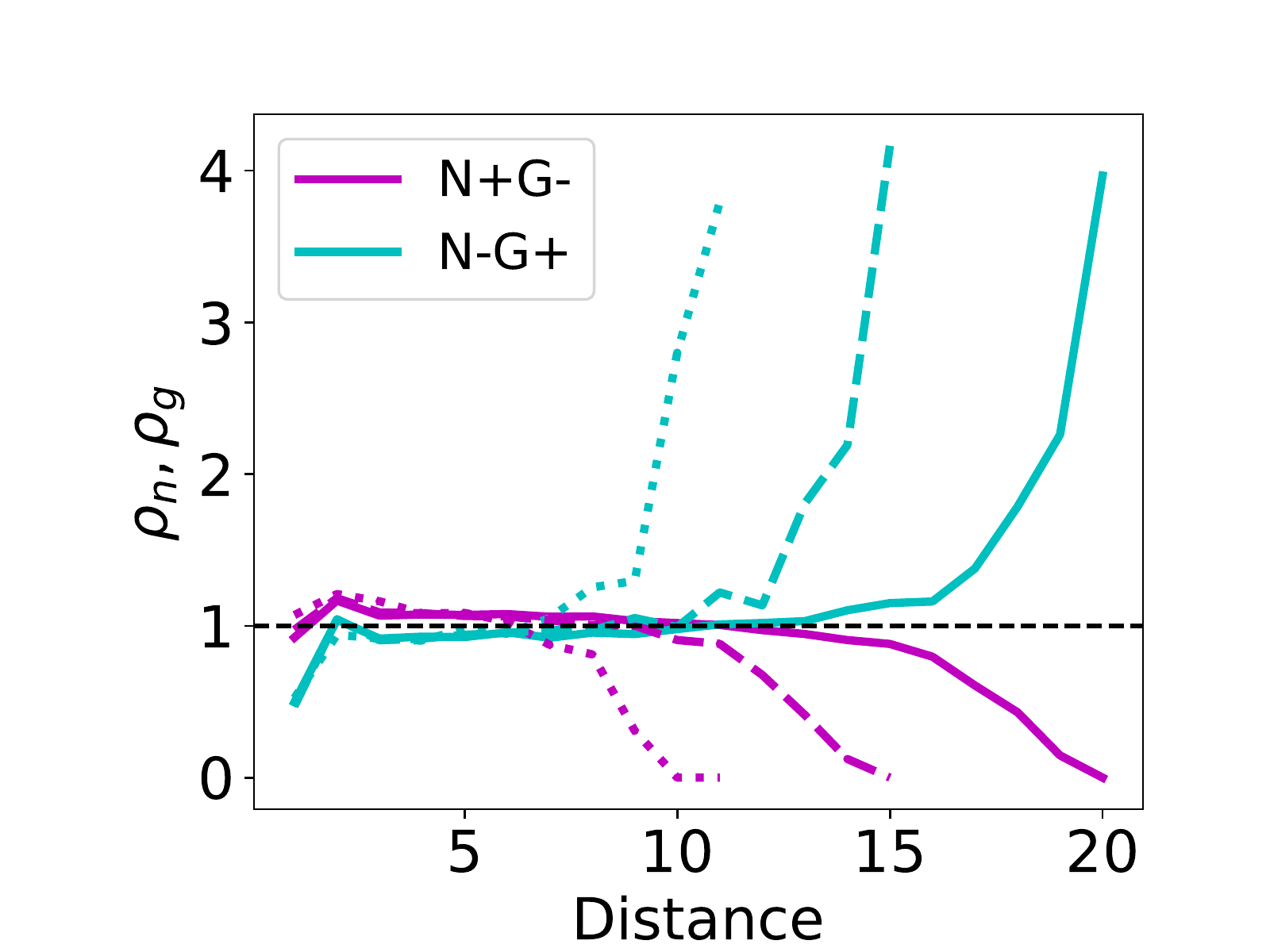}
  \subcaption{$q = 0.1$}
\end{subfigure}
\begin{subfigure}{.32\textwidth}
  \centering
  \includegraphics[width=\linewidth]{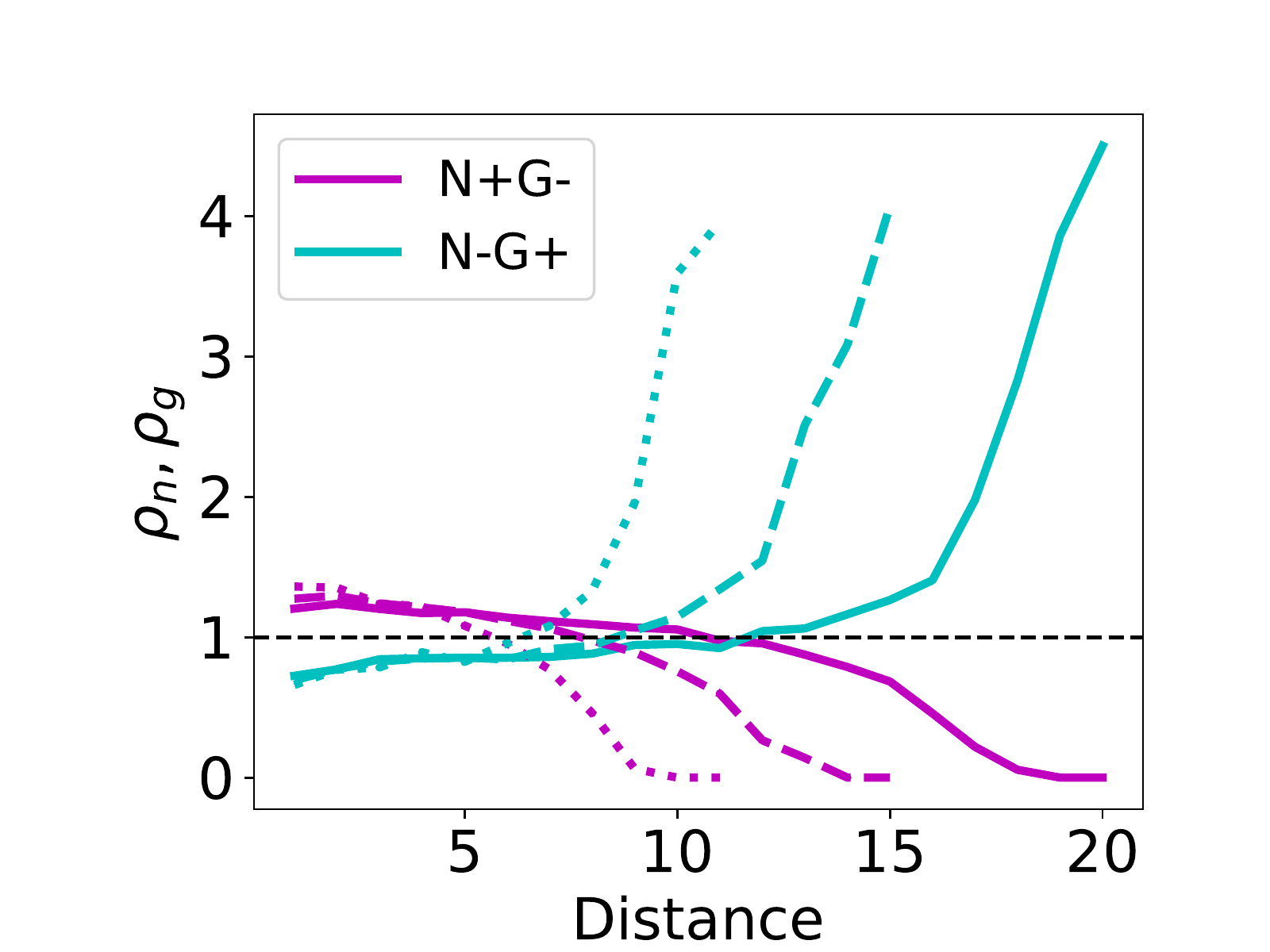}
  \subcaption{$q = 0.5$}
\end{subfigure}
\begin{subfigure}{.32\textwidth}
  \centering
  \includegraphics[width=\linewidth]{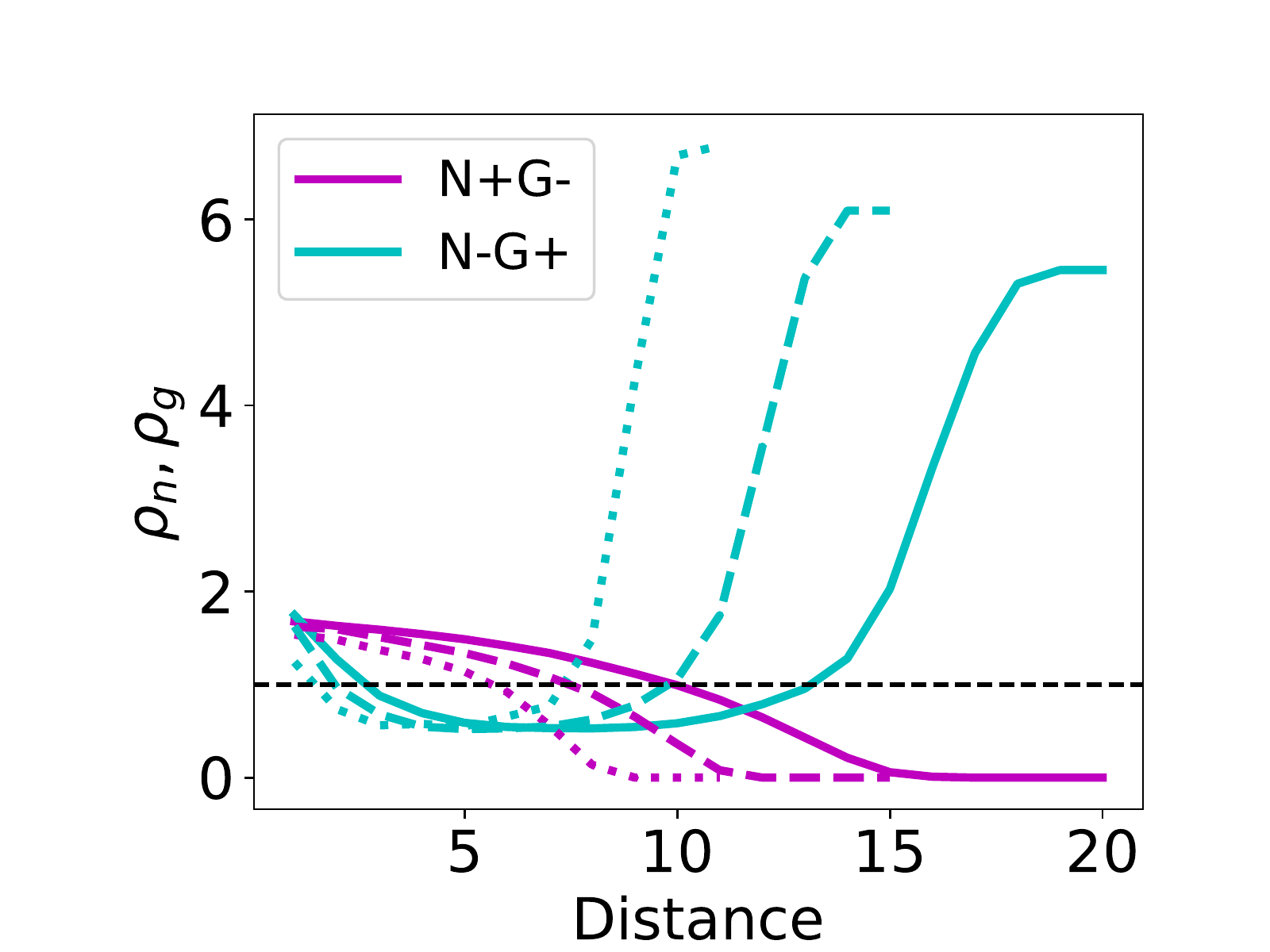}
  \subcaption{$q = 0.9$}
\end{subfigure}
\caption{PCFs $\rho_n$ (magenta) and $\rho_g$ (cyan) corresponding to the patterns in Fig. \ref{fig: non local size comparison}. Organoids with 93, 177 and 324 cells are represented by the dotted, dashed and solid lines respectively.}
\label{fig: pcf size comparison}
\end{figure}

Checkerboard patterns like in \cite{Schardt2021} can be characterized by any local cell neighborhood. Therefore, it makes no difference to the resulting pattern whether you scale up the number of cells. With the global signal \eqref{eq: signal} it is no longer possible to define the pattern at the local level. We therefore consider it important to show how different cell numbers affect pattern formation. On a global level the patterns remain identical (Fig. \ref{fig: non local size comparison}). For cell numbers of $93$, $177$ and $325$, we find again the transition from the checkerboard to the engulfing pattern when increasing $q$. For all simulations, the energy difference $-\Delta\varepsilon_g$ was set to $7$. The corresponding PCFs highlight that the general trend is conserved (Fig. \ref{fig: pcf size comparison}) over the different tissue sizes. In summary, the number of cells and therefore the size of the tissue has no influence on the resulting pattern on a global level.

\subsection{Comparison: Experiment and simulation}

\begin{figure}[htbp]
\centering
\begin{subfigure}{.32\textwidth}
  \centering
  \includegraphics[width=\linewidth]{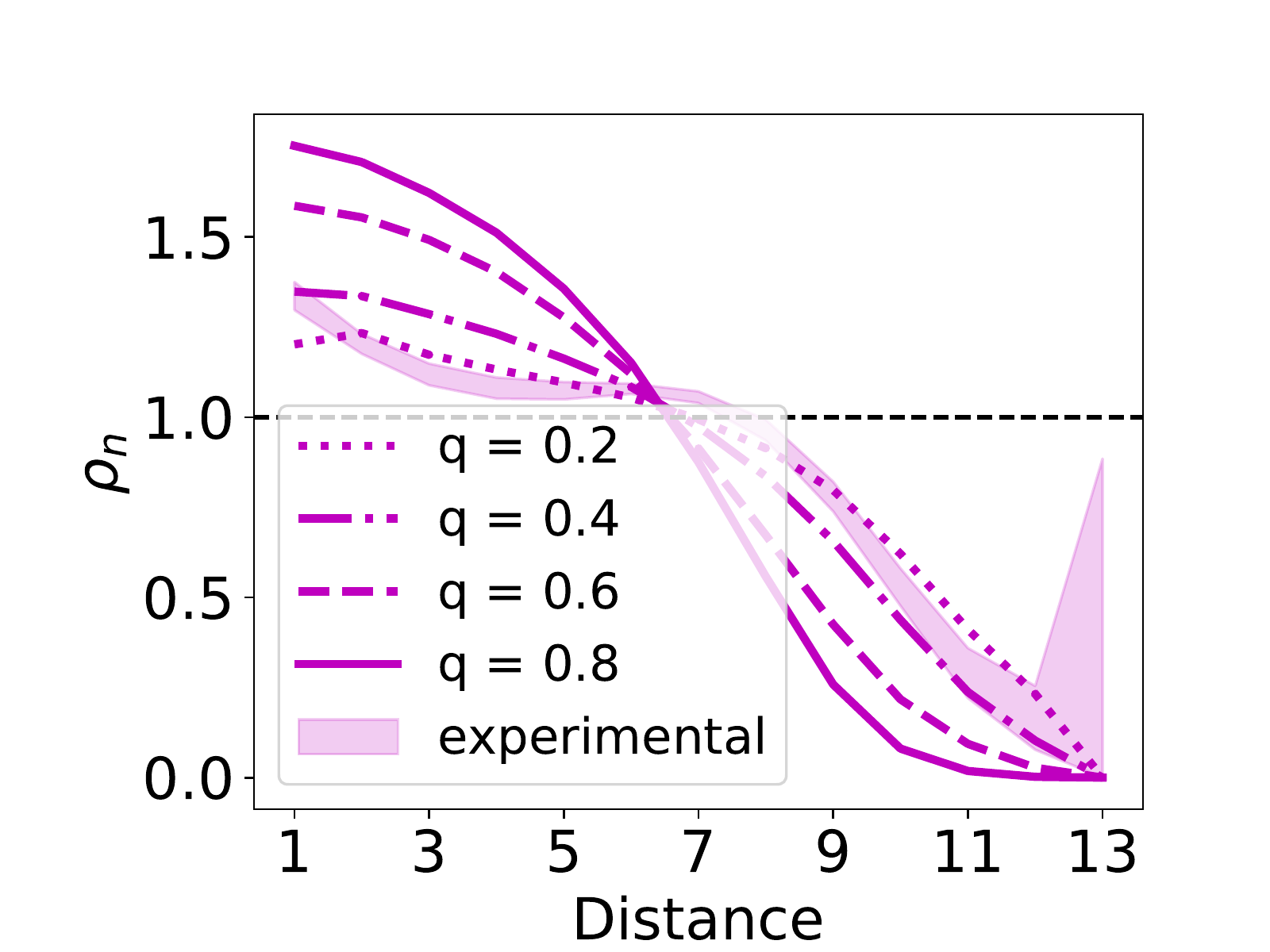}
\end{subfigure}
\begin{subfigure}{.32\textwidth}
  \centering
  \includegraphics[width=\linewidth]{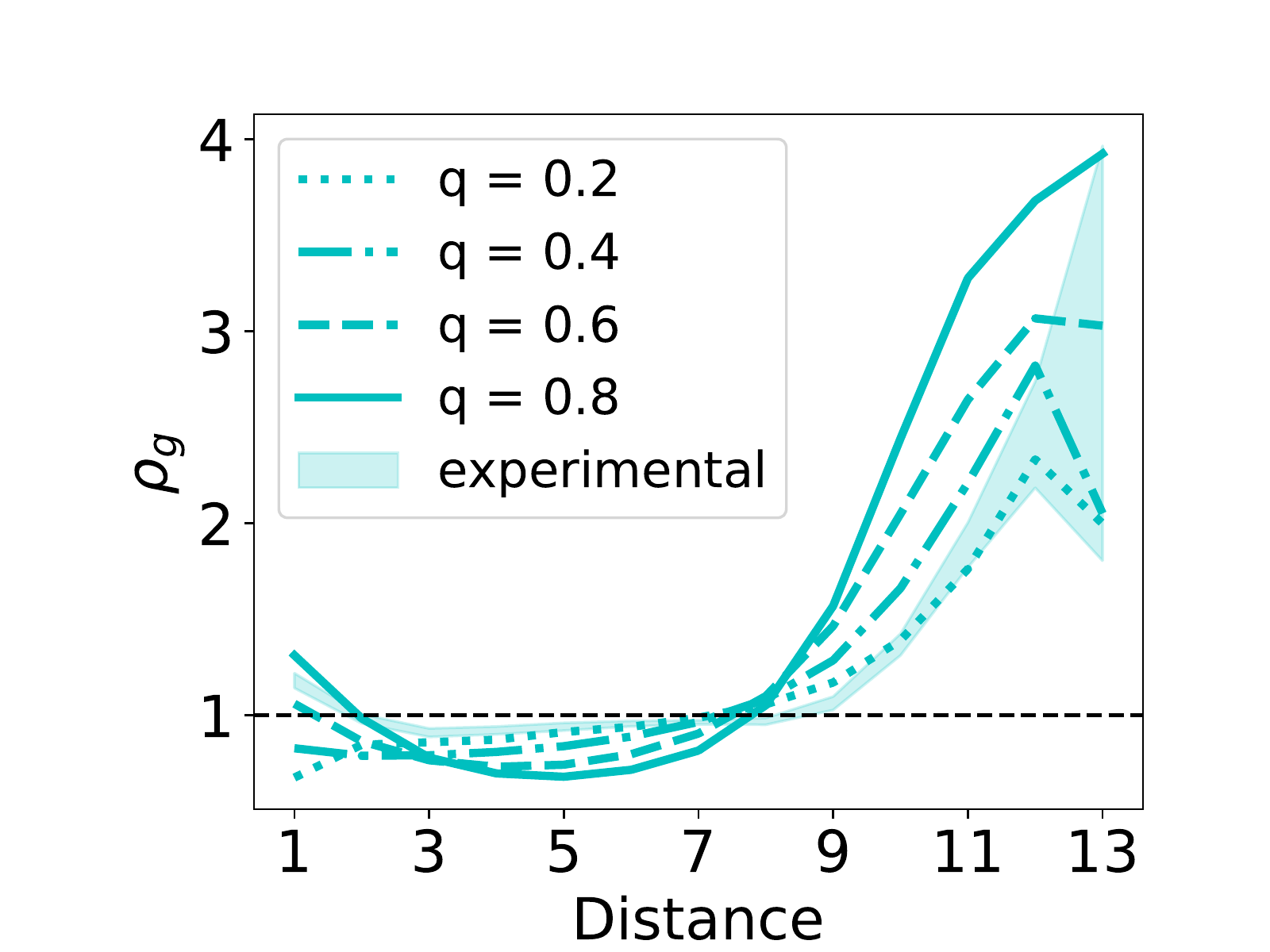}
\end{subfigure}
\begin{subfigure}{.32\textwidth}
  \centering
  \includegraphics[width=\linewidth]{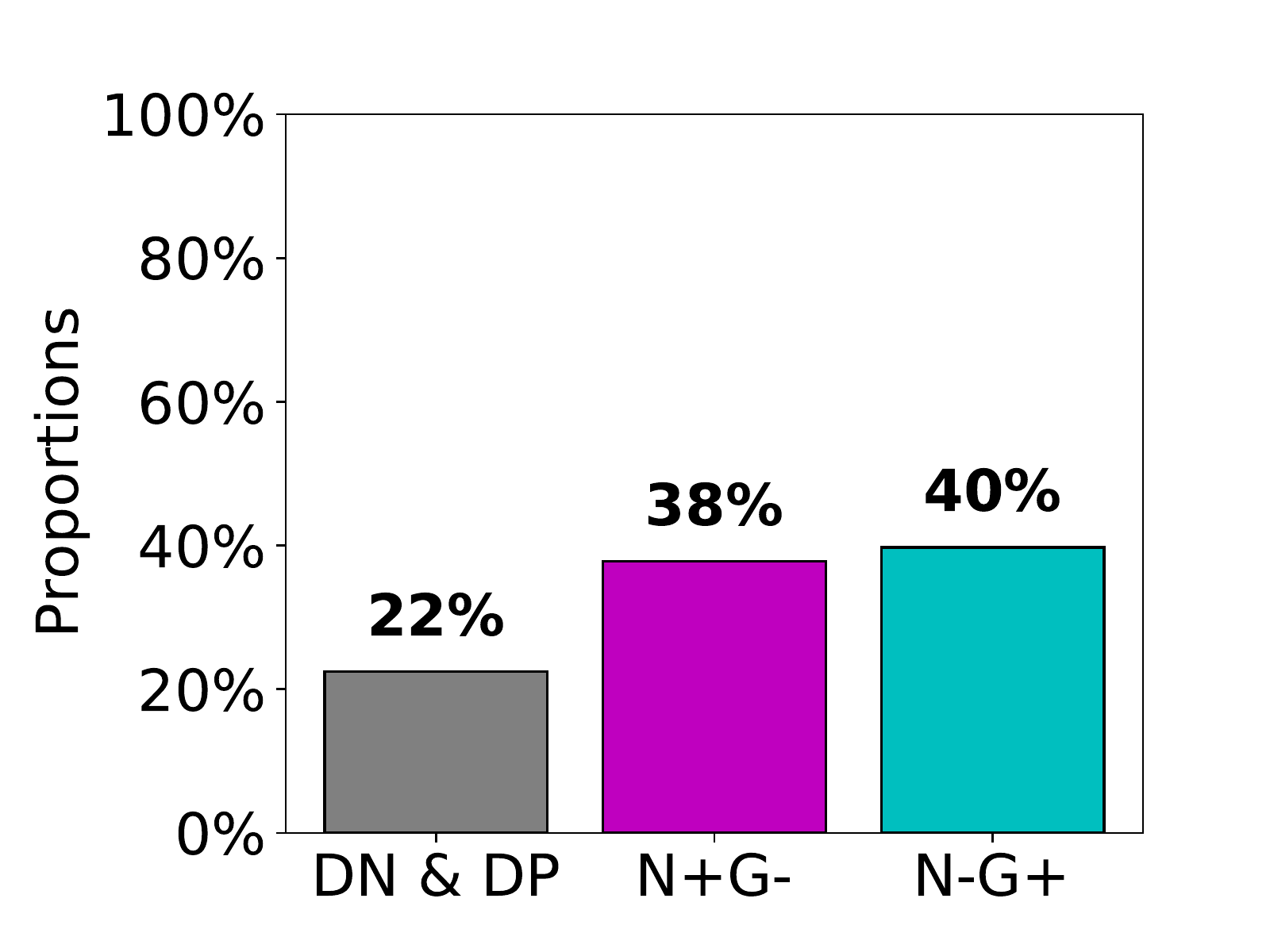}
\end{subfigure}
\begin{subfigure}{.32\textwidth}
  \centering
  \includegraphics[width=\linewidth]{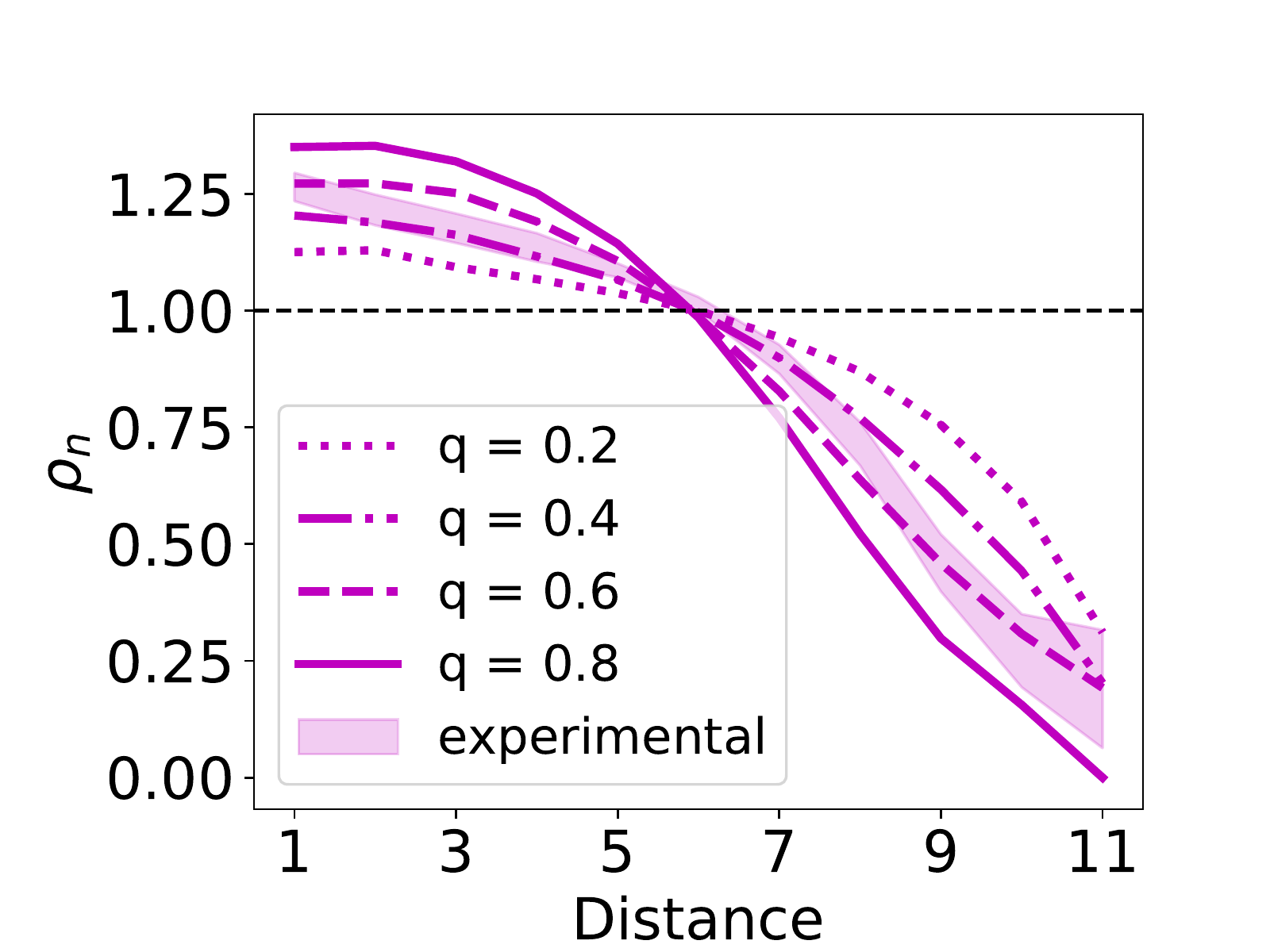}
\end{subfigure}
\begin{subfigure}{.32\textwidth}
  \centering
  \includegraphics[width=\linewidth]{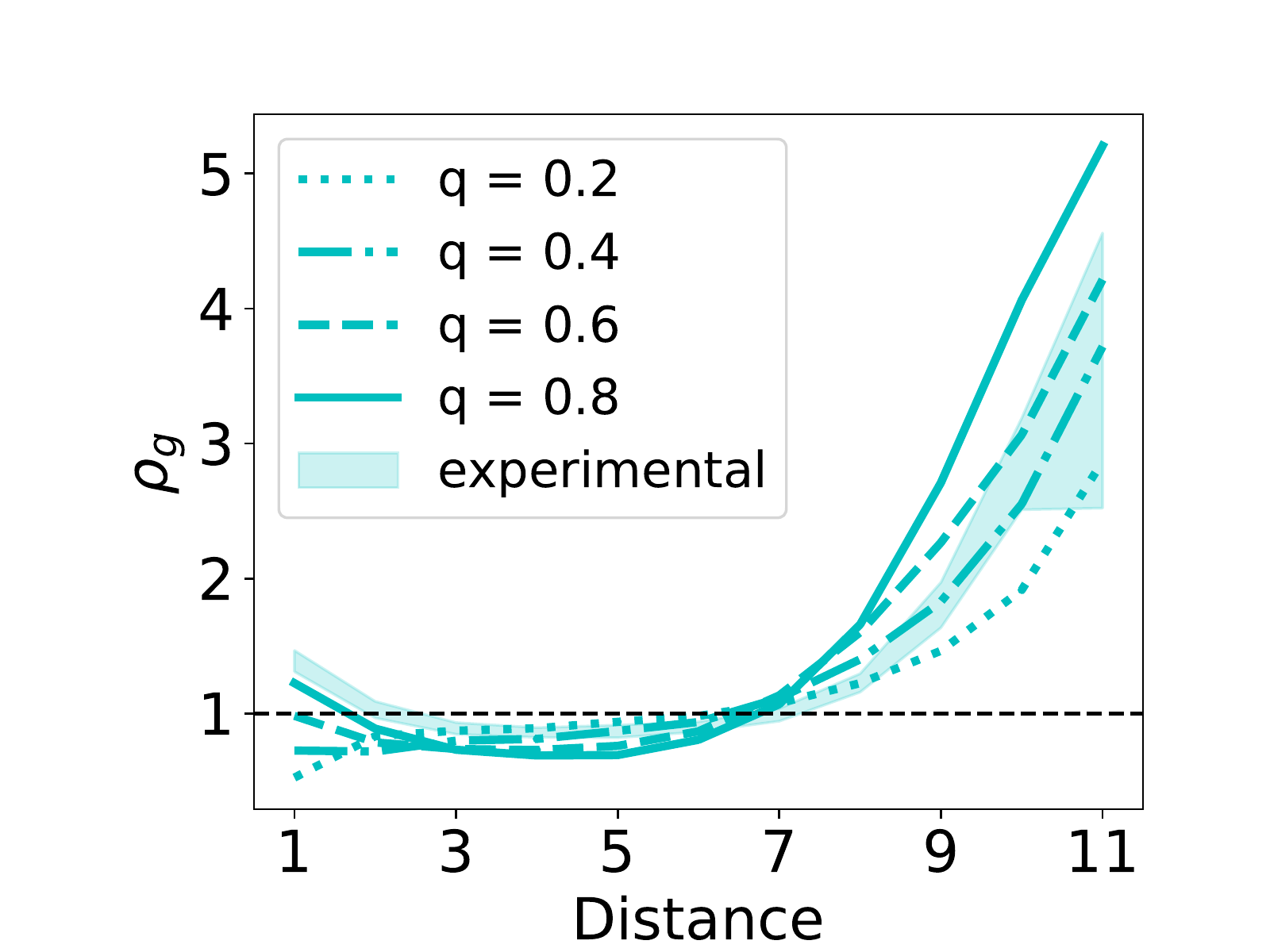}
\end{subfigure}
\begin{subfigure}{.32\textwidth}
  \centering
  \includegraphics[width=\linewidth]{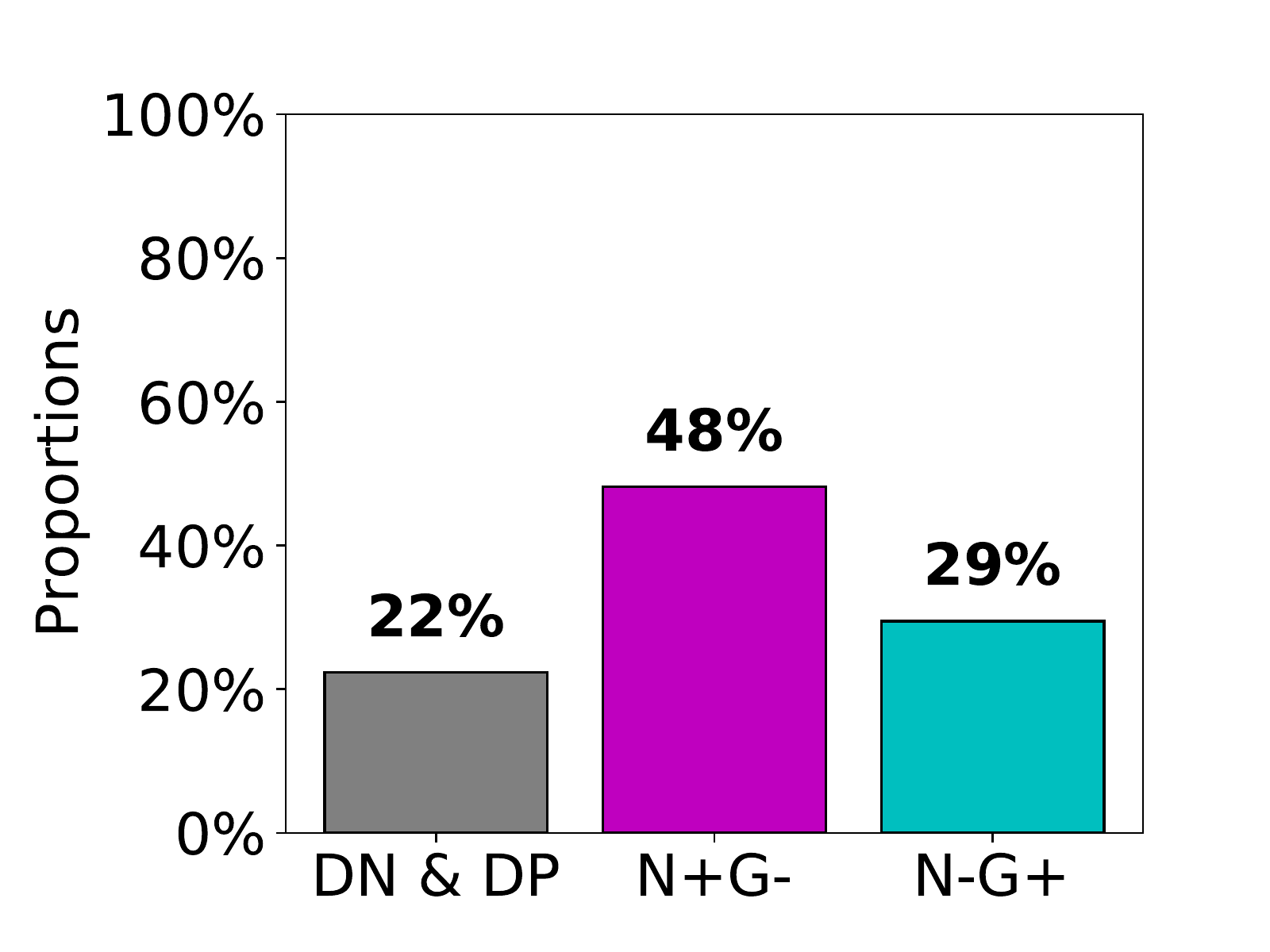}
\end{subfigure}
\begin{subfigure}{.32\textwidth}
  \centering
  \includegraphics[width=\linewidth]{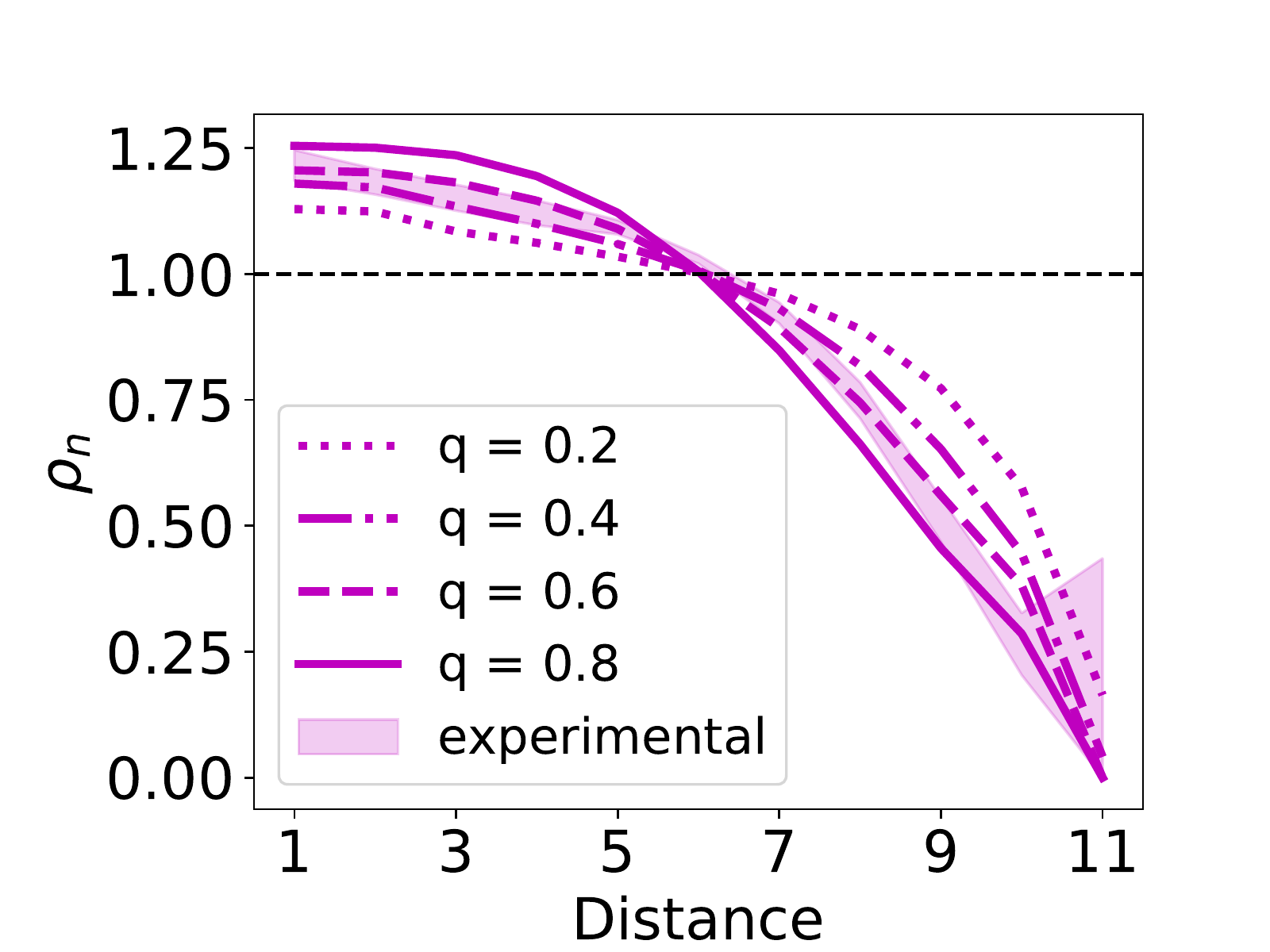}
\end{subfigure}
\begin{subfigure}{.32\textwidth}
  \centering
  \includegraphics[width=\linewidth]{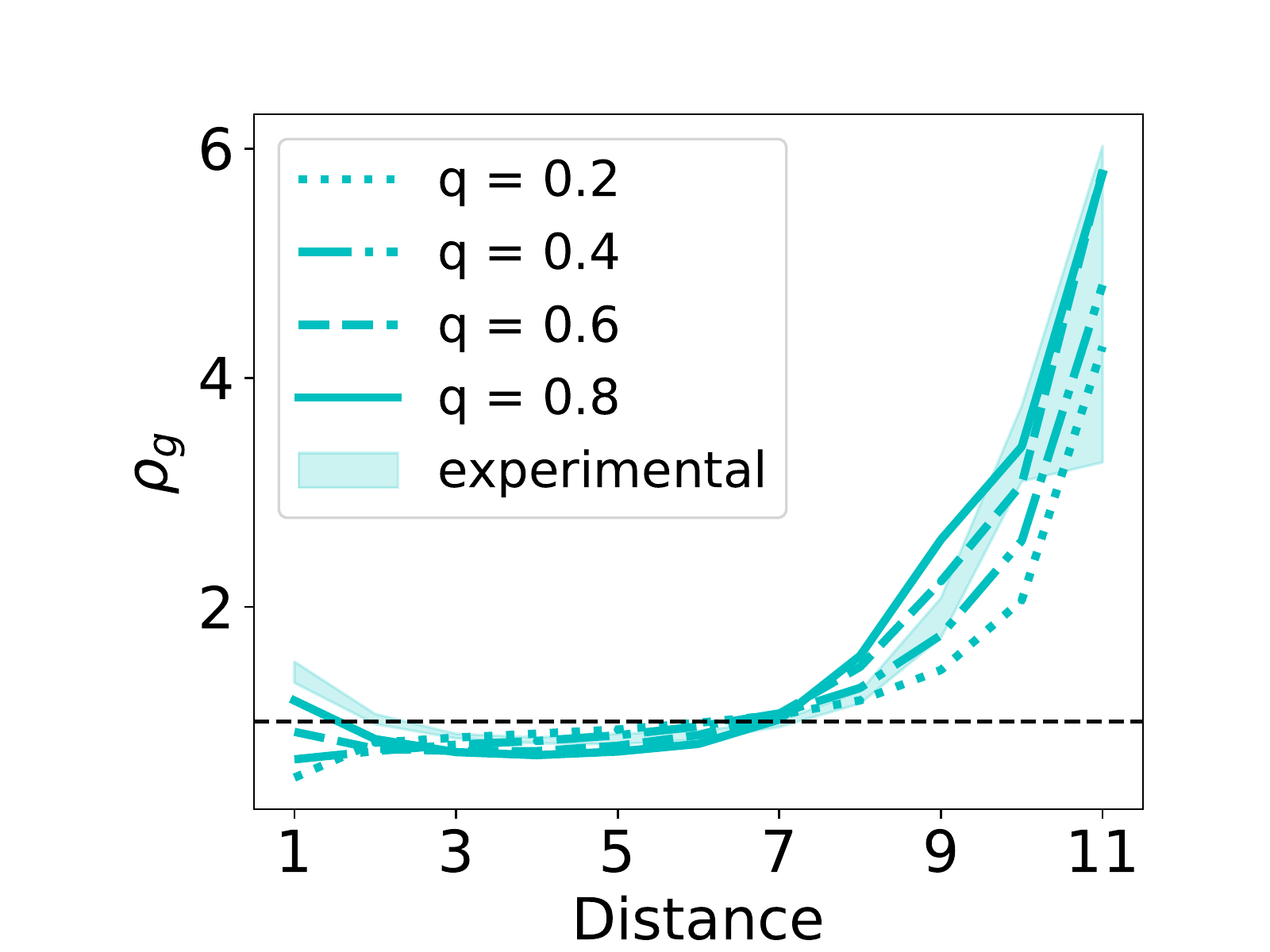}
\end{subfigure}
\begin{subfigure}{.32\textwidth}
  \centering
  \includegraphics[width=\linewidth]{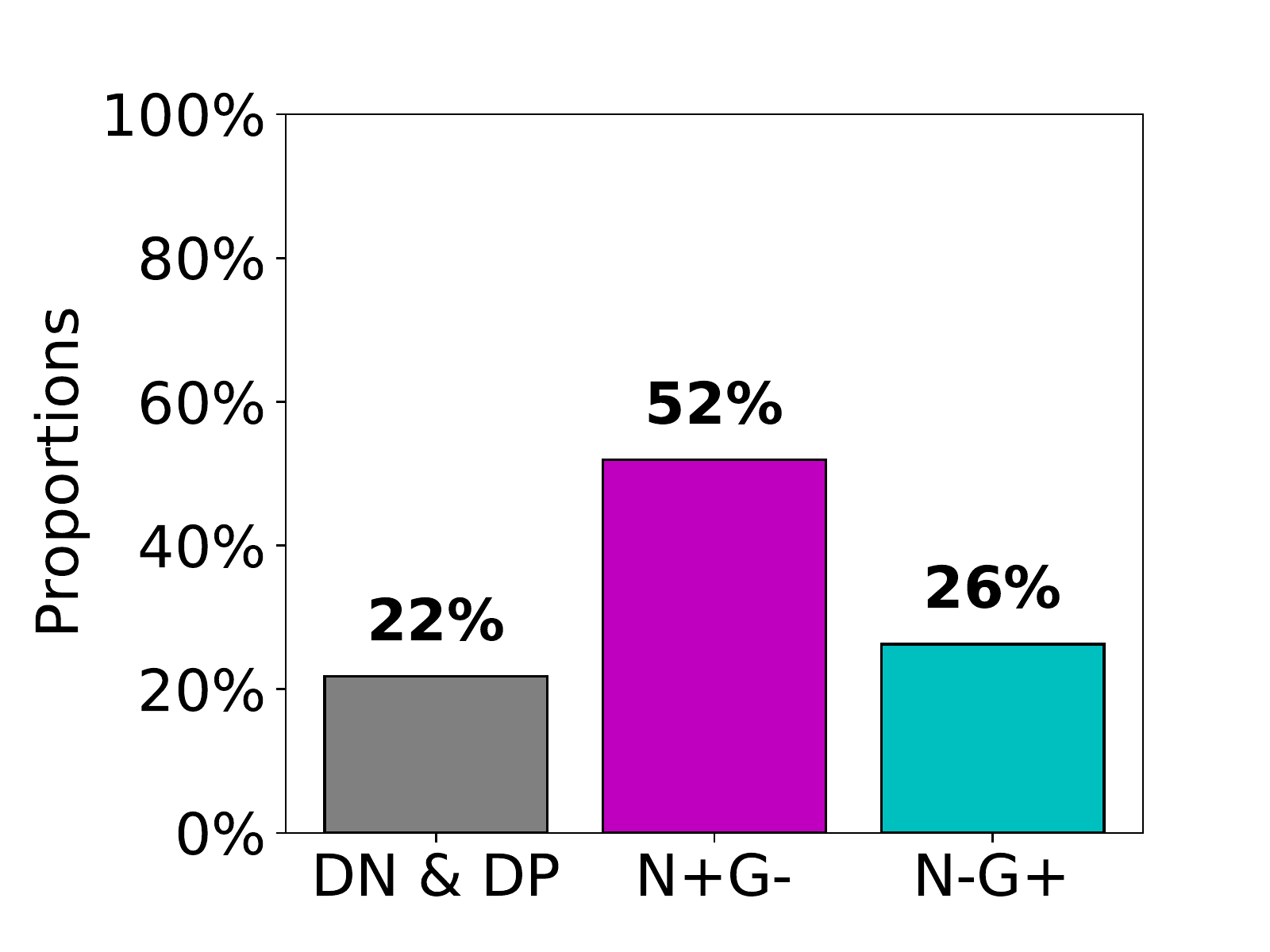}
\end{subfigure}
\caption{Three positive examples of the connection between simulation and experimental data.  In each row, we observe the data of a single ICM organoid from \cite{Mathew2019}. From left to right, the PCFs $\rho_n$ and $\rho_g$ and the cell type proportions are visualized. The PCF plots include PCFs for different dispersion parameters $q$ as lines as well as the experimental range of possibilities we achieve by randomly deciding the fates of DP and DN cells as shaded regions.}
\label{fig: PCFs good organoids}
\end{figure}

This section has two goals. First, the experimental data are characterized using the range of possible PCFs generated in section \ref{sec: data pre-processing}. Second, these are compared with the PCFs from the simulation results in section \ref{sec: simulations on data}. Going through the results, we detected a high heterogeneity of the data. We highlight two groups of organoids. The positive examples, those that show a strong similarity to each other and closely resemble the results of our simulations. The negative examples, those that show slight to extreme irregularities allowing us to highlight the difficulties when trying to characterize their patterns. Representatively, we concentrate on three specially selected organoids each. We consider this approach to be reasonable, since there is no way to analyze the patterns of all organoids as a whole. The first group of organoids exhibits a common general trend of the PCF (Fig. \ref{fig: PCFs good organoids}). For $\rho_n$, we see a monotonous decrease with values above $1$ for low to medium distances and below $1$ for medium to large distances. The trend in $\rho_g$ starts with values slightly above $1$ for direct neighbors. It is followed by values slightly below $1$ for medium distances and concluded with strongly increased values at large distances. Altogether, these trends show the characteristics of a spatial segregation of cells. To be precise, the N+G-- are being engulfed by N--G+ cells. The variability of the data, expressed by the pair correlation region, turns out to be rather small, except for large distance regions. This is due to the number of cell pairings decreasing with distance. This means that any randomly decided cell at that distance has a stronger influence on the PCF values. The simulation results paint a picture similar to the experimental data. It follows the same trends that we established above (Fig. \ref{fig: pair correlations non-local}). However, it is important to mention that those three examples correspond to different dispersion values. The first organoid shows the highest overlap to low, the second to medium and the third to simulations with high dispersion values $q$. In addition, the high values of $\rho_g(1)$ are only captured by high $q$, sometimes leading to a mismatch between low and large distance region. The cell type proportions in the third column of Fig. \ref{fig: PCFs good organoids} complete the picture. These do not show excessive proportions of certain cell types, thus leading to results that are straightforward to interpret. Overall, the first group of organoids shows promising agreement with our model and highlights the radial expression of the cell patterns.

\begin{figure}[htbp]
\centering
\begin{subfigure}{.32\textwidth}
  \centering
  \includegraphics[width=\linewidth]{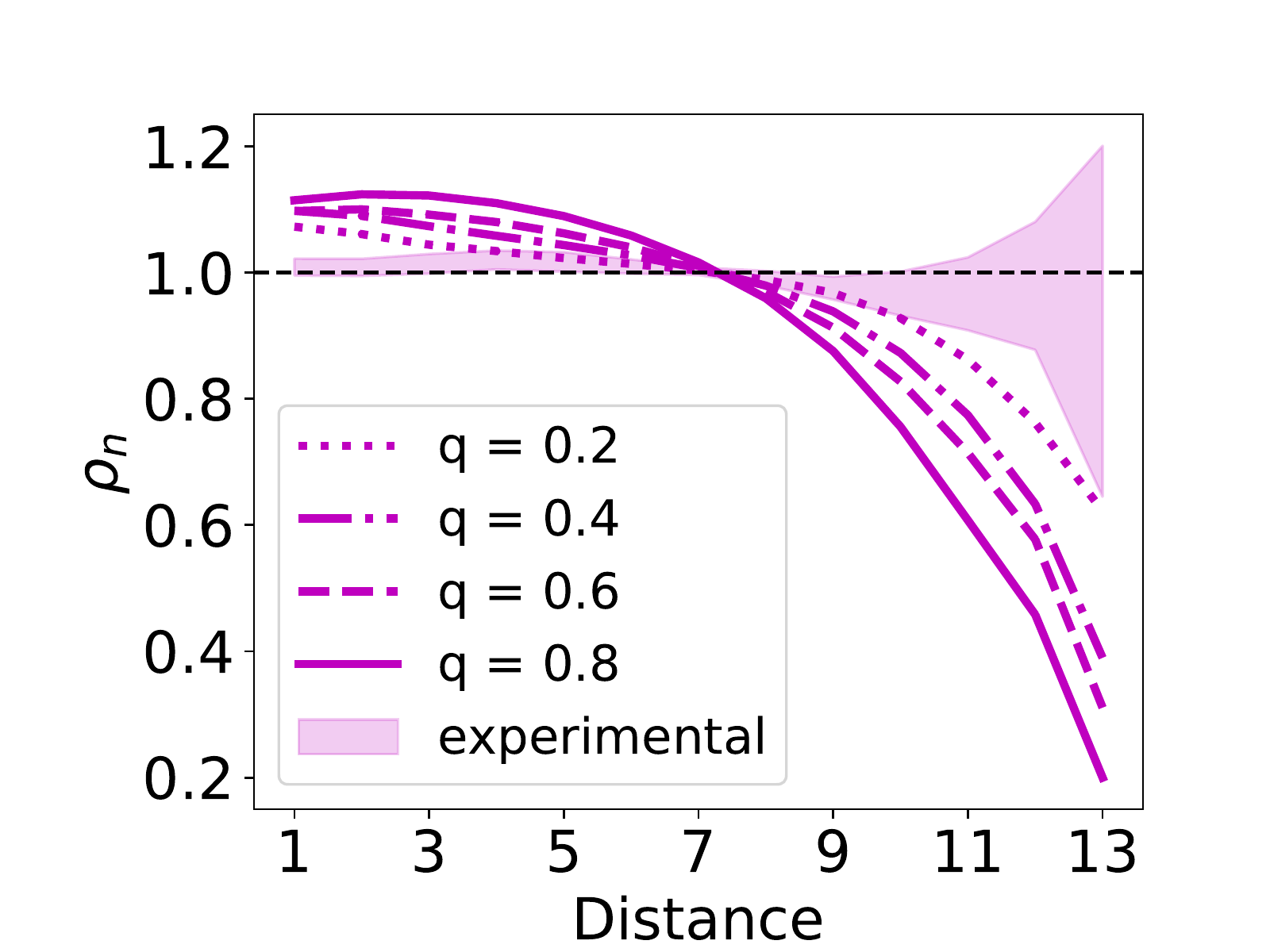}
\end{subfigure}
\begin{subfigure}{.32\textwidth}
  \centering
  \includegraphics[width=\linewidth]{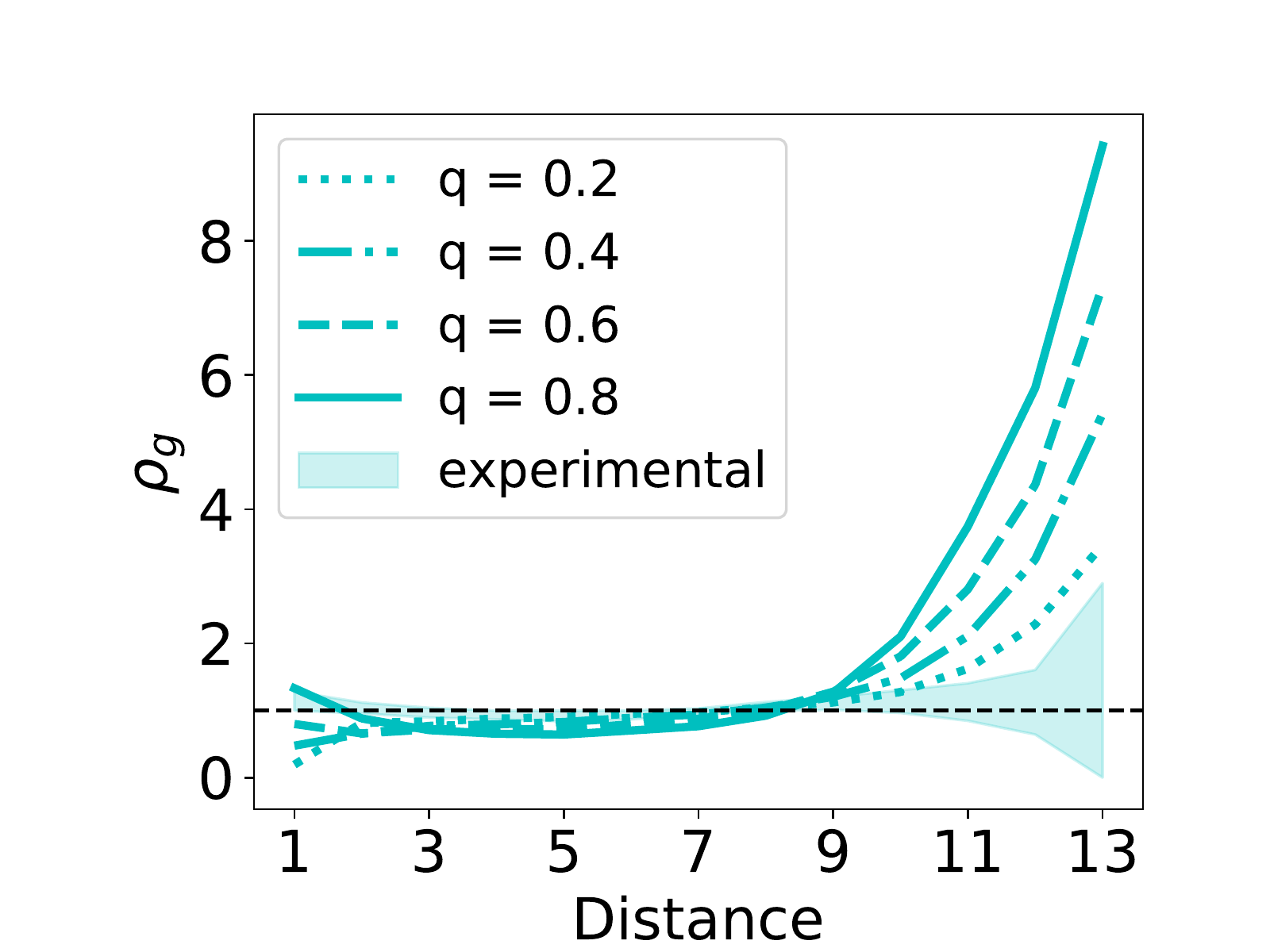}
\end{subfigure}
\begin{subfigure}{.32\textwidth}
  \centering
  \includegraphics[width=\linewidth]{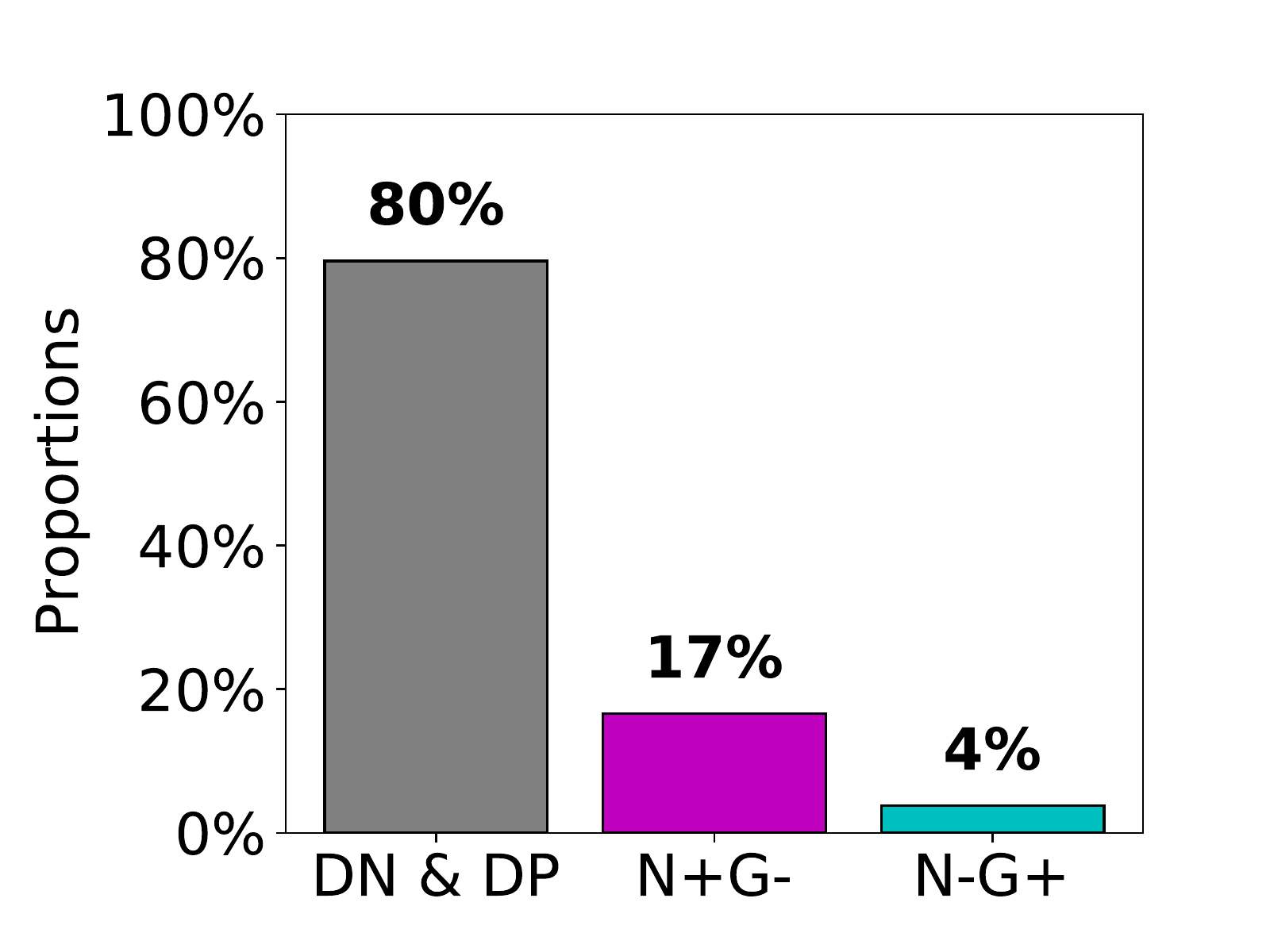}
\end{subfigure}
\begin{subfigure}{.32\textwidth}
  \centering
  \includegraphics[width=\linewidth]{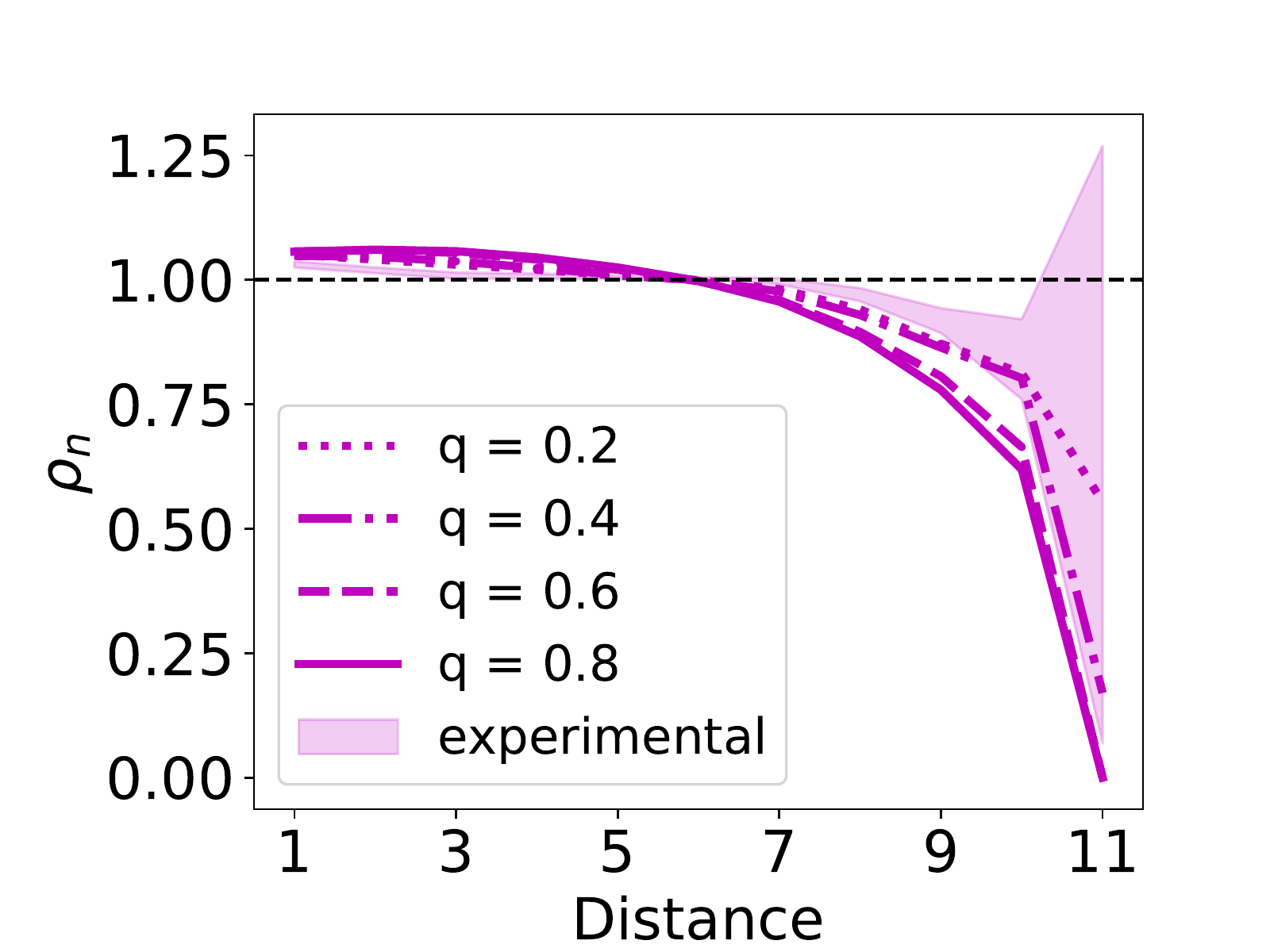}
\end{subfigure}
\begin{subfigure}{.32\textwidth}
  \centering
  \includegraphics[width=\linewidth]{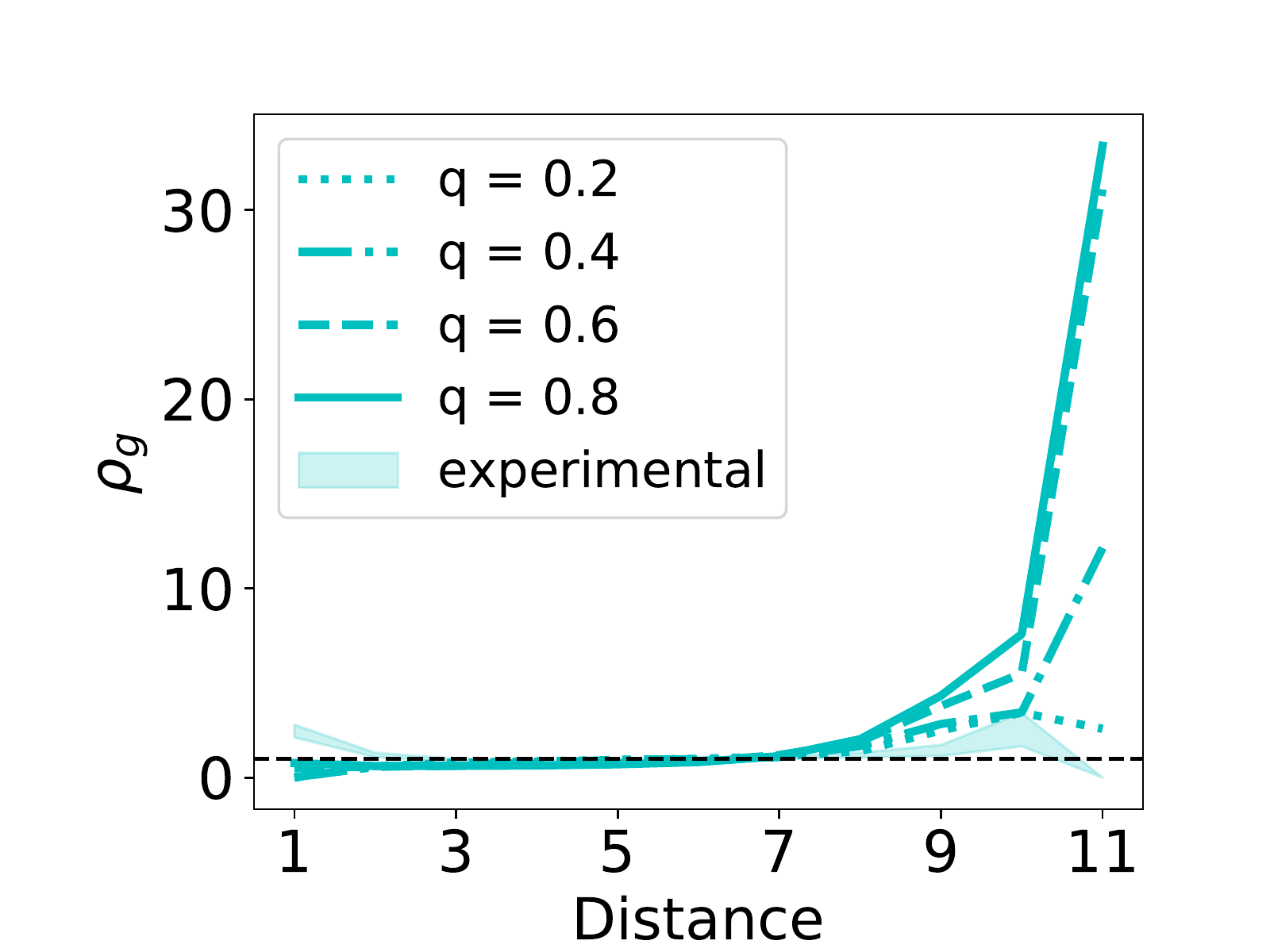}
\end{subfigure}
\begin{subfigure}{.32\textwidth}
  \centering
  \includegraphics[width=\linewidth]{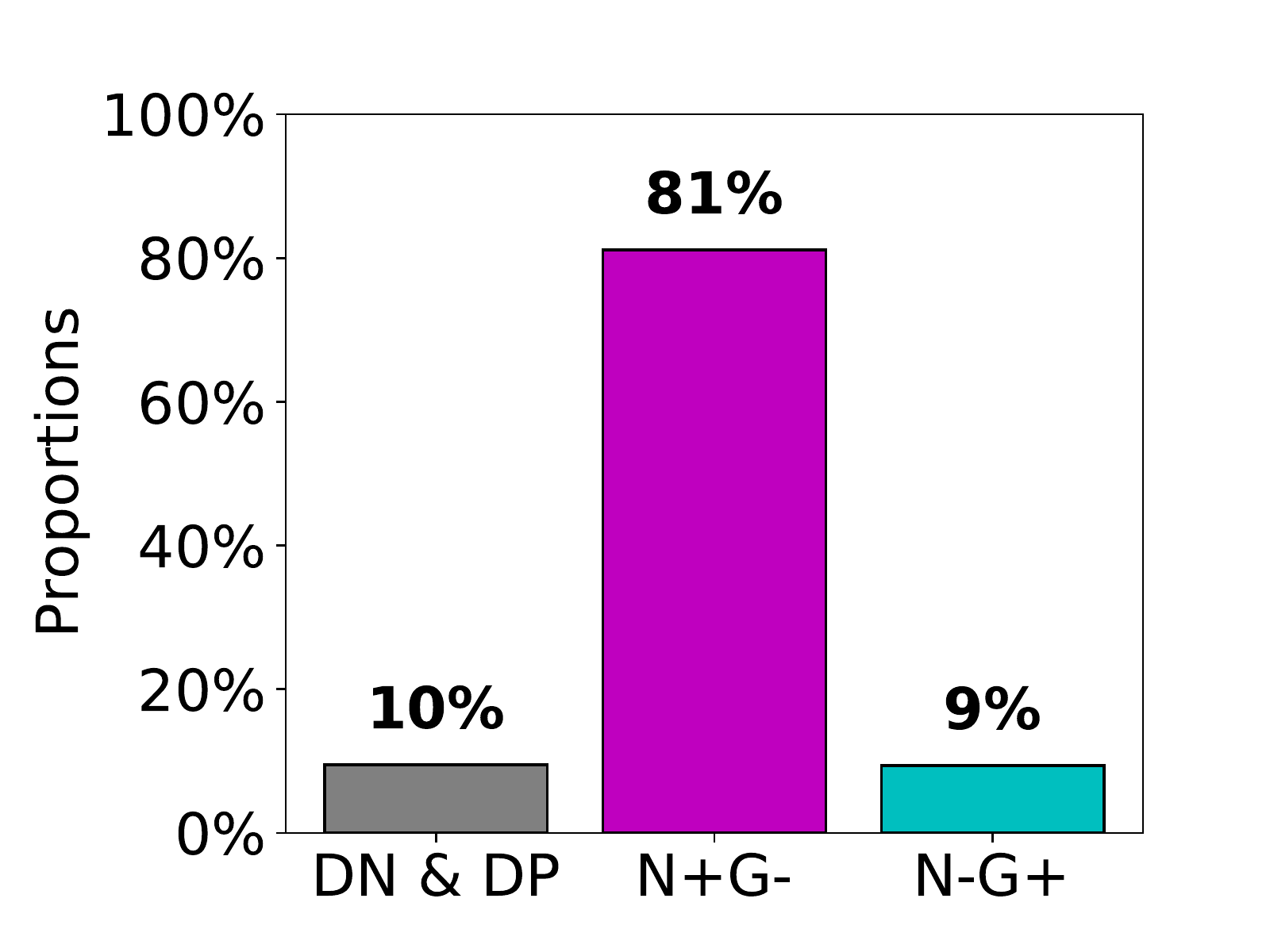}
\end{subfigure}
\begin{subfigure}{.32\textwidth}
  \centering
  \includegraphics[width=\linewidth]{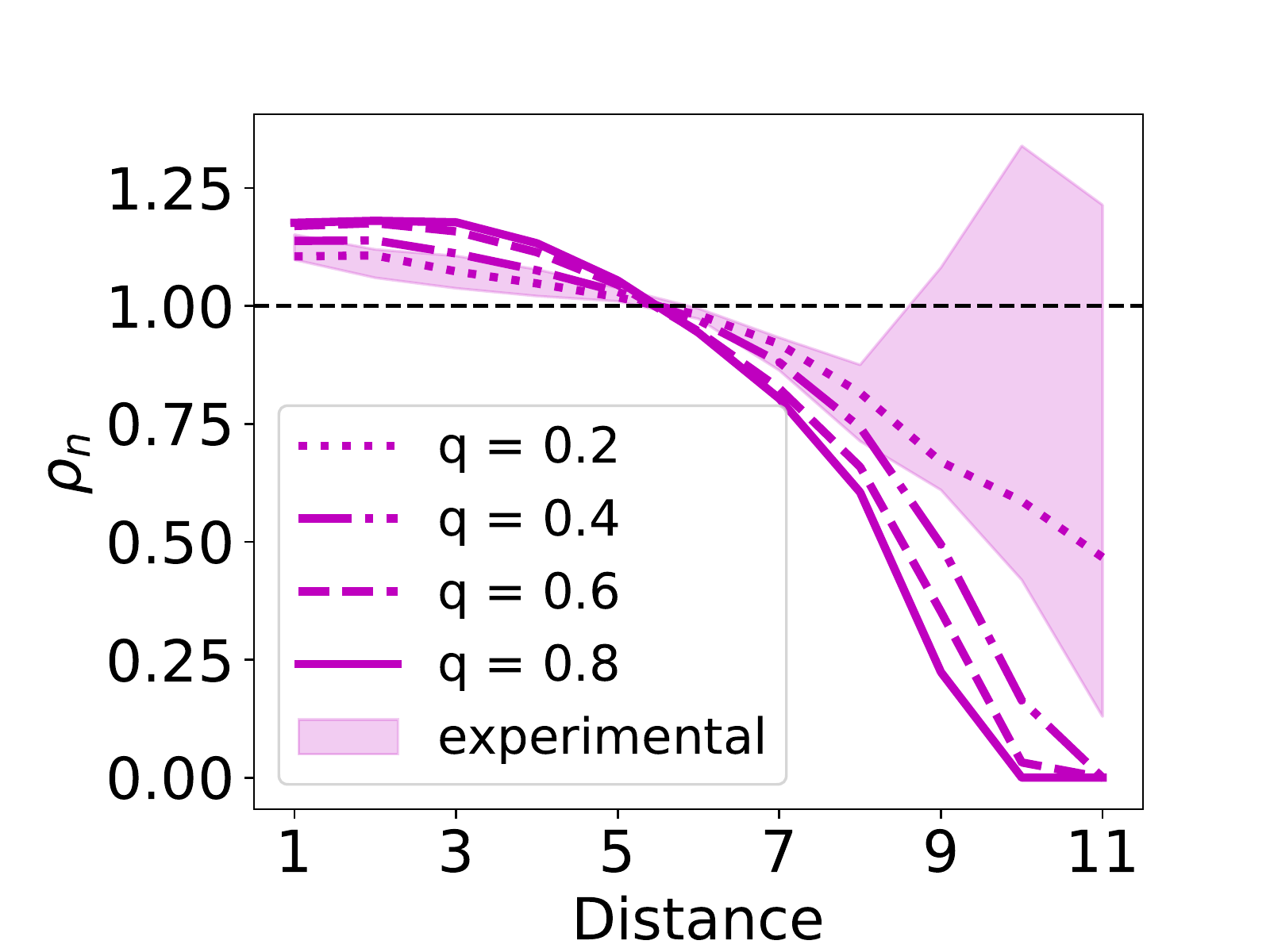}
\end{subfigure}
\begin{subfigure}{.32\textwidth}
  \centering
  \includegraphics[width=\linewidth]{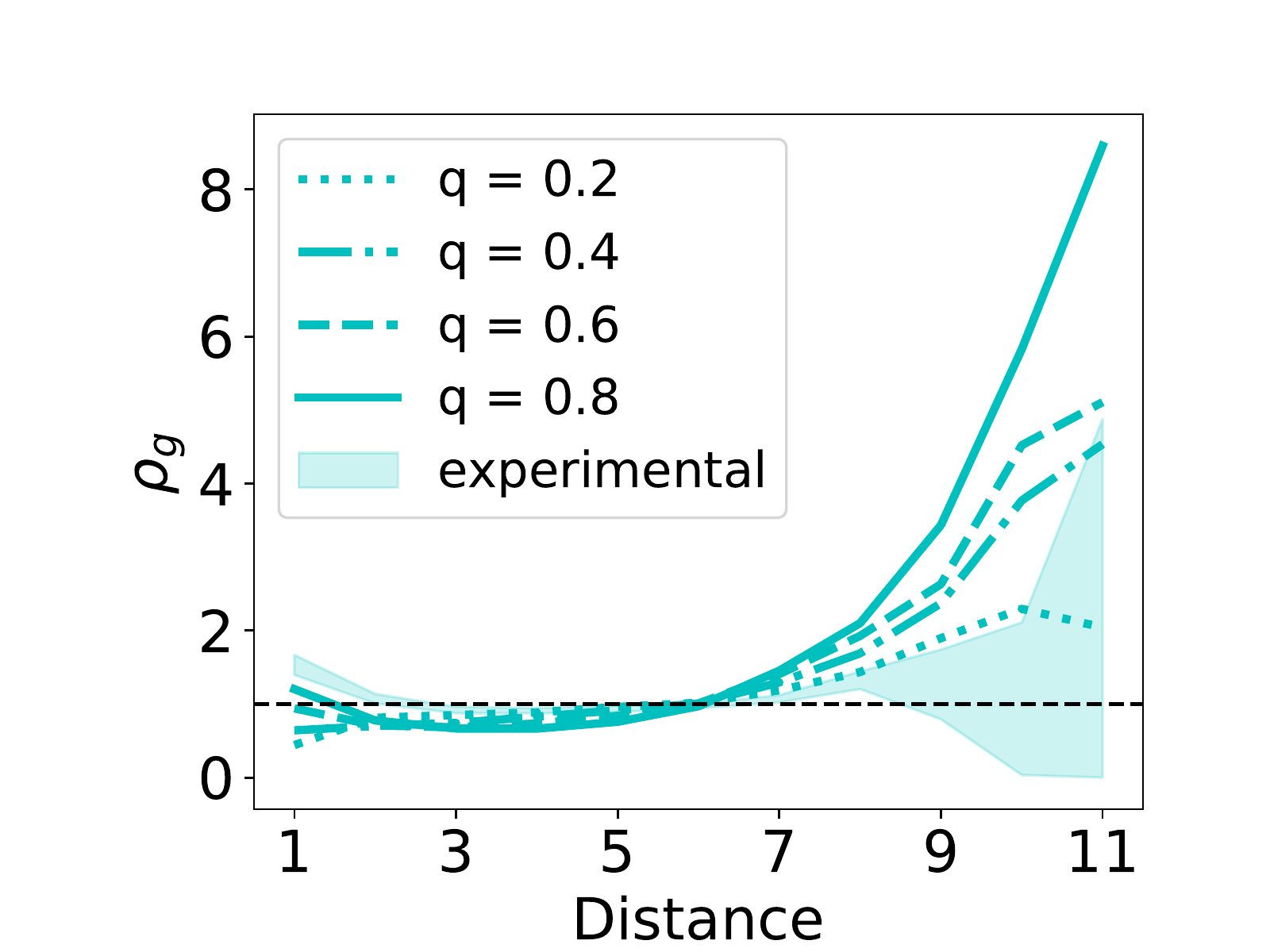}
\end{subfigure}
\begin{subfigure}{.32\textwidth}
  \centering
  \includegraphics[width=\linewidth]{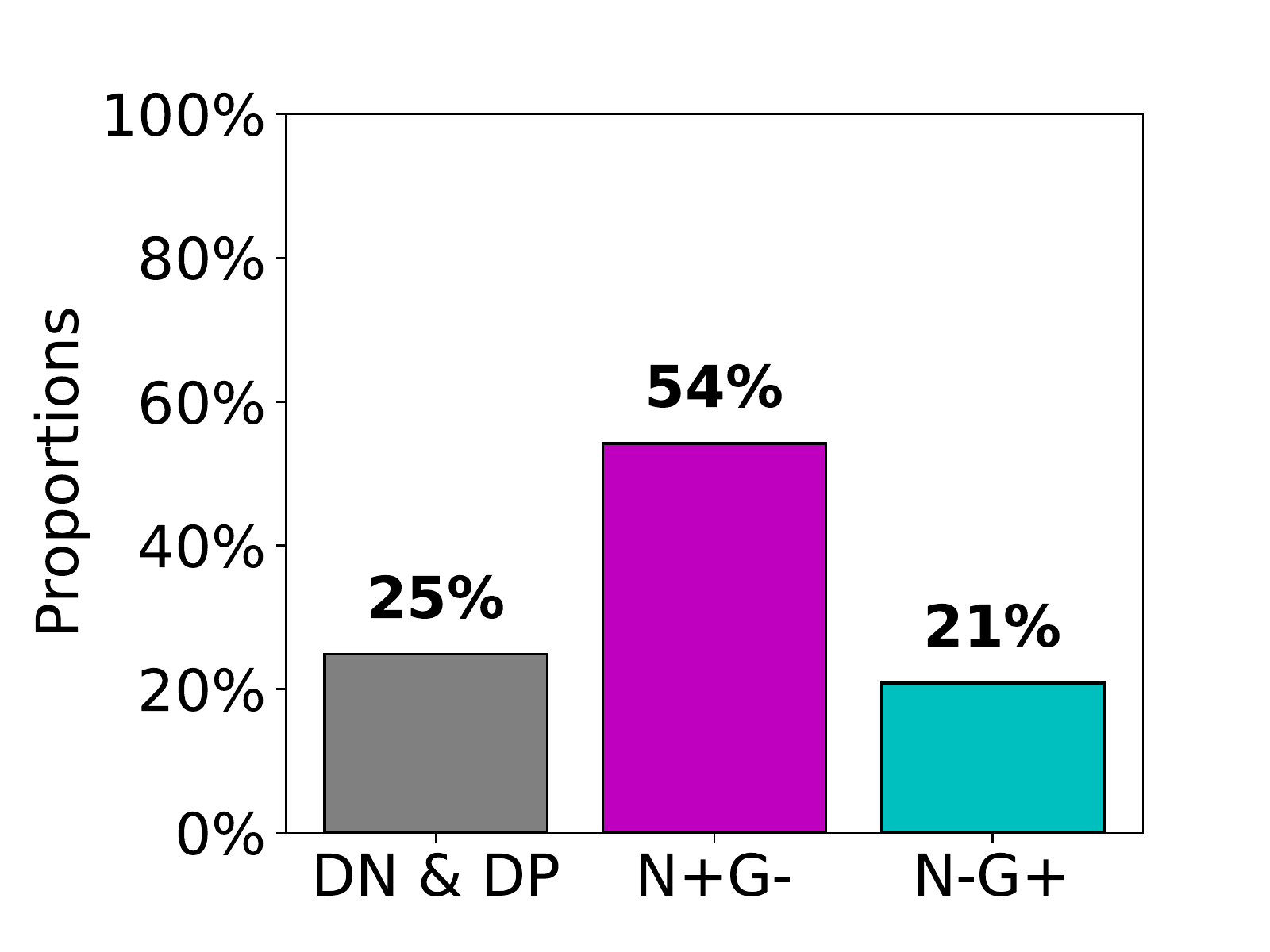}
\end{subfigure}
\caption{Three negative examples of the connection between simulation and experimental data. In each row, we observe the data of a single ICM organoid from \cite{Mathew2019}. From left to right, the PCFs $\rho_n$ and $\rho_g$ and the cell type proportions are visualized. The PCF plots include PCFs for different dispersion parameters $q$ as lines as well as the experimental range of possibilities we achieve by randomly deciding the fates of DP and DN cells as shaded regions}
\label{fig: PCFs bad organoids}
\end{figure}

We included the second group of organoids to showcase that in some cases, the experimental data and our model do not align well (Fig. \ref{fig: PCFs bad organoids}). The first organoid is characterized by its high proportion of DP and DN cells. This increases the width of the pair correlation region, which means that the trend is no longer as pronounced. Instead, for large distances the pair correlation regions for N+G-- and N--G+ show large uncertainties in both directions. This organoid consists of $1751$ cells of which $1393$ are either DP or DN. Therefore, we have to mention that our sample size of $1000$ patterns cannot nearly describe the whole amount of $2^{1393}$ possibilities. The second organoid shows extreme proportions of N+G-- and N--G+ cells. Due to the normalization \eqref{eq: PCF normalization}, this leads to $\rho_g$ being highly sensitive to N--G+ cell pairs found at any distance. This means, any N--G+ cell pair found at large distance will greatly increase $\rho_g$. In addition to that, some cells might be wrongfully connected in our cell graph $G$. The simulations are carried out on $G$ yielding the spatial segregation described before. The experimental data however, can then differ greatly when $G$ is not accurately describing the cell neighborhoods especially in the boundary region. The previously established trend is therefore no longer clearly pronounced. For small $q$, we find some overlap. The third organoid is special not due to its cell type proportions but rather its geometry. Whereas every organoid before had a sphere-like structure, this one has a tail-like appendage with large amounts of DP and DN on one of its ends. Therefore, again for large distances, we get large areas of uncertainty such that the trend is no longer clearly visible. We find some overlap for small $q$, but not enough to confidently characterize the organoid. In total, the negative examples show either large amounts of undecidable cells, extreme cell type proportions or geometrical deformations. In addition to the mismatch between simulation and experimental data for small distances, this also leads to mismatches in large distance regions. We encourage the readers to take a look at our interactive organoid visualization on GitHub (\blue{\url{https://schardts.github.io/Organoids48h}}) where especially the problem with the last organoid becomes visible using a rotatable 3D visualization of the ICM organoids. All organoids used in this publication and more can be found there. The ones presented here have the following IDs in their order of appearance: $2, 32, 36, 1, 12, 39$. In conclusion, most of the ICM organoids show characteristics of the spatial segregation of N+G-- and N--G+. These are shown by high values of $\rho_n$ for small distances, as well as low values for large distances. Together with high values for $\rho_g$ at large distances this indicates an engulfing of N+G-- by N--G+ cells.

\section{Discussion}

In this study, we demonstrated how cells can organize on a global scale using a signal that reaches beyond their nearest neighbors. Compared to \cite{Stanoev2021}, we used the model we previously established in \cite{Schardt2021}. This differs in two ways. First, the equations of NANOG and GATA6 were derived and analyzed in detail using a statistical mechanics approach. Second, the calculations are not performed on a grid but on two- and three-dimensional position data of cells. This provides us with more freedom in the tissue geometry, such that experimental cell position data can be integrated into the model. While this study focuses on differentiation of epiblast and primitive endoderm cells, the idea of long ranging signals can also be found in other systems such as the Delta/Notch mediated inhibition in \textit{Drosophila} \cite{DeJoussineau2003,Cohen2010,Chen2014}.

\subsection*{Signal dispersion is the deciding factor in forming local/global patterns}
Our simulation proved that global cell-cell communication enables a range of patterns from two cell types in a checkerboard like arrangement to one cell type engulfing the other. The signal emanating from a cell, which is used to communicate with other cells, plays a central role in this study. The introduced dispersion parameter $q$ allows us to artificially vary between a signal that only reaches the neighboring cells and a signal that spreads evenly in the tissue. Simulations on two-dimensional organoids were used to illustrate the effects of increased dispersion of the signal. We found that for low signal dispersion N+G-- and N--G+ cells tend to avoid being adjacent to the same cell type, hence the term checkerboard pattern. Furthermore, when increasing the signal dispersion, N--G+ cells accumulate more at the boundary such that overall larger clusters of equal cell types are formed. High signal dispersion leads to an ideal segregation of cells with N--G+ engulfing N+G-- cells. Cell sorting through differential adhesion provides an alternative explanation for the distribution cells. The minimization of the energy as a function of differential adhesion has already enabled the generation of both checkerboard and engulfing patterns in ICM organoid like structures \cite{Emily2007}. Simulations on embryo-like structures have also been able to replicate a spatial segregation of two different cell types \cite{Nissen2017}. To this date however, it has only been observed that the number of E-cadherin, the molecules involved in the adhesion of these cells, differs only after the cells have differentiated \cite{Filimonow2019}. Alternatively, Eph/Ephrin ligand receptor pairings have recently been proposed to account for differential adhesion \cite{Cang2021}. Another recent study has performed a quantitative analysis of the signaling range of FGF4 \cite{Raina2021}. It was found that FGF4 is able to migrate from a cell beyond its nearest neighbor in an exponentially decreasing manner. Together with the results of our model this supports the hypothesis of a global cell-cell communication as an alternative way for cells in the ICM organoid to organize.

\subsection*{Pair correlation functions enable the quantification of spatial structures within ICM organoids}
Pair correlation functions (PCFs) or alternatively radial distribution functions have traditionally been used to characterize the internal structure of fluids \cite{Barker1976}. In \cite{Binder2013}, a discrete grid-based PCF has been defined. This way, patterns that originate from a mixture of two different cell types have been quantified. In contrast to the grid-based approach, we used the cell graph to define our PCF. The PCFs confirmed quantitatively what we qualitatively observed in the visual representation of the organoids. Increasing the signal dispersion leads to a depletion of N+G-- pairs at large pairwise distances, whereas the number of N--G+ pairs rises. In simulations with larger cell numbers, the PCFs proved useful to highlight that the number of cells does not influence the global pattern of the organoids. Overall, the PCFs play a central role in characterizing patterns of any cell number allowing us to tackle the experimental data from \cite{Mathew2019}.

\subsection*{Simulations replicate patterns found in ICM organoids}
We found striking similarities between the results of our simulations and the 48h ICM organoid data provided in \cite{Mathew2019}. The range of possible PCFs for each individual organoid shows for the most part already a qualitative agreement with the results of our model. We provided interactive visualizations containing the pair correlation regions, the cell type proportions and a 3D model of the organoid (\blue{\url{https://schardts.github.io/Organoids48h}}). In three cases, we showed that our model was able to replicate the patterns found in the organoid up to a certain degree. The most pronounced mismatches of the PCFs of experimental data and simulations are found for low and large distances. We are certain that mismatches for small distances can be resolved by including cell division in our system. It has already been proven that cell division leads to a clustering of a given cell type \cite{Liebisch2020}. Since clusters are an accumulation of directly connected cells of equal type, the corresponding PCFs must increase for small distances especially at a distance of $1$. We expect the mismatch for higher distances to be harder to correct. For the largest distances, there are always the fewest cell pairs to be found, i.e. the impact of a cell pair at these distances on the PCF is much higher than anywhere else. In addition to that, the graph that connects the different cells has been generated using only a fixed cutoff distance for the edges. This in turn leads to some mistakenly connected cells at the boundary influencing the amount of cell pairs found at large distances.

\subsection*{Conclusion}
We have extended our established model by global cell-cell communication via cell signaling. The provided simulation results yield cell differentiation patterns that closely resemble the ones found in experiments on ICM organoid data. This renders global cell-cell communication a possible explanation for the spatial segregation of PrE and Epi cells in ICM organoids. So far, the global signaling obeys a simple mathematical rule depending on the number of cells it has to travel across in order to reach its destination. A detailed description of the signaling transport mechanism, possibly including diffusion and advection mechanism, provides room for further research. Additionally, signal production and uptake of cells play a crucial role in how effective different means of signal transport might be. Another perspective can be achieved by incorporating cell growth and cell division into the model and analyzing their effect on the resulting patterns. With this in mind, our study paves the way for numerous subsequent studies regarding signal-based pattern formation in the preimplantation embryo and other developmental systems.

\bibliography{mybib}

\end{document}

%% file: GRN.tex
\begin{tikzpicture}

\newcommand\W{1.2}

\begin{scope}[%
every node/.style={anchor=west, regular polygon, 
regular polygon sides=6,
draw,
minimum width=\W cm,
outer sep=0,shape border rotate=90,blur shadow={shadow blur steps=5}
},
      transform shape]
    \node[fill=white] (A) at (0*\W,0) {}; \fill (A) circle (0.08*\W);
    \node[fill=white] (B) at ($(A)+(0.433*\W,0)$) {}; \fill (B) circle (0.08*\W);
    \node[fill=white] (C) at ($(B)+(0.433*\W,0)$) {}; \fill (C) circle (0.08*\W);
    \node[fill=white] (D) at ($(C)+(0.433*\W,0)$) {}; \fill (D) circle (0.08*\W);
    \node[fill=white] (E) at ($(D)+(0.433*\W,0)$) {}; \fill (E) circle (0.08*\W);
    \node[fill=white] (F) at ($(E)+(0.433*\W,0)$) {}; \fill (F) circle (0.08*\W);
    \node[fill=white] (G) at ($(F)+(0.433*\W,0)$) {}; \fill (G) circle (0.08*\W);
\end{scope}

\begin{scope}[%
every node/.style={anchor=west, regular polygon, 
regular polygon sides=6,
draw,
minimum width=2*\W cm,
outer sep=0,shape border rotate=90,blur shadow={shadow blur steps=5}
},
      transform shape]
\node[fill=white] (I) at (-3.5*\W, -2*\W) {};
\end{scope}

\draw [->] ($(B)+(0,0.5*\W)$) to [out=140,in=40] ($(A)+(0,0.5*\W)$);
\draw [->] ($(C)+(0,0.5*\W)$) to [out=140,in=40] ($(A)+(0,0.5*\W)$);
\draw [->] ($(D)+(0,0.5*\W)$) to [out=140,in=40] ($(A)+(0,0.5*\W)$);
\draw [->] ($(E)+(0,0.5*\W)$) to [out=140,in=40] ($(A)+(0,0.5*\W)$);
\draw [->] ($(F)+(0,0.5*\W)$) to [out=140,in=40] ($(A)+(0,0.5*\W)$);
\draw [->] ($(G)+(0,0.5*\W)$) to [out=140,in=40] ($(A)+(0,0.5*\W)$);

\draw[dashed] (A) circle (0.6*\W); 
\draw[dashed] (I) circle (1.5*\W);
\draw[dashed] ($(A) + (-0.49613*0.6*\W, 0.86824*0.6*\W)$) -- ($(I) + (-0.3*1.5*\W, 0.954*1.5*\W)$);
\draw[dashed] ($(A) - (-0.49613*0.6*\W, 0.86824*0.6*\W)$) -- ($(I) - (-0.7*1.5*\W, 0.714*1.5*\W)$);

\node (NANOG) at ($(I) + (-0.5*\W, 0)$) {N};
\node (GATA6) at ($(I) + (0.5*\W, 0)$) {G};
\node (S) at ($(I) + (-0.5*\W, 1*\W)$) {S};

\draw[->] (S) -- (NANOG);
\draw[-|] ($(NANOG)+(0.25*\W,0.1*\W)$) -- ($(GATA6)+(-0.25*\W,0.1*\W)$);
\draw[|-] ($(NANOG)+(0.25*\W,-0.1*\W)$) -- ($(GATA6)+(-0.25*\W,-0.1*\W)$);

\node () at ($(A)+(0,-1.25*\W)$) {$q$};
\node () at ($(D)+(0.25*\W,-1.25*\W)$) {signal influence};

\node () at ($(A)+(0,-2*\W)$) {$0.1$};
\node () at ($(B)+(0,-2*\W)$) {$90\%$};
\node () at ($(C)+(0,-2*\W)$) {$9\%$};
\node () at ($(D)+(0,-2*\W)$) {$1\%$};
\node () at ($(E)+(0,-2*\W)$) {$0\%$};
\node () at ($(F)+(0,-2*\W)$) {$0\%$};
\node () at ($(G)+(0,-2*\W)$) {$0\%$};

\node () at ($(A)+(0,-2.5*\W)$) {$0.5$};
\node () at ($(B)+(0,-2.5*\W)$) {$51\%$};
\node () at ($(C)+(0,-2.5*\W)$) {$25\%$};
\node () at ($(D)+(0,-2.5*\W)$) {$13\%$};
\node () at ($(E)+(0,-2.5*\W)$) {$6\%$};
\node () at ($(F)+(0,-2.5*\W)$) {$3\%$};
\node () at ($(G)+(0,-2.5*\W)$) {$2\%$};

\node () at ($(A)+(0,-3*\W)$) {$0.9$};
\node () at ($(B)+(0,-3*\W)$) {$21\%$};
\node () at ($(C)+(0,-3*\W)$) {$19\%$};
\node () at ($(D)+(0,-3*\W)$) {$17\%$};
\node () at ($(E)+(0,-3*\W)$) {$16\%$};
\node () at ($(F)+(0,-3*\W)$) {$15\%$};
\node () at ($(G)+(0,-3*\W)$) {$13\%$};

\draw ($(A)+(0.433*\W,-0.25*\W)$) -- ($(A)+(0.433*\W,-0.75*\W)$);
\draw ($(B)+(0.433*\W,-0.25*\W)$) -- ($(B)+(0.433*\W,-0.75*\W)$);
\draw ($(C)+(0.433*\W,-0.25*\W)$) -- ($(C)+(0.433*\W,-0.75*\W)$);
\draw ($(D)+(0.433*\W,-0.25*\W)$) -- ($(D)+(0.433*\W,-0.75*\W)$);
\draw ($(E)+(0.433*\W,-0.25*\W)$) -- ($(E)+(0.433*\W,-0.75*\W)$);
\draw ($(F)+(0.433*\W,-0.25*\W)$) -- ($(F)+(0.433*\W,-0.75*\W)$);

\draw ($(A)+(0.433*\W,-1.75*\W)$) -- ($(A)+(0.433*\W,-3.25*\W)$);
\draw ($(B)+(0.433*\W,-1.75*\W)$) -- ($(B)+(0.433*\W,-3.25*\W)$);
\draw ($(C)+(0.433*\W,-1.75*\W)$) -- ($(C)+(0.433*\W,-3.25*\W)$);
\draw ($(D)+(0.433*\W,-1.75*\W)$) -- ($(D)+(0.433*\W,-3.25*\W)$);
\draw ($(E)+(0.433*\W,-1.75*\W)$) -- ($(E)+(0.433*\W,-3.25*\W)$);
\draw ($(F)+(0.433*\W,-1.75*\W)$) -- ($(F)+(0.433*\W,-3.25*\W)$);

\end{tikzpicture}

%% file: experimental.tex
\begin{tikzpicture}

\definecolor{NANOG}{rgb}{0.75,0,0.75}
\definecolor{GATA6}{rgb}{0,0.75,0.75}
\definecolor{DPDN}{rgb}{0.75,0.75,0.75}

\newcommand\W{0.8}

\begin{scope}[%
every node/.style={anchor=west, regular polygon, 
regular polygon sides=6,
draw,
minimum width=\W cm,
outer sep=0,shape border rotate=90,blur shadow={shadow blur steps=5}
},
      transform shape]
    \node[fill=NANOG] (A1) at (0*\W,0) {}; \fill (A1) circle (0.08*\W);
    \node[fill=DPDN] (B1) at ($(A1)+(0.433*\W,0)$) {}; \fill (B1) circle (0.08*\W);
    \node[fill=GATA6] (C1) at ($(B1)+(0.433*\W,0)$) {}; \fill (C1) circle (0.08*\W);
    
    \node[fill=GATA6] (A2) at ($(A1)-(2*0.433*\W,3/4*\W)$) {}; \fill (A2) circle (0.08*\W);
    \node[fill=GATA6] (B2) at ($(A2)+(0.433*\W,0)$) {}; \fill (B2) circle (0.08*\W);
    \node[fill=NANOG] (C2) at ($(B2)+(0.433*\W,0)$) {}; \fill (C2) circle (0.08*\W);
    \node[fill=DPDN] (D2) at ($(C2)+(0.433*\W,0)$) {}; \fill (D2) circle (0.08*\W);
    
    \node[fill=GATA6] (A3) at ($(A2)-(2*0.433*\W,3/4*\W)$) {}; \fill (A3) circle (0.08*\W);
    \node[fill=DPDN] (B3) at ($(A3)+(0.433*\W,0)$) {}; \fill (B3) circle (0.08*\W);
    \node[fill=NANOG] (C3) at ($(B3)+(0.433*\W,0)$) {}; \fill (C3) circle (0.08*\W);
    \node[fill=GATA6] (D3) at ($(C3)+(0.433*\W,0)$) {}; \fill (D3) circle (0.08*\W);
    \node[fill=NANOG] (E3) at ($(D3)+(0.433*\W,0)$) {}; \fill (E3) circle (0.08*\W);
    
    \node[fill=NANOG] (A4) at ($(A3)-(0,3/4*\W)$) {}; \fill (A4) circle (0.08*\W);
    \node[fill=GATA6] (B4) at ($(A4)+(0.433*\W,0)$) {}; \fill (B4) circle (0.08*\W);
    \node[fill=NANOG] (C4) at ($(B4)+(0.433*\W,0)$) {}; \fill (C4) circle (0.08*\W);
    \node[fill=NANOG] (D4) at ($(C4)+(0.433*\W,0)$) {}; \fill (D4) circle (0.08*\W);
    
    \node[fill=GATA6] (A5) at ($(A4)-(0,3/4*\W)$) {}; \fill (A5) circle (0.08*\W);
    \node[fill=NANOG] (B5) at ($(A5)+(0.433*\W,0)$) {}; \fill (B5) circle (0.08*\W);
    \node[fill=DPDN] (C5) at ($(B5)+(0.433*\W,0)$) {}; \fill (C5) circle (0.08*\W);
\end{scope}

\begin{scope}[%
every node/.style={anchor=west, regular polygon, 
regular polygon sides=6,
draw,
minimum width=\W cm,
outer sep=0,shape border rotate=90,blur shadow={shadow blur steps=5}
},
      transform shape]
    \node[fill=DPDN] (A1) at (5.65*\W,-1.5*\W) {}; \fill (A1) circle (0.08*\W);
    \node[fill=NANOG] (B1) at ($(A1)+(2*\W,1.5*\W)$) {}; \fill (B1) circle (0.08*\W);
    \node[fill=GATA6] (C1) at ($(A1)+(2*\W,-1.5*\W)$) {}; \fill (C1) circle (0.08*\W);
\end{scope}
\draw (A1) -- ($(A1)+(0,1.5*\W)$);
\draw[->] ($(A1)+(0,1.5*\W)$) -- (B1) node[midway, above] {$p$};
\draw (A1) -- ($(A1)+(0,-1.5*\W)$);
\draw[->] ($(A1)+(0,-1.5*\W)$) -- (C1) node[midway, below] {$1-p$};

\begin{scope}[%
every node/.style={anchor=west, regular polygon, 
regular polygon sides=6,
draw,
minimum width=\W cm,
outer sep=0,shape border rotate=90,blur shadow={shadow blur steps=5}
},
      transform shape]
    \node[fill=NANOG] (A1) at (12.5*\W,2.5) {}; \fill (A1) circle (0.08*\W);
    \node[fill=NANOG] (B1) at ($(A1)+(0.433*\W,0)$) {}; \fill (B1) circle (0.08*\W);
    \node[fill=GATA6] (C1) at ($(B1)+(0.433*\W,0)$) {}; \fill (C1) circle (0.08*\W);
    
    \node[fill=GATA6] (A2) at ($(A1)-(2*0.433*\W,3/4*\W)$) {}; \fill (A2) circle (0.08*\W);
    \node[fill=GATA6] (B2) at ($(A2)+(0.433*\W,0)$) {}; \fill (B2) circle (0.08*\W);
    \node[fill=NANOG] (C2) at ($(B2)+(0.433*\W,0)$) {}; \fill (C2) circle (0.08*\W);
    \node[fill=NANOG] (D2) at ($(C2)+(0.433*\W,0)$) {}; \fill (D2) circle (0.08*\W);
    
    \node[fill=GATA6] (A3) at ($(A2)-(2*0.433*\W,3/4*\W)$) {}; \fill (A3) circle (0.08*\W);
    \node[fill=NANOG] (B3) at ($(A3)+(0.433*\W,0)$) {}; \fill (B3) circle (0.08*\W);
    \node[fill=NANOG] (C3) at ($(B3)+(0.433*\W,0)$) {}; \fill (C3) circle (0.08*\W);
    \node[fill=GATA6] (D3) at ($(C3)+(0.433*\W,0)$) {}; \fill (D3) circle (0.08*\W);
    \node[fill=NANOG] (E3) at ($(D3)+(0.433*\W,0)$) {}; \fill (E3) circle (0.08*\W);
    
    \node[fill=NANOG] (A4) at ($(A3)-(0,3/4*\W)$) {}; \fill (A4) circle (0.08*\W);
    \node[fill=GATA6] (B4) at ($(A4)+(0.433*\W,0)$) {}; \fill (B4) circle (0.08*\W);
    \node[fill=NANOG] (C4) at ($(B4)+(0.433*\W,0)$) {}; \fill (C4) circle (0.08*\W);
    \node[fill=NANOG] (D4) at ($(C4)+(0.433*\W,0)$) {}; \fill (D4) circle (0.08*\W);
    
    \node[fill=GATA6] (A5) at ($(A4)-(0,3/4*\W)$) {}; \fill (A5) circle (0.08*\W);
    \node[fill=NANOG] (B5) at ($(A5)+(0.433*\W,0)$) {}; \fill (B5) circle (0.08*\W);
    \node[fill=GATA6] (C5) at ($(B5)+(0.433*\W,0)$) {}; \fill (C5) circle (0.08*\W);
\end{scope}

\begin{scope}[%
every node/.style={anchor=west, regular polygon, 
regular polygon sides=6,
draw,
minimum width=\W cm,
outer sep=0,shape border rotate=90,blur shadow={shadow blur steps=5}
},
      transform shape]
    \node[fill=NANOG] (A1) at (12.5*\W,-2.5) {}; \fill (A1) circle (0.08*\W);
    \node[fill=GATA6] (B1) at ($(A1)+(0.433*\W,0)$) {}; \fill (B1) circle (0.08*\W);
    \node[fill=GATA6] (C1) at ($(B1)+(0.433*\W,0)$) {}; \fill (C1) circle (0.08*\W);
    
    \node[fill=GATA6] (A2) at ($(A1)-(2*0.433*\W,3/4*\W)$) {}; \fill (A2) circle (0.08*\W);
    \node[fill=GATA6] (B2) at ($(A2)+(0.433*\W,0)$) {}; \fill (B2) circle (0.08*\W);
    \node[fill=NANOG] (C2) at ($(B2)+(0.433*\W,0)$) {}; \fill (C2) circle (0.08*\W);
    \node[fill=GATA6] (D2) at ($(C2)+(0.433*\W,0)$) {}; \fill (D2) circle (0.08*\W);
    
    \node[fill=GATA6] (A3) at ($(A2)-(2*0.433*\W,3/4*\W)$) {}; \fill (A3) circle (0.08*\W);
    \node[fill=GATA6] (B3) at ($(A3)+(0.433*\W,0)$) {}; \fill (B3) circle (0.08*\W);
    \node[fill=NANOG] (C3) at ($(B3)+(0.433*\W,0)$) {}; \fill (C3) circle (0.08*\W);
    \node[fill=GATA6] (D3) at ($(C3)+(0.433*\W,0)$) {}; \fill (D3) circle (0.08*\W);
    \node[fill=NANOG] (E3) at ($(D3)+(0.433*\W,0)$) {}; \fill (E3) circle (0.08*\W);
    
    \node[fill=NANOG] (A4) at ($(A3)-(0,3/4*\W)$) {}; \fill (A4) circle (0.08*\W);
    \node[fill=GATA6] (B4) at ($(A4)+(0.433*\W,0)$) {}; \fill (B4) circle (0.08*\W);
    \node[fill=NANOG] (C4) at ($(B4)+(0.433*\W,0)$) {}; \fill (C4) circle (0.08*\W);
    \node[fill=NANOG] (D4) at ($(C4)+(0.433*\W,0)$) {}; \fill (D4) circle (0.08*\W);
    
    \node[fill=GATA6] (A5) at ($(A4)-(0,3/4*\W)$) {}; \fill (A5) circle (0.08*\W);
    \node[fill=NANOG] (B5) at ($(A5)+(0.433*\W,0)$) {}; \fill (B5) circle (0.08*\W);
    \node[fill=GATA6] (C5) at ($(B5)+(0.433*\W,0)$) {}; \fill (C5) circle (0.08*\W);
\end{scope}

\fill (13.8*\W, -0.9) circle (0.065*\W);
\fill (13.8*\W, -1.25) circle (0.065*\W);
\fill (13.8*\W, -1.6) circle (0.065*\W);

\draw [decorate,decoration={brace,amplitude=10pt,mirror,raise=4pt}]
(16.2*\W,-6*\W) -- (16.2*\W,3*\W) node [black,midway,xshift=1.25cm,align=center] {$1000$ \\ patterns};

\draw[dashed] (4.6*\W , 2.5*\W) -- (4.6*\W , -5.5*\W);
\draw[dashed] (10.25*\W , 2.5*\W) -- (10.25*\W , -5.5*\W);
\end{tikzpicture}

%% file: patterns.tex
\begin{tikzpicture}
\newcommand\W{4}
\newcommand\width{0.4}
\centering
\node (a) at (-\W,\W)
    {\includegraphics[width=\width\textwidth]{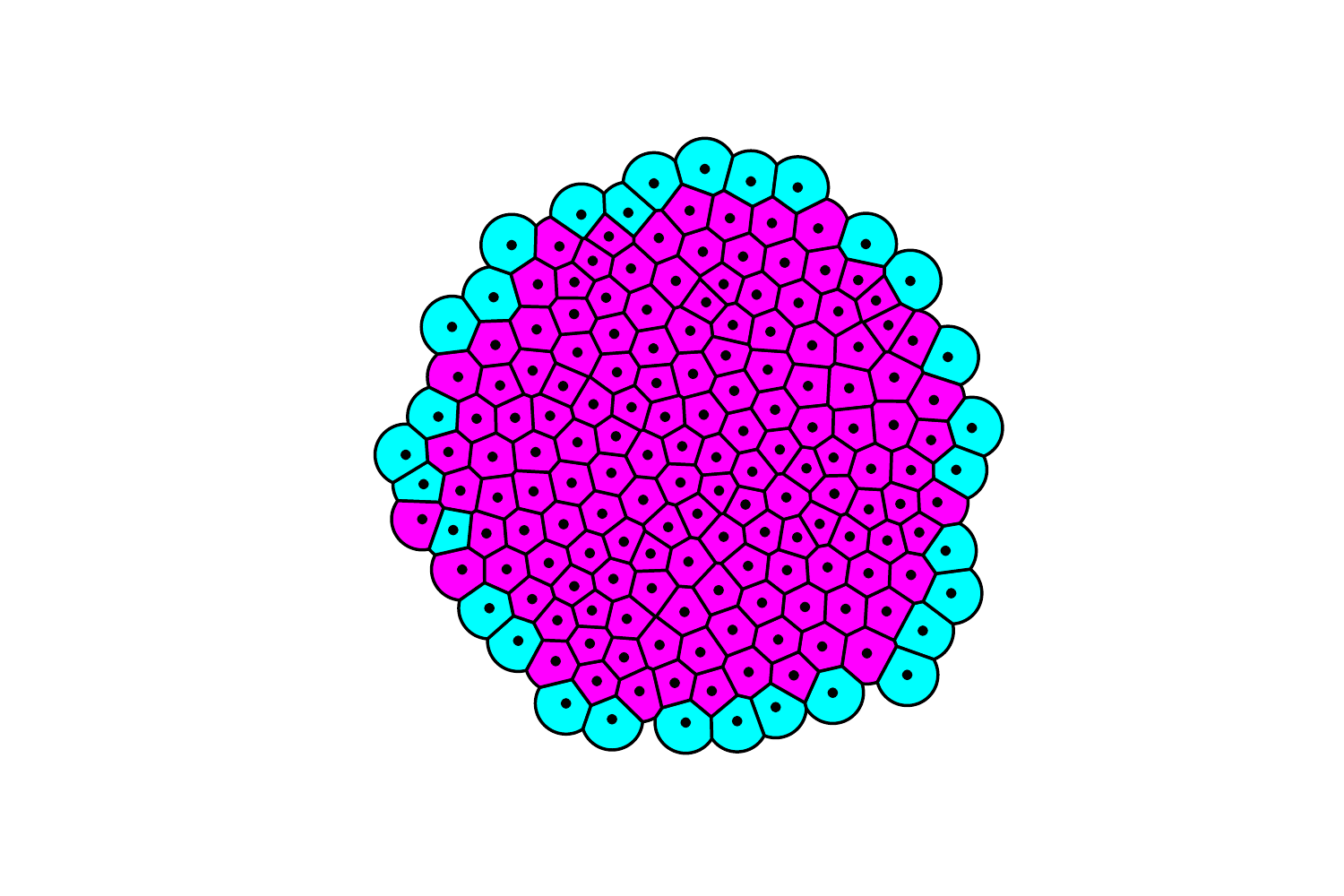}};
\node (b) at (0,\W)
    {\includegraphics[width=\width\textwidth]{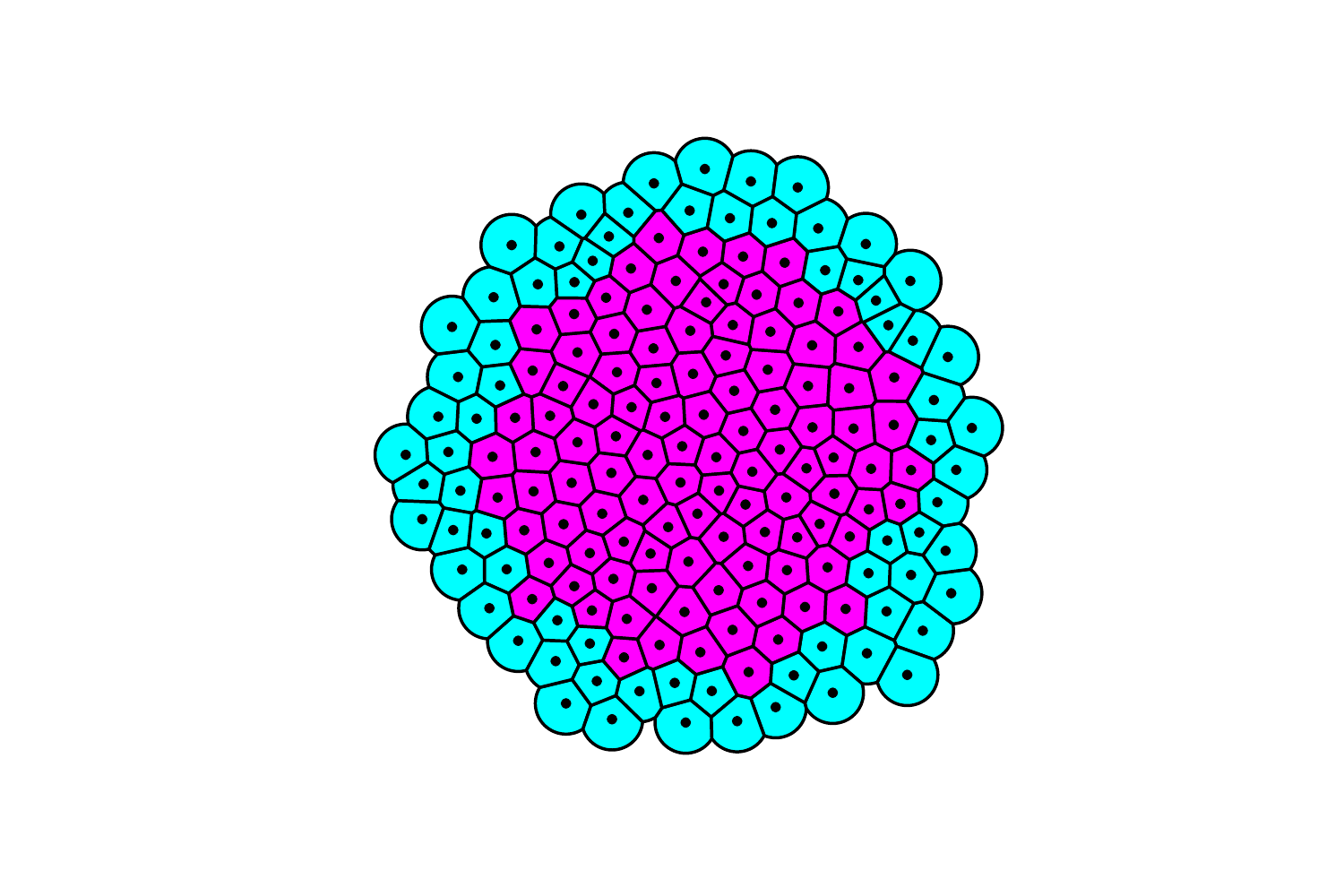}};
\node (b) at (\W,\W)
    {\includegraphics[width=\width\textwidth]{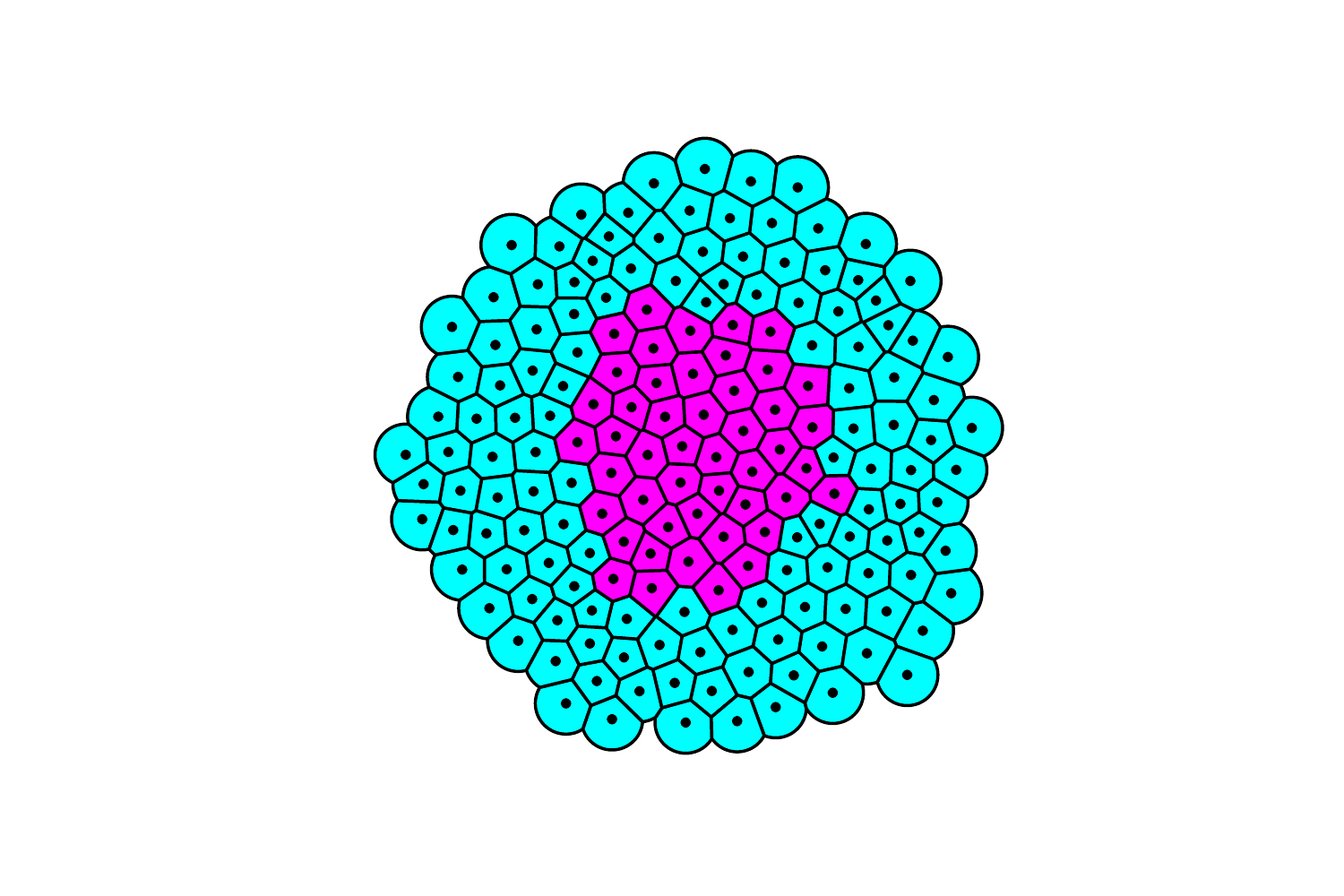}};
\node (a) at (-\W,0)
    {\includegraphics[width=\width\textwidth]{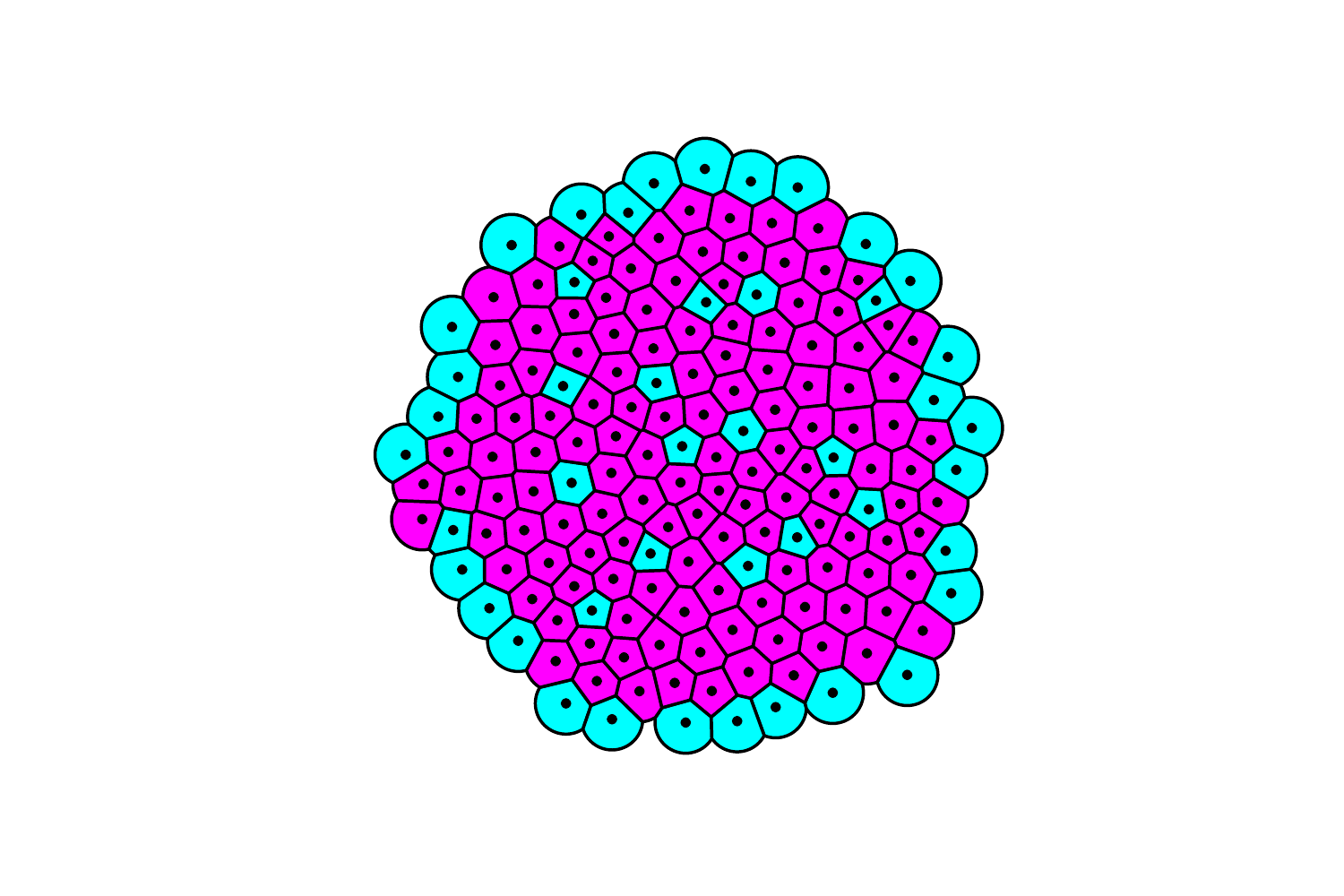}};
\node (b) at (0,0)
    {\includegraphics[width=\width\textwidth]{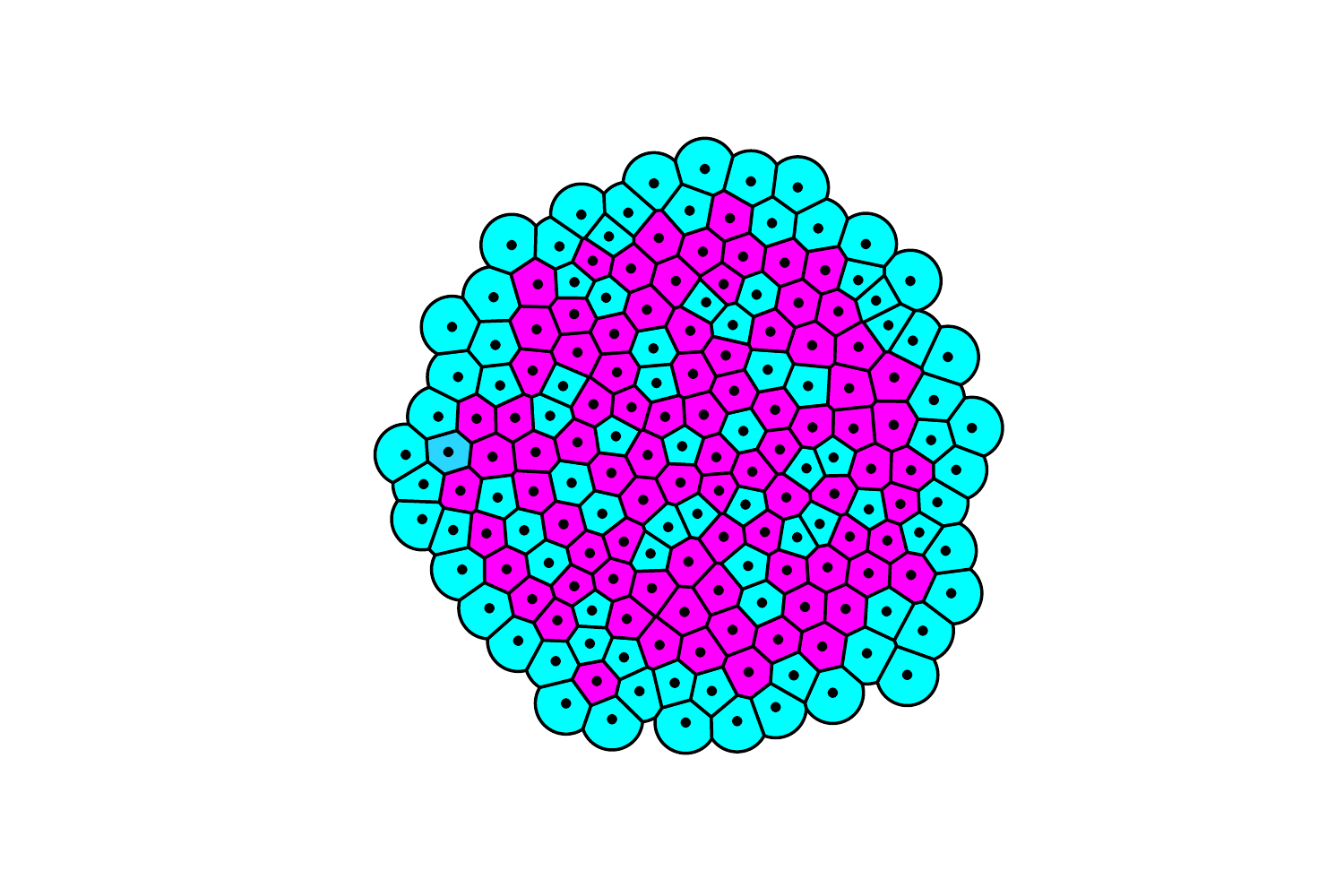}};
\node (b) at (\W,0)
    {\includegraphics[width=\width\textwidth]{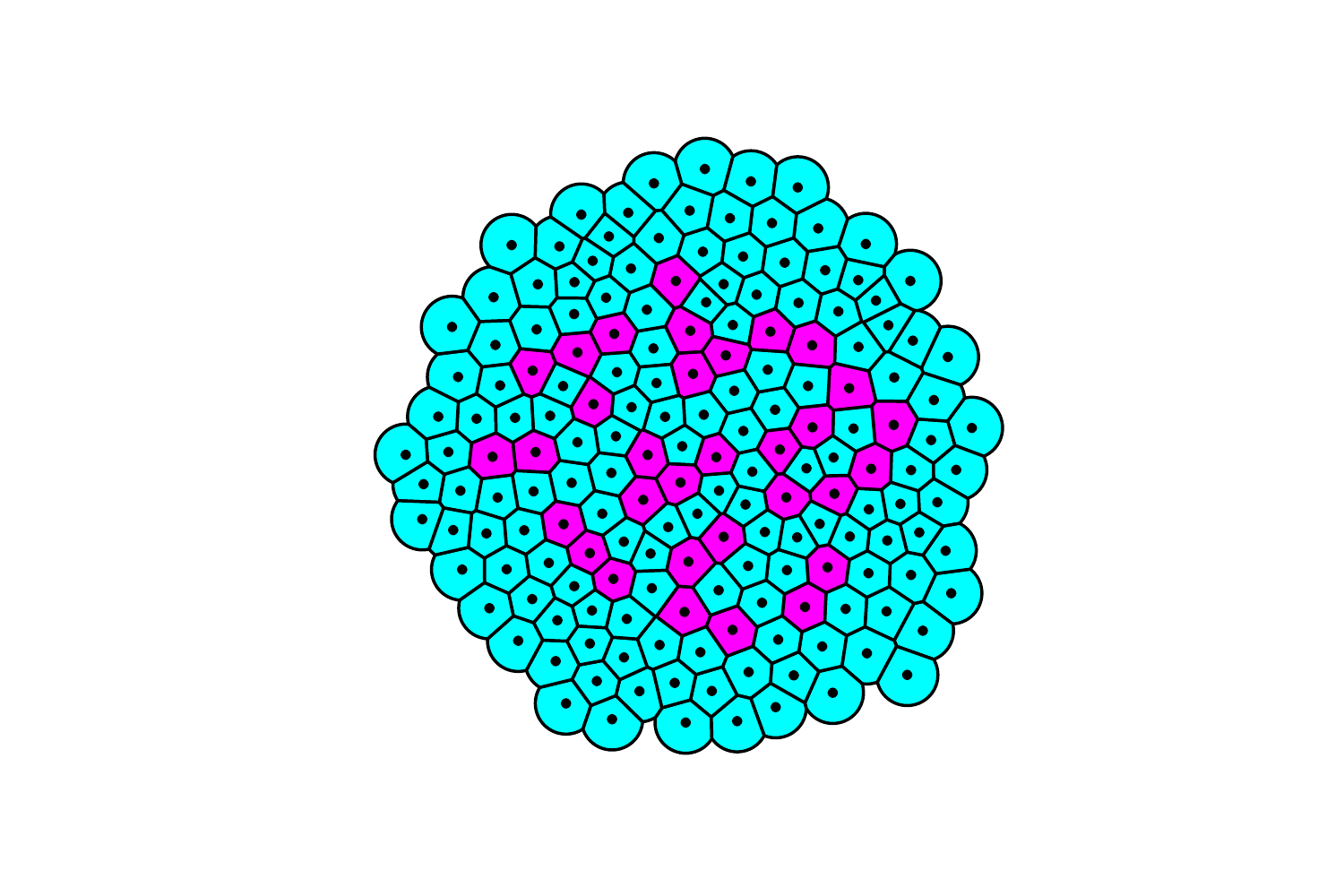}};
\node (a) at (-\W,-\W)
    {\includegraphics[width=\width\textwidth]{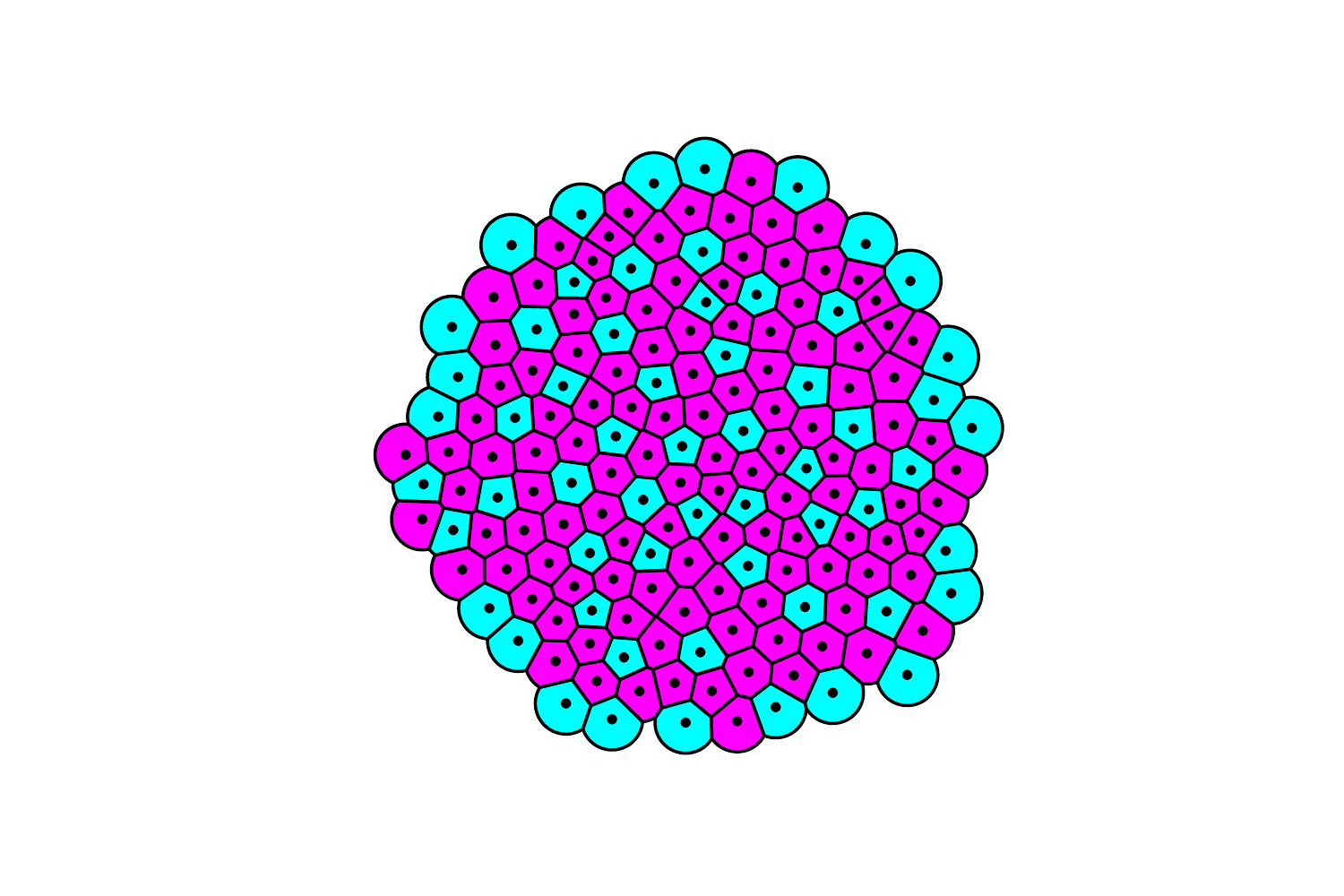}};
\node (b) at (0,-\W)
    {\includegraphics[width=\width\textwidth]{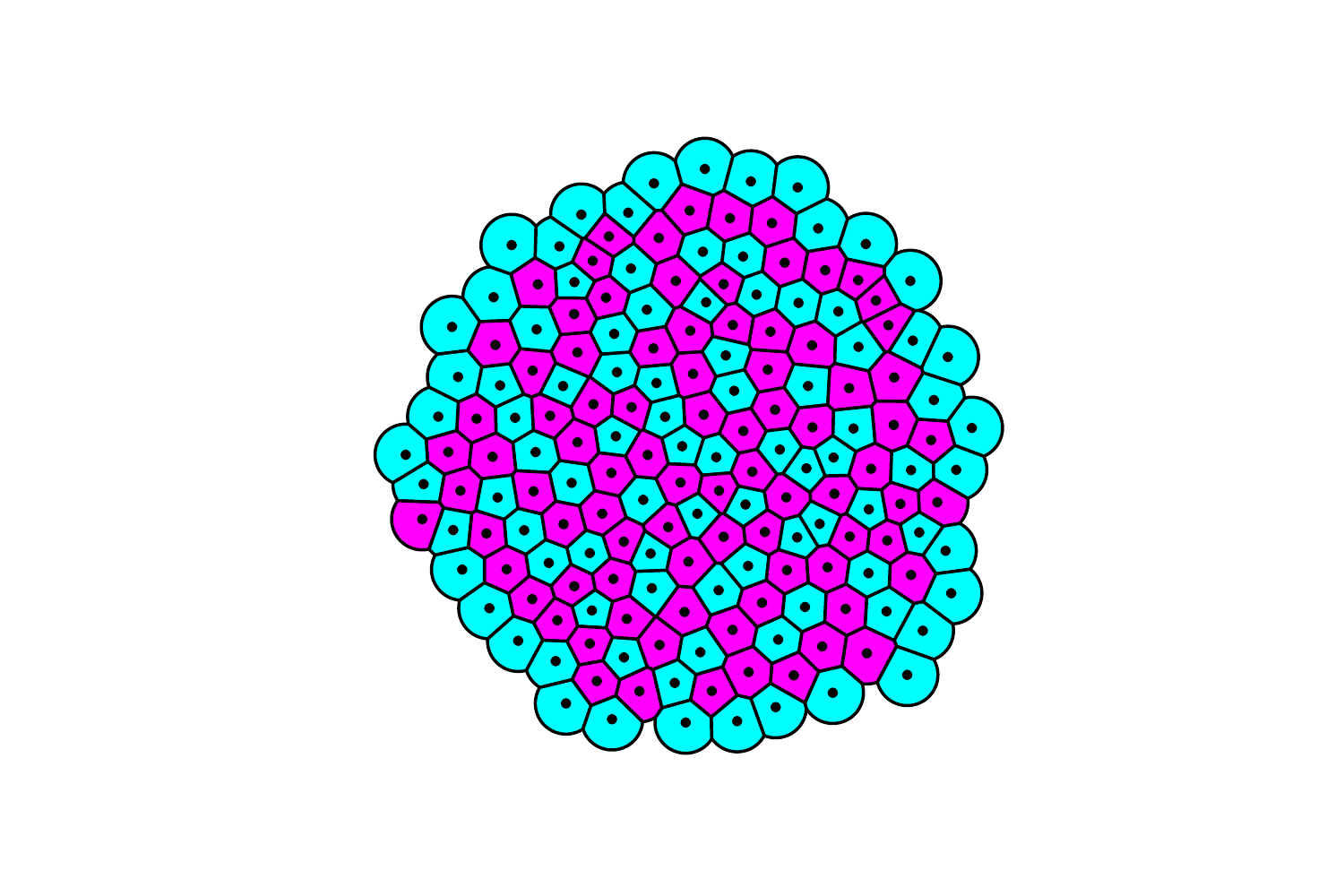}};
\node (b) at (\W,-\W)
    {\includegraphics[width=\width\textwidth]{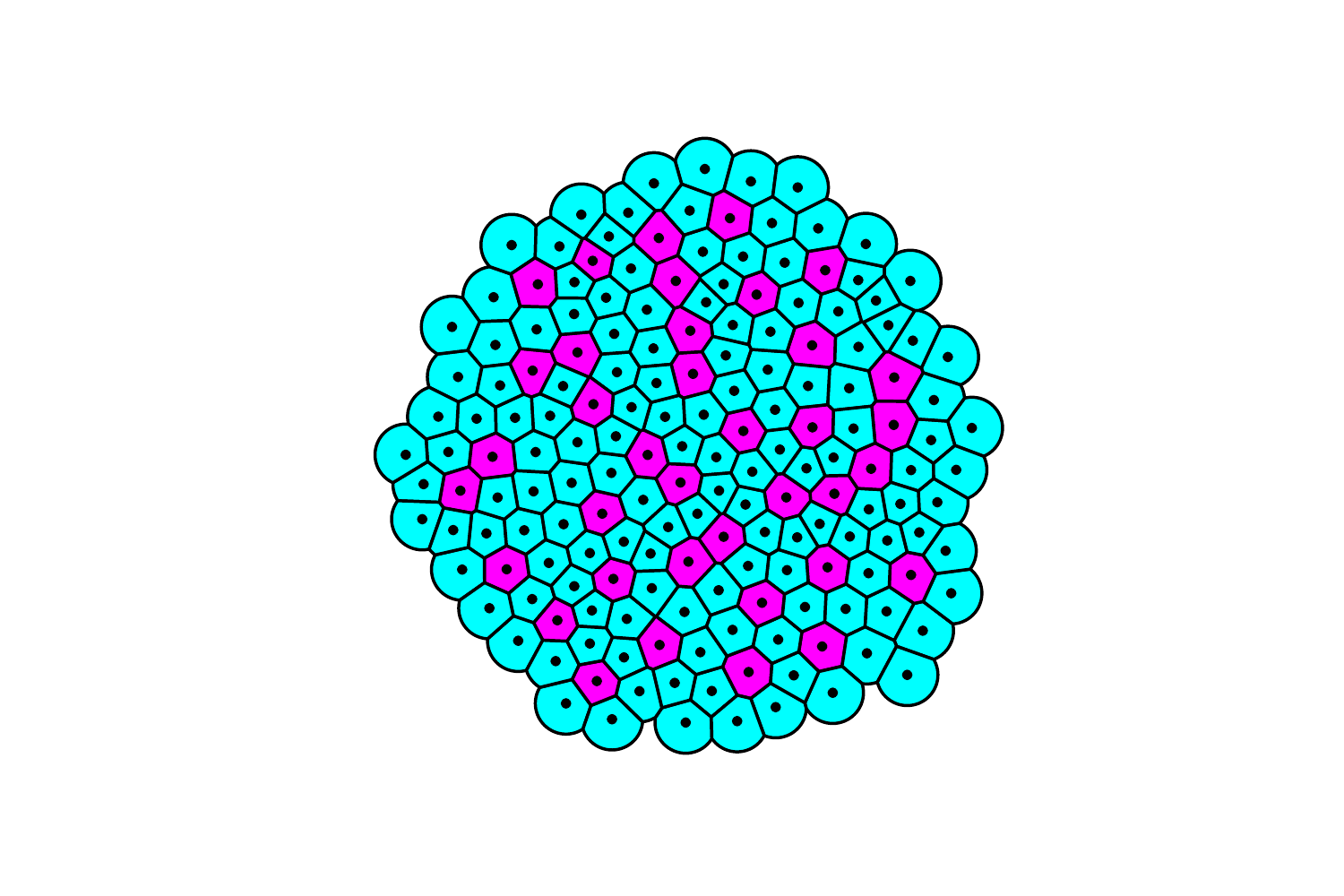}};
    
\draw[->] (-\W-0.5*\W,-\W-0.5*\W) -- (\W+0.5*\W,-\W-0.5*\W) node[below=0.2*\W] {$-\Delta\varepsilon_g$};
\draw[->] (-\W-0.5*\W,-\W-0.5*\W) -- (-\W-0.5*\W,+\W+0.5*\W) node[left=0.2*\W] {$q$};

\fill (-\W-0.5*\W,-\W) circle (0.1) node[left=0.1*\W] {$0.1$};
\fill (-\W-0.5*\W,0) circle (0.1) node[left=0.1*\W] {$0.5$};
\fill (-\W-0.5*\W,\W) circle (0.1) node[left=0.1*\W] {$0.9$};
\fill (-\W,-\W-0.5*\W) circle (0.1) node[below=0.1*\W] {$6.5$};
\fill (0,-\W-0.5*\W) circle (0.1) node[below=0.1*\W] {$7$};
\fill (\W,-\W-0.5*\W) circle (0.1) node[below=0.1*\W] {$7.5$};

\end{tikzpicture}

%% file: patterns_cell_number.tex
\begin{tikzpicture}
\newcommand\W{4}
\newcommand\width{0.4}
\centering
\node (a) at (-\W,\W)
    {\includegraphics[width=\width\textwidth]{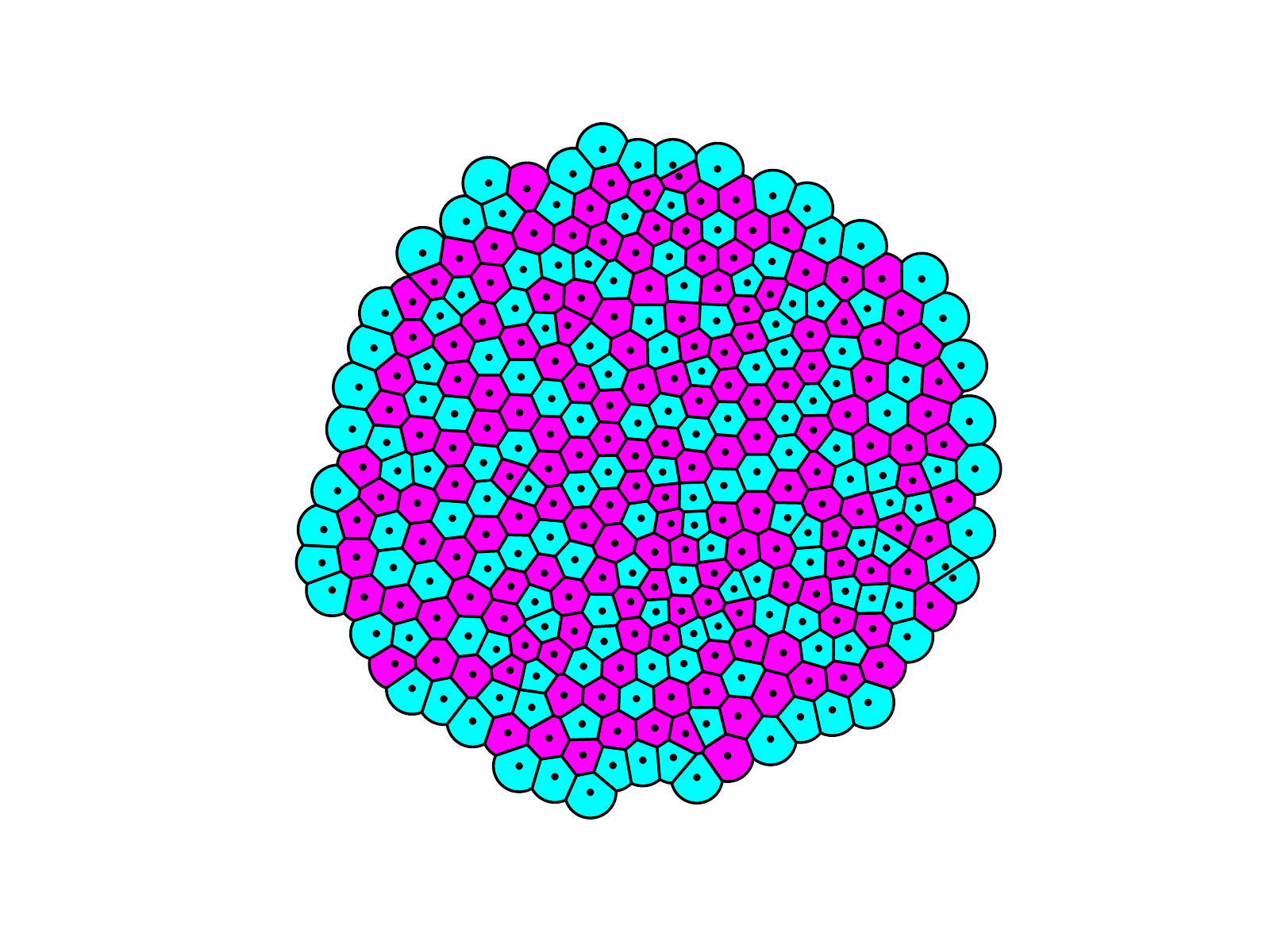}};
\node (b) at (0,\W)
    {\includegraphics[width=\width\textwidth]{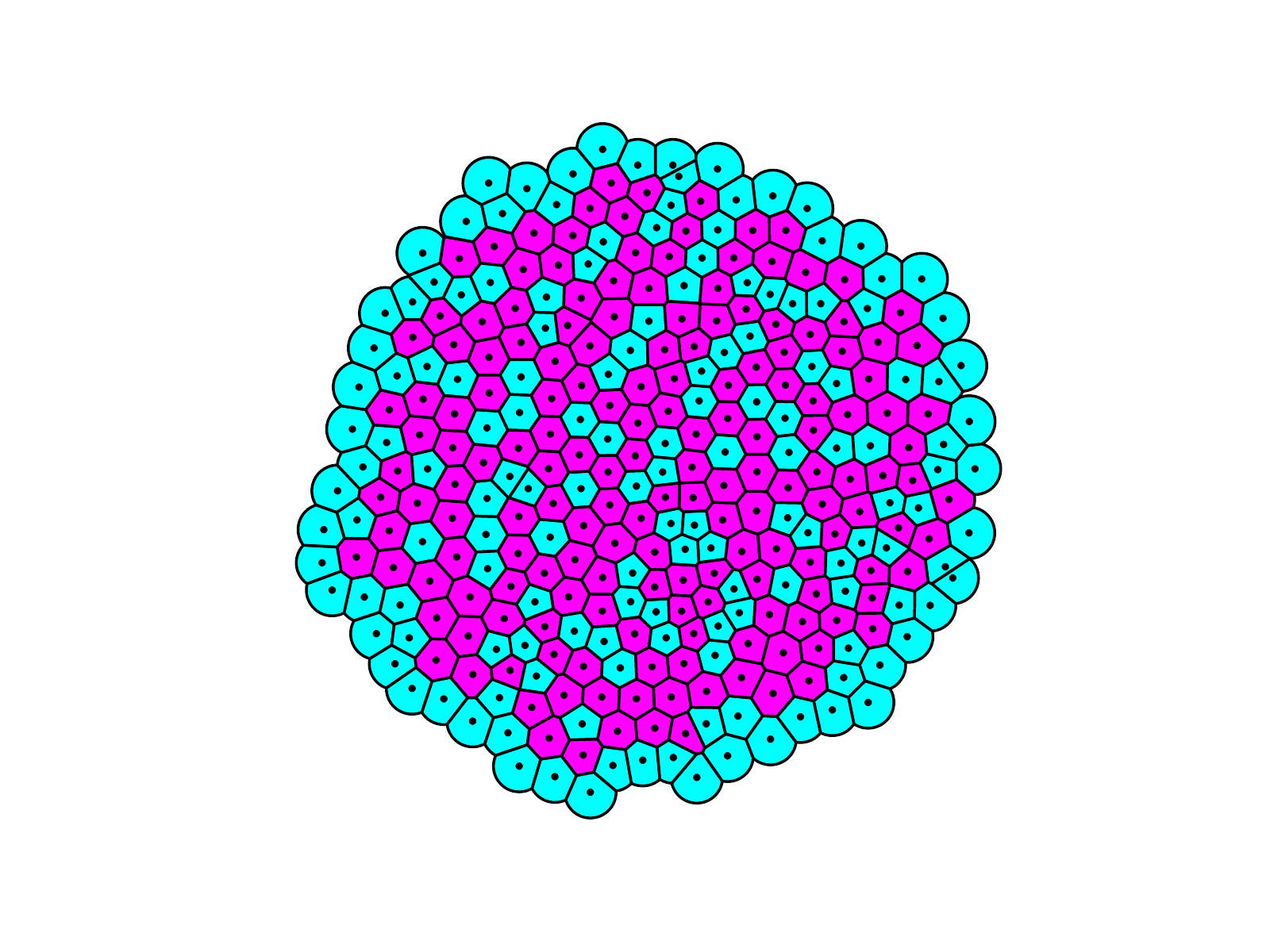}};
\node (b) at (\W,\W)
    {\includegraphics[width=\width\textwidth]{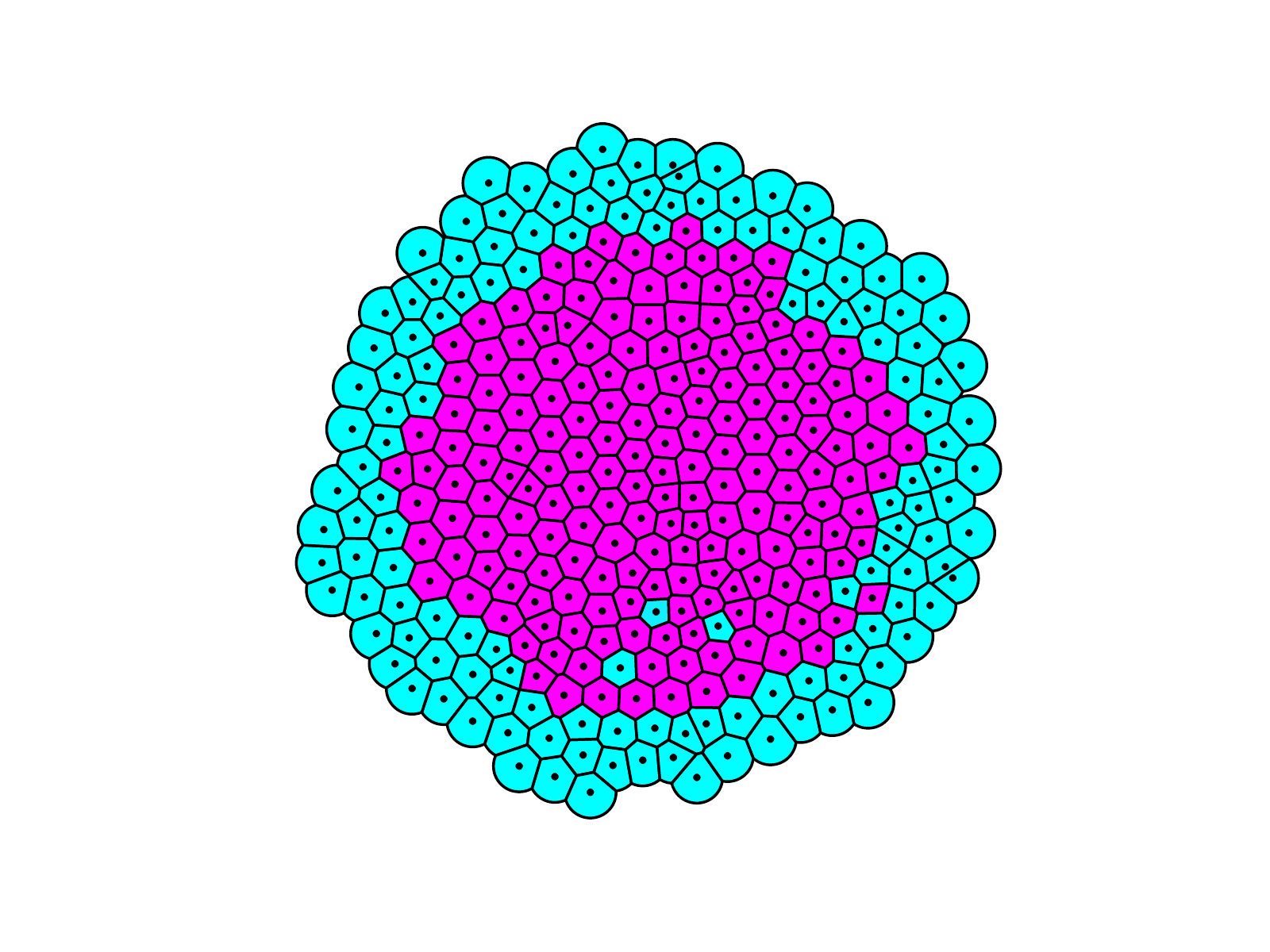}};
\node (a) at (-\W,0)
    {\includegraphics[width=\width\textwidth]{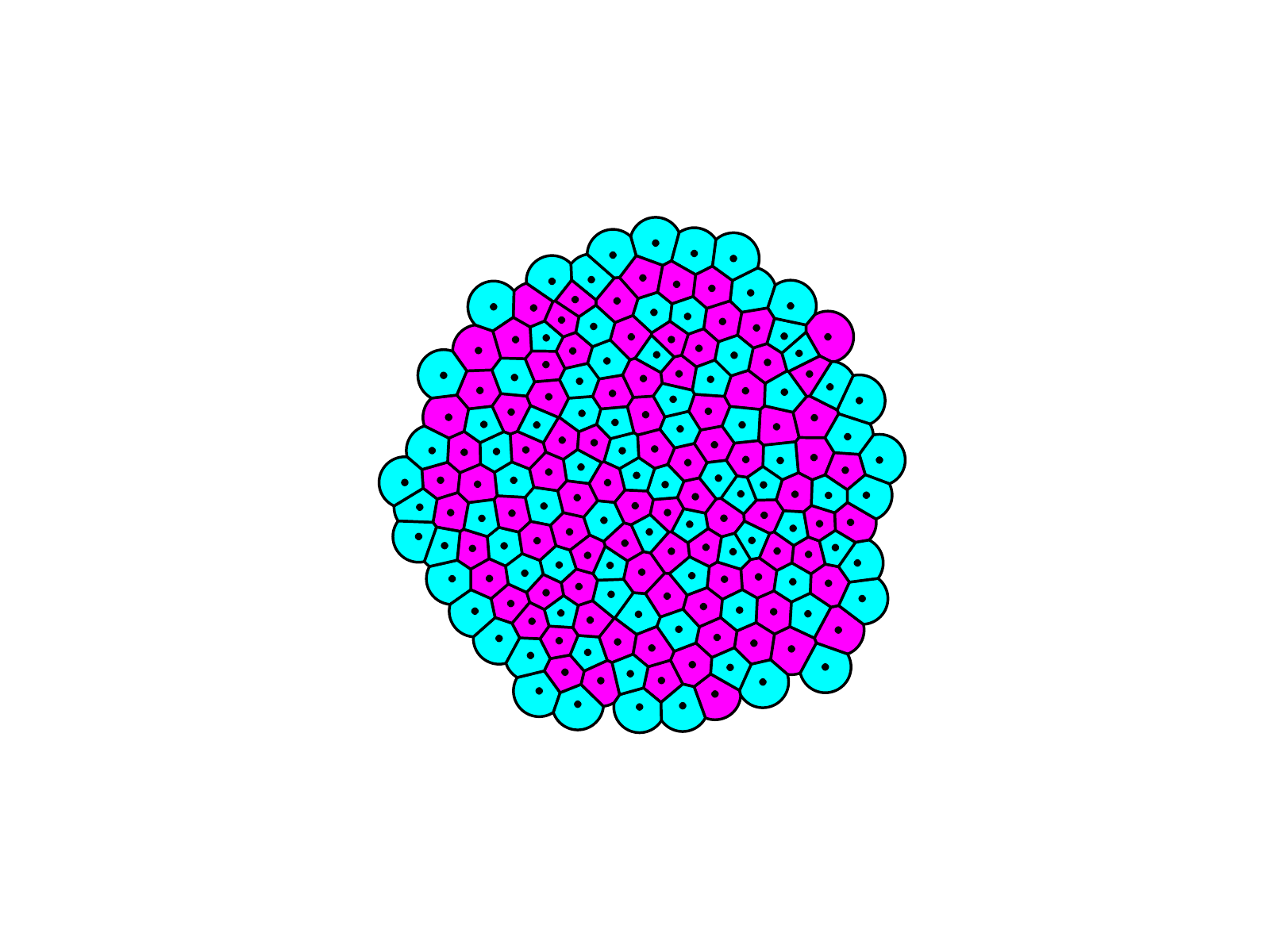}};
\node (b) at (0,0)
    {\includegraphics[width=\width\textwidth]{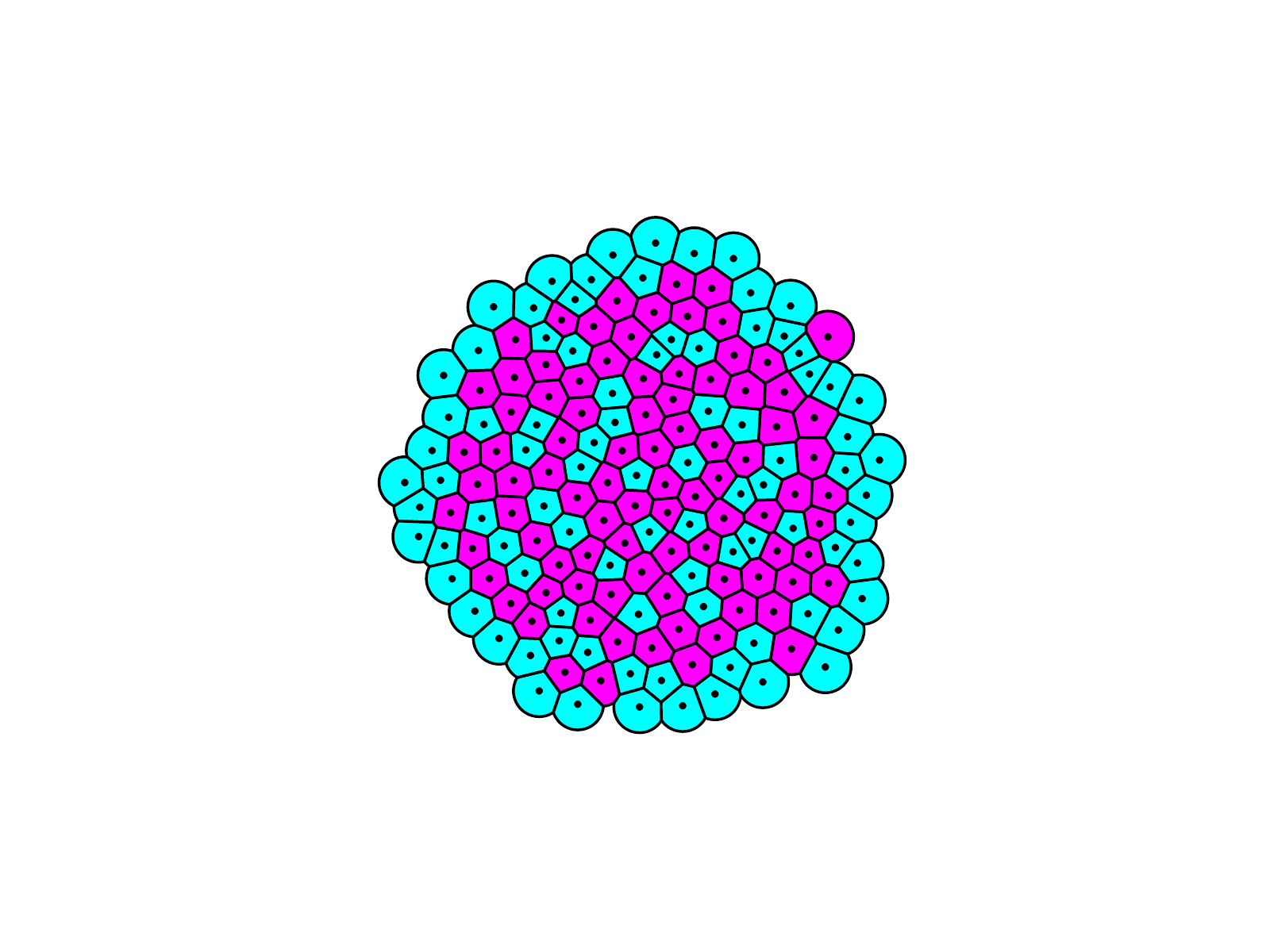}};
\node (b) at (\W,0)
    {\includegraphics[width=\width\textwidth]{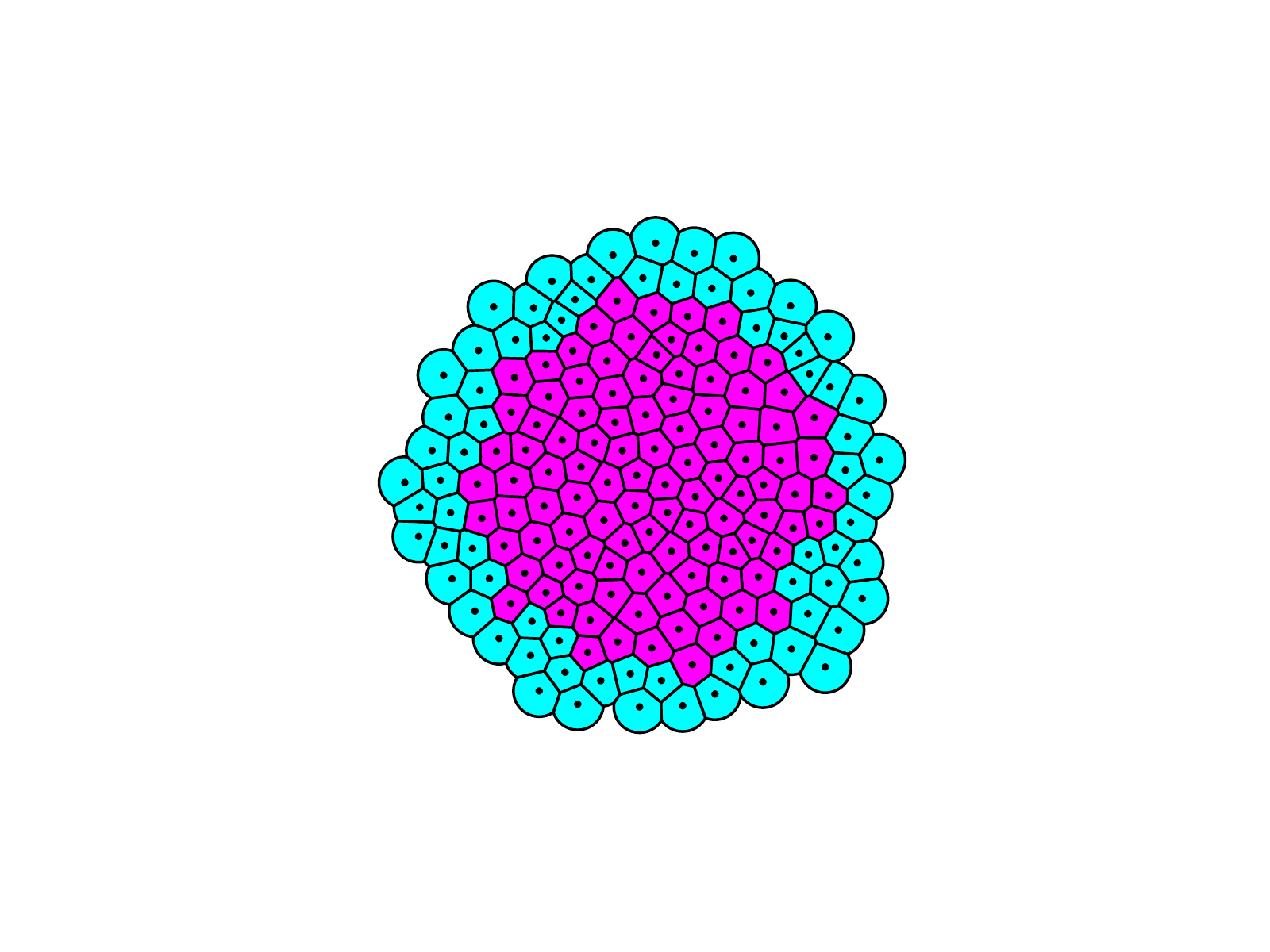}};
\node (a) at (-\W,-\W)
    {\includegraphics[width=\width\textwidth]{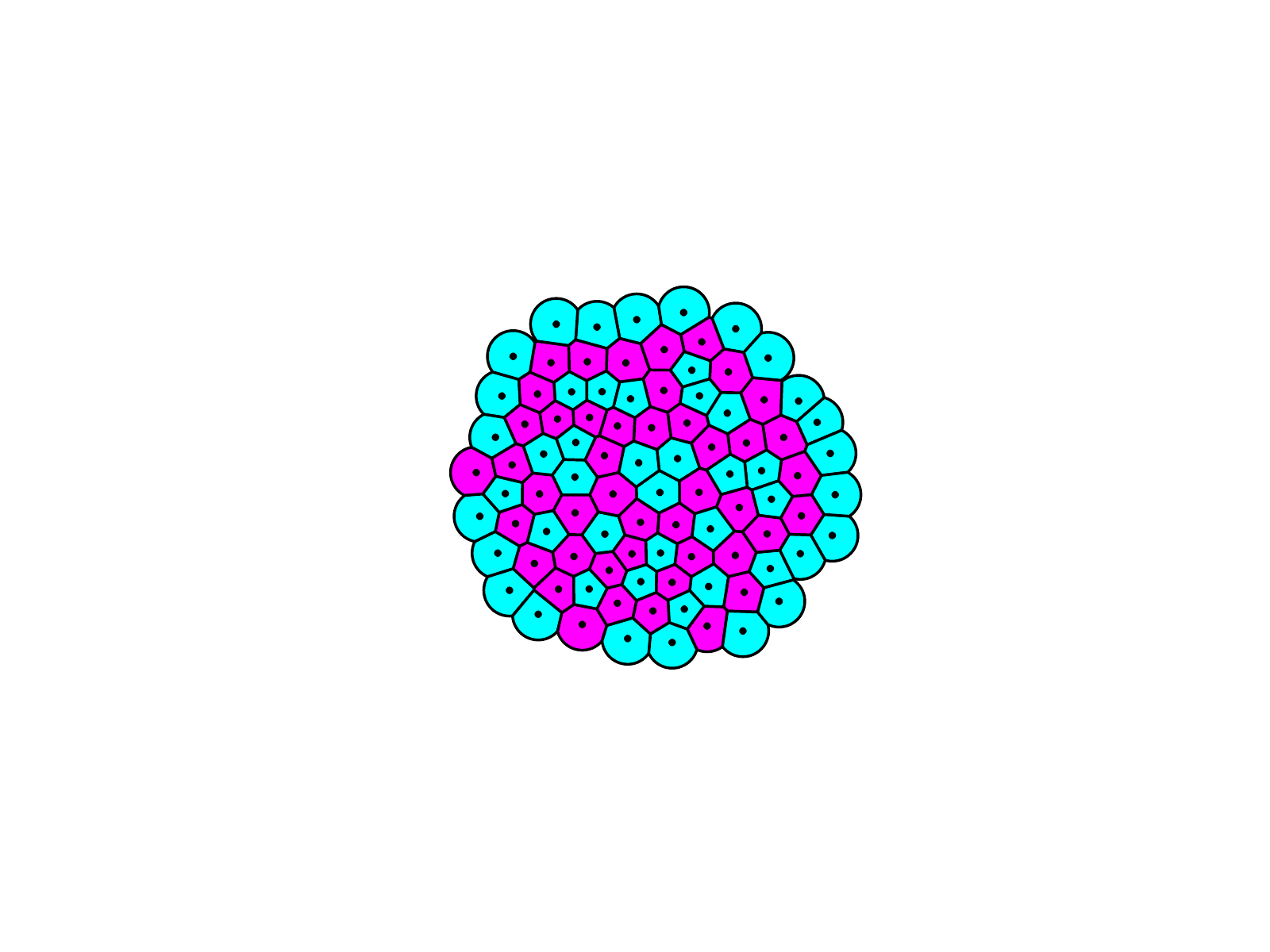}};
\node (b) at (0,-\W)
    {\includegraphics[width=\width\textwidth]{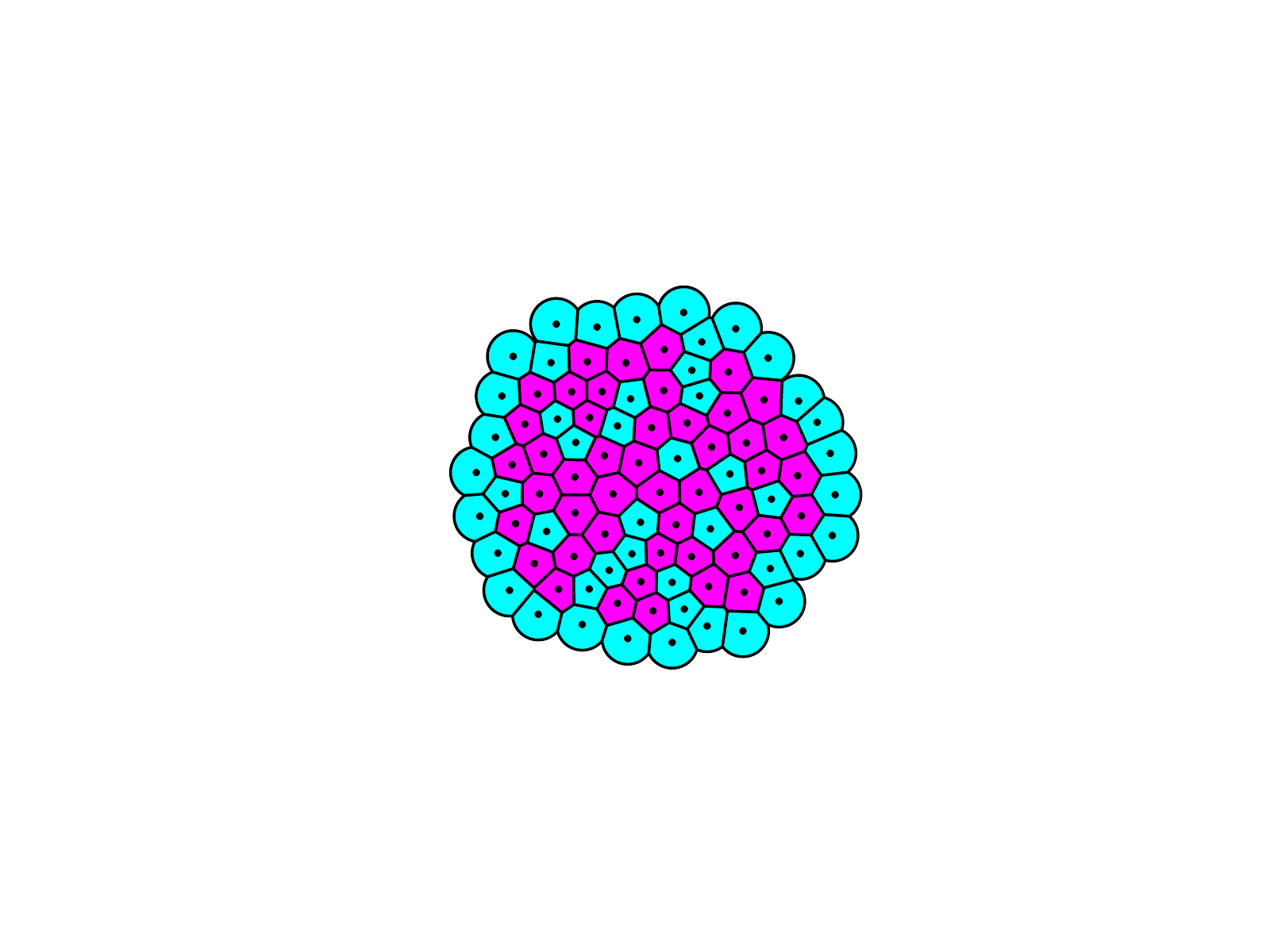}};
\node (b) at (\W,-\W)
    {\includegraphics[width=\width\textwidth]{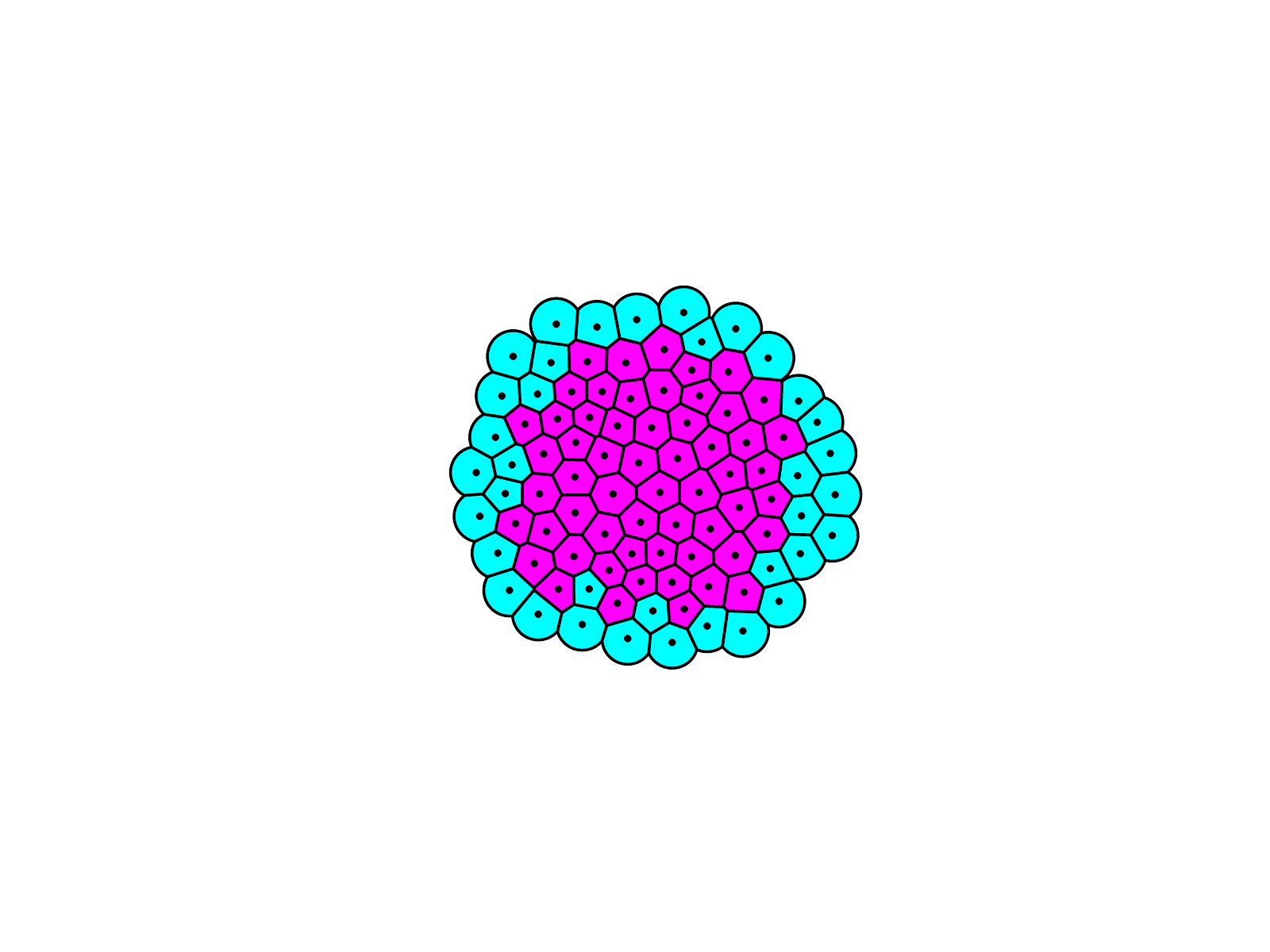}};
    
\draw[->] (-\W-0.5*\W,-\W-0.5*\W) -- (\W+0.5*\W,-\W-0.5*\W) node[below=0.2*\W] {$q$};
\draw[->] (-\W-0.5*\W,-\W-0.5*\W) -- (-\W-0.5*\W,+\W+0.5*\W) node[left=0.35*\W, rotate=90] {cell number};

\fill (-\W-0.5*\W,-\W) circle (0.1) node[left=0.1*\W] {$93$};
\fill (-\W-0.5*\W,0) circle (0.1) node[left=0.1*\W] {$177$};
\fill (-\W-0.5*\W,\W) circle (0.1) node[left=0.1*\W] {$324$};
\fill (-\W,-\W-0.5*\W) circle (0.1) node[below=0.1*\W] {$0.1$};
\fill (0,-\W-0.5*\W) circle (0.1) node[below=0.1*\W] {$0.5$};
\fill (\W,-\W-0.5*\W) circle (0.1) node[below=0.1*\W] {$0.9$};

\end{tikzpicture}